\documentclass[a4paper,11pt]{article}
\pdfoutput=1
\usepackage{jheppub}
\usepackage[T1]{fontenc}
\usepackage{graphicx}
\usepackage{epsfig}
\usepackage{rotating}
\usepackage{amssymb}
\usepackage{subfigure}
\usepackage{dsfont}
\usepackage{psfrag}
\usepackage{amsmath,euscript,array,mathrsfs}
\usepackage{bbold}
\usepackage{epsf}

\allowdisplaybreaks

\newcommand{\comm}[1]{} 

\newcommand{\beq}{\begin{equation}}
\newcommand{\eeq}{\end{equation}}
\newcommand{\beqs}{\begin{eqnarray}}
\newcommand{\eeqs}{\end{eqnarray}}

\newcommand{\lsim}{\lesssim}

\newcommand{\Tr}{{\rm Tr}}

\def\hbar{\hspace{0pt}\raisebox{1pt}{$-$} \hspace{-7pt} h}

\def\di{\mbox{d}}
\def\r{\rho}

\newcommand{\be}{\begin{equation}}
\newcommand{\ee}{\end{equation}}
\newcommand{\bea}{\begin{eqnarray}}
\newcommand{\eea}{\end{eqnarray}}

\def\lbldef#1#2{\expandafter\gdef\csname #1\endcsname {#2}}

\def\href#1#2{#2}

\newcommand{\ber}{\begin{eqnarray}}
\newcommand{\eer}{\end{eqnarray}}

\newcommand{\beqar}{\begin{eqnarray}}

\newcommand{\eeqar}{\end{eqnarray}}

\newcommand{\dsl}{\kern.06em\hbox{\raise.15ex\hbox{$/$}\kern-.56em\hbox{$\partial$}}}

\newcommand{\eeqarr}{\end{eqnarray}}
\newcommand{\ZZ}{{\rm \kern 0.275em Z \kern -0.92em Z}\;}

\def\CC{{\mathchoice
{\rm C\mkern-8mu\vrule height1.45ex depth-.05ex
width.05em\mkern9mu\kern-.05em}
{\rm C\mkern-8mu\vrule height1.45ex depth-.05ex
width.05em\mkern9mu\kern-.05em}
{\rm C\mkern-8mu\vrule height1ex depth-.07ex
width.035em\mkern9mu\kern-.035em}
{\rm C\mkern-8mu\vrule height.65ex depth-.1ex
width.025em\mkern8mu\kern-.025em}}}

\def\RR{{\rm I\kern-1.6pt {\rm R}}}

\def\ZZ{{\rm Z}\kern-3.8pt {\rm Z} \kern2pt}
\def\IB{\relax{\rm I\kern-.18em B}}
\def\ID{\relax{\rm I\kern-.18em D}}
\def\II{\relax{\rm I\kern-.18em I}}
\def\IP{\relax{\rm I\kern-.18em P}}

\newcommand{\bear}{\begin{eqnarray}}
\newcommand{\eear}{\end{eqnarray}}

\def\a{\alpha}
\def\b{\beta}

\def\d{\delta}

\def\r{\rho}

\def\z{\zeta}

\def\tt{\tilde{\theta}}

\def\6{\partial}

\def\bea{\begin{eqnarray}}
\def\eea{\end{eqnarray}}

\def\beqx{\begin{displaymath}}
\def\eeqx{\end{displaymath}}

\newcommand{\bmat}{\left(\begin{array}}
\newcommand{\emat}{\end{array}\right)}

\def\a{\alpha}
\def\b{\beta}

\def\d{\delta}

\def\r{\rho}

\def\z{\zeta}

\def\bo{{\raise-.3ex\hbox{\large$\Box$}}}               
\def\face{{\raise.2ex\hbox{$\displaystyle \bigodot$}\mskip-2.2mu \llap {$\ddot
        \smile$}}}                                   
\def\>{\rangle}                                      
\def\<{\langle}                                      

\def\leftrightarrowfill{$\mathsurround=0pt \mathord\leftarrow \mkern-6mu
        \cleaders\hbox{$\mkern-2mu \mathord- \mkern-2mu$}\hfill
        \mkern-6mu \mathord\rightarrow$}        
\def\dvec#1{\vbox{\ialign{##\crcr
        \leftrightarrowfill\crcr\noalign{\kern-1pt\nointerlineskip}
        $\hfil\displaystyle{#1}\hfil$\crcr}}}           
\def\Tr{{\rm Tr \,}}                                    

\def\-{\hphantom{-}}
\newcommand{\dd}{\mbox{d}}

\title{
Towards top-down holographic composite Higgs: 
 minimal coset from maximal supergravity
}

\author[a]{Daniel Elander}
\affiliation[a]{Laboratoire Charles Coulomb (L2C), University of Montpellier, CNRS, Montpellier, France}

\author[b]{and Maurizio Piai}
\affiliation[b]{Department of Physics, College of Science, Swansea University,
Singleton Park, SA2 8PP, Swansea, Wales, UK}

\date{\today}

\abstract{
Within the context of top-down holography, we study a one-parameter family of regular background solutions of maximal gauged supergravity in seven dimensions, dimensionally reduced on a 2-torus. The dual, four-dimensional confining field theory realises the global (spontaneous as well as explicit) symmetry breaking pattern $SO(5)\rightarrow SO(4)$. We compute the complete mass spectrum for the fluctuations of the 128 bosonic degrees of freedom of the five-dimensional gravity theory, which correspond to scalar, pseudoscalar, vector, axial-vector, and tensor bound states of the dual field theory, and includes particles with exotic $SO(4)$ quantum numbers. We confirm the existence of tachyonic instabilities near the boundaries of the parameter space.

We discuss the interplay between explicit and spontaneous symmetry breaking. The $SO(5)/SO(4)$ coset might provide a first step towards the realisation of a calculable framework and ultraviolet completion of minimal composite Higgs models, if the four pseudo-Nambu-Goldstone bosons are identified with the real components of the Higgs doublet in the standard model (SM), and a subgroup of $SO(4)$ with the $SU(2)\times U(1)$ SM gauge group. We exhibit an example with an additional localised boundary term that mimics the effect of a weakly-coupled external sector.
}

\begin{document}
\maketitle

\newpage
\allowdisplaybreaks

\section{Introduction}
\label{Sec:Introduction}

After the discovery of the Higgs boson~\cite{Aad:2012tfa,Chatrchyan:2012xdj}, and in preparation for
the restart of the experimental programme of the Large Hadron Collider (LHC), 
renewed attention in the literature has focused  on 
Composite Higgs 
Models (CHMs)~\cite{Kaplan:1983fs,Georgi:1984af,Dugan:1984hq}.
In such scenarios, the standard-model (SM) Higgs doublet scalar fields would arise as
pseudo-Nambu-Goldstone Bosons (pNGBs)  in the low-energy 
 Effective Field Theory (EFT) description of
a more fundamental theory.
It is usually  assumed that the underlying dynamics be strongly coupled,
leading to the formation of  infinitely many bound states.
An overview of the field is provided by the reviews in Refs.~\cite{Panico:2015jxa,Witzel:2019jbe,Cacciapaglia:2020kgq}, 
and by the useful summary tables in Refs.~\cite{Ferretti:2013kya,Ferretti:2016upr,Cacciapaglia:2019bqz}
(see also the selection of useful papers in Refs.~\cite{Contino:2003ve,Agashe:2004rs,
Katz:2005au,Agashe:2005dk,Agashe:2006at,Contino:2006qr,Barbieri:2007bh,Lodone:2008yy,Falkowski:2008fz,Gripaios:2009pe,Mrazek:2011iu,Contino:2011np,
Marzocca:2012zn,Grojean:2013qca,Barnard:2013zea,Cacciapaglia:2014uja,Ferretti:2014qta,Arbey:2015exa,vonGersdorff:2015fta,Cacciapaglia:2015eqa,Feruglio:2016zvt,DeGrand:2016pgq,Fichet:2016xvs,Galloway:2016fuo,Agugliaro:2016clv,Belyaev:2016ftv,Bizot:2016zyu,Csaki:2017cep,
Chala:2017sjk,Golterman:2017vdj,Csaki:2017jby,Alanne:2017rrs,Alanne:2017ymh,Sannino:2017utc,
Alanne:2018wtp,Bizot:2018tds,Cai:2018tet,Agugliaro:2018vsu,
Cacciapaglia:2018avr,Gertov:2019yqo,Ayyar:2019exp,
Cacciapaglia:2019ixa,BuarqueFranzosi:2019eee,Cacciapaglia:2019dsq,Cacciapaglia:2020vyf},
 and references therein).

We want to study the strong-coupling dynamics underlying CHMs,
and address calculability (from first principles), in particular for
 observable quantities of phenomenological relevance,
such as the mass spectra of bound states.
One possible approach to this endeavour is provided by lattice gauge theory.
There exist lattice studies of $SU(2)$~\cite{Hietanen:2014xca,Detmold:2014kba,
Arthur:2016dir,Arthur:2016ozw,Pica:2016zst,Lee:2017uvl,Drach:2017btk,Drach:2020wux}
and $Sp(4)$~\cite{Bennett:2017kga,Bennett:2019jzz,Bennett:2019cxd} 
gauge theories that explore the fundamental origin of CHMs
based on the $SU(4)/Sp(4)$ coset~\cite{Barnard:2013zea}, as well as
lattice explorations with $SU(4)$ gauge group~\cite{Ayyar:2017qdf,Ayyar:2018zuk,Ayyar:2018ppa,Ayyar:2018glg,
Cossu:2019hse} aimed at capturing some of the dynamical features of models based on the $SU(5)/SO(5)$ coset.
 Ref.~\cite{Appelquist:2020bqj} shows  that salient features
of CHMs based upon  $SU(N_f)\times SU(N_f)/SU(N_f)$ cosets (see also Refs.~\cite{Vecchi:2015fma,
Ma:2015gra,BuarqueFranzosi:2018eaj}) are captured by the lattice studies of the $SU(3)$ gauge theory
with $N_f=8$ Dirac fermions in the fundamental representation~\cite{Aoki:2014oha,Appelquist:2016viq,Aoki:2016wnc,
Gasbarro:2017fmi,Appelquist:2018yqe}, via the application of
 the dilaton EFT~\cite{Matsuzaki:2013eva,
Golterman:2016lsd,Kasai:2016ifi,
Hansen:2016fri, Golterman:2016cdd,
Appelquist:2017wcg,Appelquist:2017vyy,Golterman:2018mfm,
Cata:2019edh,Appelquist:2019lgk,Golterman:2020tdq}.\footnote{The basic principles  underlying the dilaton EFT date
 far back in the literature~\cite{Migdal:1982jp,Coleman:1985rnk},
and have previously been applied in the context of 
dynamical symmetry breaking~\cite{Leung:1985sn,
Bardeen:1985sm,Yamawaki:1985zg},
as well as  in the more recent development of 
dilaton-Higgs models (see, for example, Refs.~\cite{Goldberger:2007zk,
Hong:2004td,Dietrich:2005jn,Hashimoto:2010nw,
Appelquist:2010gy,Vecchi:2010gj,Elander:2012fk,Chacko:2012sy,Bellazzini:2012vz,
Abe:2012eu,Eichten:2012qb,Bellazzini:2013fga,Hernandez-Leon:2017kea} and references therein).}

The $SO(5)/SO(4)$ minimal coset has dimension four and includes  custodial symmetry.
It is natural to identify the pNGBs, transforming as $4$ of $SO(4)$, with the Higgs 
doublet in the standard model~\cite{Agashe:2004rs}. Yet, this coset 
 is conspicuous by its absence in the list of current lattice explorations, as summarised 
in the previous paragraph.
It is not trivial to provide a dynamical origin 
for CHMs based on this coset,
in terms of familiar gauge theories with fermionic matter.
An example is provided by Ref.~\cite{Caracciolo:2012je}, which beautifully 
exploits supersymmetry and Seiberg duality, though subject to the usual limitations in terms of calculability within this approach.

A radically different approach to address calculability in strongly coupled gauge theories
adopts gauge-gravity 
dualities (holography)~\cite{Maldacena:1997re,Gubser:1998bc,Witten:1998qj,Aharony:1999ti}.
The large-$N$ limit of special  strongly-coupled gauge theories
is equivalent to a weakly-coupled theory of gravity in extra dimensions, 
amenable to perturbative treatment. Interesting observables are recovered by
applying the dictionary of the correspondence---which requires the implementation
  of holographic renormalisation~\cite{Bianchi:2001kw,
Skenderis:2002wp,Papadimitriou:2004ap}.
Global symmetries of the 
field theory are generically realised as gauge symmetries in the bulk of the gravity theory.
Hence, the starting point for the realisation of the dynamics of CHMs 
with an $SO(5)/SO(4)$ coset should be a higher-dimensional  theory of 
gravity supplemented by $SO(5)$ gauge symmetry.

An extensive body of work (see for instance Refs.~\cite{Contino:2003ve,Agashe:2004rs,Agashe:2005dk,Agashe:2006at,Contino:2006qr,Falkowski:2008fz,Contino:2011np})
demonstrate the reach of the  bottom-up approach to holography 
in building a realisation of the CHMs with $SO(5)/SO(4)$ coset.
By choosing
a fixed curved background in $D=5$ dimensions, and a set of fields
propagating in this background, it is possible to capture the salient features of the
dynamics of the pNGBs and 
some of the spin-1/2 and spin-1 composite  particles.
The spontaneous symmetry breaking $SO(5)\rightarrow SO(4)$ is implemented 
by means of appropriate choices of the field content, background, and boundary conditions.
While calculability is greatly improved by this approach, important properties of the long distance dynamics
are not captured by bottom-up models: for example, demonstrating confinement 
requires embedding the gravity theory in string theory, hence allowing for the holographic
treatment of the Wilson loops~\cite{Rey:1998ik,Maldacena:1998im} (see also Refs.~\cite{Kinar:1998vq,
Brandhuber:1999jr,Avramis:2006nv,Nunez:2009da,Faedo:2013ota}) to recover the
area law, as suggested in Ref.~\cite{Witten:1998zw}. Furthermore, top-down models fix the relation between masses of the resonances in a predictive way, as it emerges from the dual strong dynamics.

Recently, inspired by  sophisticated
holographic models exhibiting QCD-like properties (such as the Sakai-Sugimoto 
model~\cite{Sakai:2004cn,Sakai:2005yt} and its precursors~\cite{Karch:2002sh,Kruczenski:2003be,
Erdmenger:2007cm}),
new  bottom-up holographic realisations of CHMs
have been developed~\cite{Erdmenger:2020lvq,Erdmenger:2020flu}.
The same coset structures studied on the lattice are here realised 
by bulk fields,
 inspired by the DBI  action that
 describes  extended objects probing
the curved background (and captures the quenched approximation for matter fields
in the large-$N$ limit of the field theory).
In a parallel direction, Ref.~\cite{Elander:2020nyd} includes 
 deformations to mimic the  backreaction due to many such extended objects 
(edging towards the Veneziano limit of the large-$N$ field theory).

With this paper, we propose
an alternative approach, closer in spirit to the original  holographic dualities.
Known supergravity theories  can provide
 complete top-down holographic models, in which the strong-coupling field-theory
 dynamics is fully captured.
We consider supergravities that include at their core the 
interesting (gauged) symmetry group, without the need for additional
 matter fields and/or  stacks of $Dp$-branes.
 The resulting enhanced calculability is a double-edged sword: while the fact that the field content and phenomenology are rigidly dictated by supergravity leads to greater predictability, it also makes the task of identifying realistic models a more challenging endeavour. We exemplify this programme
with the maximal $SO(5)$ gauged supergravity theory in seven dimensions, within which
we realise the $SO(5)/SO(4)$ coset of a minimal CHM.

In the rest of this introduction, we summarise the main features of the model, 
as they emerge from the relevant technical literature.
Our starting point is the established fact that maximal supergravity in seven dimensions
has a gauged $SO(5)$ symmetry~\cite{Pilch:1984xy,Nastase:1999cb,
Pernici:1984xx,Pernici:1984zw,Cvetic:1999xp,Lu:1999bc,Campos:2000yu,Cvetic:2000ah,Samtleben:2005bp}.
It can be obtained by dimensional reduction on $S^4$ of maximal supergravity in $D=11$ dimensions, and the $SO(5)$ originates from the isometry group of the $S^4$.
 The background solution with AdS$_7\times S^4$ geometry is interpreted,
 in the language of gauge-gravity dualities, in terms of the somewhat mysterious ${\cal N}=(2,0)$
 strongly coupled gauge theory living on a stack of $M5$-branes, which is a superconformal chiral theory 
 with extended supersymmetry. While the microscopic details of this field theory
 are not known in general terms, many of its properties are established in the literature---for
  example $SO(5)$ is indeed its global R-symmetry---and an incomplete
  selection of interesting studies can be found in Refs.~\cite{Ganor:1996nf,Witten:1996hc,Seiberg:1997ax,Aharony:1997an,
 Harvey:1998bx,Corrado:1999pi,Intriligator:2000eq,Cordova:2015vwa}.

One reason why this obsure field theory, and its emergence as the dual of a special solution in supergravity, is of interest, in the literature on
   gauge/gravity dualities, is the observation in Ref.~\cite{Witten:1998zw}
 that one can identify background solutions in which two of the dimensions are
 compactified on circles, one of which 
  connects the lifts to eleven-dimensional supergravity  and to type-IIA supergravity in ten dimensions.
Solutions in which the other circle---along which fermions have anti-periodic boundary 
conditions, breaking supersymmetry---shrinks to zero size, at some finite value of
 the radial direction ($\r\rightarrow \r_o$) in the geometry, yield in the dual field theory the
 long distance behaviour of a confining  theory in four dimensions. Not only is there a mass gap,
but furthermore
 the holographic calculation of the appropriate rectangular Wilson 
 loops~\cite{Rey:1998ik,Maldacena:1998im} (see also Refs.~\cite{Kinar:1998vq,
Brandhuber:1999jr,Avramis:2006nv,Nunez:2009da,Faedo:2013ota}), yields a static potential 
 that grows linearly with the displacement between two heavy quarks.
This class of  curved backgrounds underlies 
 the $D4/D8$ system exploited in the aforementioned Sakai-Sugimoto holographic model
of quenched QCD~\cite{Sakai:2004cn,Sakai:2005yt}.

As a distinctive feature, we do not add probe branes to the smooth geometries that we study, but instead implement the global symmetry breaking pattern by considering the supergravity theory on its own.
We recall some of the salient aspects of this supergravity theory here, deferring 
the technical details to the appropriate sections of the main body of the paper.
The scalar manifold of the supergravity theory
describes the 14-dimensional (right) coset $SL(5, \mathbb{R})/SO(5)_c$, and 
the scalars transform (on the left) as a 2-index symmetric traceless representation of the gauged $SO(5)$.
It has been known from the onset~\cite{Pernici:1984zw} of the study of this theory
 that  background solutions exist in which one scalar $\phi$ has a non-trivial radial profile,
 breaking  conformal symmetry,  supersymmetry, and $SO(5)$ to its $SO(4)$ subgroup.
The dynamics of $\phi$, in particular in relation to the dilatation operator in the dual field theory,
has been discussed extensively, for example  in
Refs.~\cite{Elander:2013jqa,Elander:2020csd,Elander:2020fmv} and references therein, 
for a wide variety of admissible backgrounds.
The symmetric traceless  representation of $SO(5)$ decomposes as 
$14\rightarrow 1\oplus 4\oplus 9$ of $SO(4)$. The $4$ pNGBs describe the $SO(5)/SO(4)$ coset,
and hence these backgrounds provide a dynamical realisation of the coset structure 
underlying the minimal CHM in Ref.~\cite{Agashe:2004rs}, as desired.
The analysis of the UV asymptotics of the background shows that 
$SO(5)$  is broken both explicitly as well as spontaneously.

Besides observing that this structure matches the one postulated in the minimal CHM, 
our main contribution is the calculation of the spectra of fluctuations of all bosonic fields obtained after
dimensional reduction on a 2-torus, performed by adopting the formalism developed in Refs.~\cite{
Bianchi:2003ug,Berg:2005pd,Berg:2006xy,
Elander:2009bm,Elander:2010wd}. We further generalise the
results of Refs.~\cite{Brower:2000rp} to a whole class of backgrounds, and
to include  the whole spectrum of $p$-forms, for which we make use of the $R_{\xi}$ gauge,
along the lines of Ref.~\cite{Elander:2018aub}. There are $23$
independent towers of such bosonic eigenstates, with various degrees of degeneracy
governed by the unbroken $SO(4)$ symmetry,
that make up the 128 bosonic degrees of freedom of maximal supergravity.

For the time being, we ignore two model-dependent  aspects
of the theory, that are not central to the construction: we  consider only the bosonic field content,
while ignoring completely the fermions, and we also disregard the interactions descending from the Chern-Simons terms, as they do not affect the calculation of the spectra of bound states.
The spectrum of fermionic bound states in the field theory depends on how the
compactification breaks supersymmetry.\footnote{To be precise, our background solutions break
 supersymmetry locally, as they do not satisfy the first-order equations mandated by supersymmetry, 
 but also globally, because of the anti-periodic boundary conditions obeyed by the fermions along the shrinking circle.}
We focus instead on model-independent parts of the spectrum, 
directly related to the underlying symmetries.
The embedding of the standard model $SU(2)\times U(1)$ group must be anomaly free,
a requirement that, as in other CHMs, may be subtle, but not hard to satisfy. We postpone to future studies any phenomenological considerations, 
starting from the  detailed implementation of the procedure allowing to 
(weakly) gauge the $SU(2)\times U(1) $ subgroup of $SO(4)$---which 
involves some subtlety in the process of holographic renormalisation---as 
well as the coupling to the standard-model
 matter fields,
and the vacuum (mis-)alignment features leading to electroweak symmetry breaking.

Finally, we notice  that similar constructions can in principle be applied also in 
other classes of supergravities. For example, in the context of
the half-maximal supergravity in six  dimensions, coupled to $n$ 
vector multiplets~\cite{Romans:1985tw,Romans:1985tz,
 Brandhuber:1999np,Cvetic:1999un,Samtleben:2005bp} (see also 
 Refs.~\cite{Hong:2018amk,Jeong:2013jfc,DAuria:2000afl,
 Andrianopoli:2001rs,Nishimura:2000wj,Ferrara:1998gv,Gursoy:2002tx,Nunez:2001pt,Karndumri:2012vh,
 Lozano:2012au,Karndumri:2014lba,Chang:2017mxc,Gutperle:2018axv,Suh:2018tul,Suh:2018szn,Kim:2019fsg,
 Chen:2019qib,Hoyos:2020fjx}), and compactified on a circle~\cite{Wen:2004qh,
 Kuperstein:2004yf,Elander:2013jqa,Elander:2018aub}.
Solutions to these systems
exist that admit an interpretation of the long distance dynamics
 in terms of confining field theories in four dimensions.
 The scalar manifold describes the non-trivial  and non-compact $O(1,1)\times O(4,n)/(SO(4)\times O(n))$ 
 coset---though it  is not immediately apparent how to embed an interesting CHM into this coset. Alternatively, in the case of these six-dimensional supergravities it has been shown that the existence of different lifts to ten dimensions can lead to interesting structures in the dual field theory, with additional flavor symmetry emerging non-trivially~\cite{Legramandi:2021aqv}.

The paper is organised as follows.
In Sec.~\ref{Sec:model} we define the supergravity theory in $D=7$ dimensions, discussing its
field content, couplings, and free parameters. Most of the material is lifted from the literature, but we 
find it useful to fix the notation
 used in the rest of the paper.
In Sec.~\ref{Sec:torus} we perform the reduction on a torus and write the resulting
action in $D=5$ dimensions, including all the bosonic degrees of freedom, 
substantially extending available results in the literature.
In Sec.~\ref{Sec:solutions} we display the class of solutions we are interested in,
by analysing their asymptotic behaviours and hence illustrating the process that allows to generate them numerically. We demonstrate that the solutions are regular, and discuss what we mean by stating that the dual theory confines.
In Sec.~\ref{Sec:spectrum}, we compute the mass spectrum.
We start with the linearised equations of gauge invariant combinations of fluctuations of the
metric and of the background scalars in Sec.~\ref{Sec:spectrum1}. This section extends and completes some of the results 
presented in Refs.~\cite{Elander:2013jqa,Elander:2020csd,Elander:2020fmv}.
In Secs.~\ref{Sec:spectrum2}, \ref{Sec:spectrum-1}, and~\ref{Sec:spectrum-2}
we complete the study of the bosonic mass spectrum, by
looking at the fields that are trivial in the background. We 
discuss the spectra of $p$-forms, with $p=0, 1, 2$. We adopt a generalisation 
of the $R_{\xi}$ gauge
to gauge-fix the theory (see also Ref.~\cite{Elander:2018aub}), and present our numerical results, 
commenting on the treatment of numerical artefacts, where appropriate.
In Sec.~\ref{Sec:towards}, after summarising the main features of the mass spectrum,
 we study its dependence on
the free adjustment of boundary-localised terms that preserve the SO(4) symmetry of the system, but may introduce additional explicit breaking of SO(5). We focus on one particular example of such an admissible term,
that affects the pNGBs and the lightest spin-1 composite states in the dual theory.
In Sec.~\ref{Sec:conclusion} we summarise our main findings and outline 
an extensive programme of future investigations.

We relegate to the Appendix
 an extensive selection of technical details, that can be skipped 
at first reading of the paper, but would be useful to reproduce our results, or
 to apply the same formalism to other classes of background solutions.
We devote Appendix~\ref{Sec:3} to discussing the self-duality condition
of the $3$-forms in seven dimensions, and its consequences for the $2$-forms in five dimensions.
Appendix~\ref{Sec:Formalism} contains technical details about the background solutions and 
the formalism in which we treat the fluctuations, including the asymptotic expansions of the 
fluctuations.
Appendix~\ref{Sec:Lift} summarises some elements of the lift to ten- and eleven-dimensional supergravity,
borrowed from the literature. We also observe that the result of the  holographic calculation of the string tension
suggests the existence of a discontinuity, the energetically favoured string configurations qualitatively differing, 
depending on the sign of $\phi$. We leave this observation for future investigations.

We close the paper with Appendix~\ref{Sec:Brower}, in which we summarise our numerical results for the whole spectrum of
the toroidal compactification, but now in  the case of unbroken $SO(5)$. We critically compare our results to those
of Ref.~\cite{Brower:2000rp}, who considered  the same $SO(5)$-invariant background,
but used a different formalism and restricted their analysis to a different set of  supergravity modes.
We find good agreement between the two sets of results, in the common sector, thus providing a useful cross check. Unsurprisingly, we also find that several of the states in our study,
that represent  non-trivial $SO(5)$ multiplets, 
are lighter than the singlets one would retain in a consistent truncation.

\section{Maximal gauged supergravity in seven dimensions}
\label{Sec:model}

The field content and action of  maximal supergravity in $D=7$ dimensions
are discussed for example in Refs.~\cite{Nastase:1999cb,Pernici:1984xx,Pernici:1984zw,Cvetic:1999xp,Lu:1999bc,Cvetic:2000ah}
(see also Refs.~\cite{Samtleben:2005bp,Campos:2000yu,Elander:2013jqa,Cowdall:1998rs}).
The seven-dimensional space-time indexes are denoted by $\hat{M}=0,\,1,\,2,\,3,\,5,\,6,\,7$,
while we use the Greek indexes $\alpha=1\,,\cdots\,,5$ to denote the components 
of the fundamental representation of the $SO(5)$ gauge group.
The bosonic fields in the weakly-coupled action are the following (counting on-shell degrees of freedom): 
$14$ real scalars, $10$ (massless) vectors ${\cal A}_{\hat{M}\,\alpha}^{\,\,\,\,\,\,\,\,\,\,\beta}$ 
(each propagating $5$ real degrees of freedom),
$5$ (massive self-dual) $3$-forms $S_{\hat{M}\hat{N}\hat{P}\,\alpha}$ (each propagating $10$ 
real degrees of freedom, rather than $20$, because their equation of motion is the first-order  self-duality condition), 
and the metric ($14$ degrees of freedom).
The $128$ degrees of freedom match the fermionic field content, which consists of
 $16$ gauginos ($4$ complex components each) and $4$ gravitinos ($(D-3) 2^{[D/2]}/2=16$ degrees of freedom each).

The scalar manifold describes the $SL(5,\mathbb{R})/SO(5)_c$ right coset,
and we label by $i,j=1,\,\cdots\,,5$ the fundamental representation of the global $SO(5)_c$ symmetry,
to retain the distinction with the indexes of the gauged $SO(5)$.
The $14$ scalars parametrising the coset 
are written in terms of a $5\times 5$ real unit-determinant matrix $\Pi_{\alpha}^{\,\,\,\,i}$ 
(with ${\rm det}(\Pi_{\alpha}^{\,\,\,\,i})=1$),
that transforms under the action of $SO(5)_c$ on the right, and of  $SO(5)$ on the left.\footnote{We could as 
well use the $SO(5)$ gauge freedom
in such a way as to represent each element of the coset with a symmetric matrix, hence making manifest the
fact that the coset is equivalent to the $14$ of $SO(5)_c$.}
The $10$ vectors transform as the adjoint of $SO(5)$, and the $3$-forms in the fundamental.
Because  $SO(5)\simeq Sp(4)$, the gravitinos transform in the
spinorial $4$-dimensional representation of $SO(5)_c$, while the gauginos transform 
as the irreducible part of the product of the vectorial and the
spinorial representation of $SO(5)_c$
(as $5\otimes 4 \rightarrow 16\oplus 4$ in $Sp(4)$~\cite{Yamatsu:2015npn}).
We ignore fermions from here on.

The bosonic part of the action is the following:\footnote{The 
action here is $\frac{1}{2}$ of the action in Refs.~\cite{Pernici:1984zw,Elander:2013jqa},
which amounts to a harmless overall rescaling of the Planck constant in seven dimensions.}
\beqs
{\cal S}_7&=&\int \di^7 x \left\{\sqrt{-\hat{g}_7}\left[\frac{{\cal R}_7}{4}\nonumber
-\frac{1}{4}\hat{g}^{\hat{M}\hat{N}} \delta^{ik}\delta_{j\ell} 
P_{\hat{M}\,i}^{\,\,\,\,\,\,\,\,\,\,j} P_{\hat{N}k}^{\,\,\,\,\,\,\,\,\,\ell}
-\frac{m^2}{8}\left(2\delta^{i}_{\,\,\,k}\delta^{j}_{\,\,\,\ell} -\delta^{ij}\delta_{k\ell} \right)T_{ij}T^{k\ell}+
\right.\right.\label{Eq:Pernici}
\\
&&\left.\left.\frac{}{}
+\frac{1}{8}\left(\Pi_{\alpha}^{\,\,\,i}\Pi_{\beta}^{\,\,\,j}{\cal F}^{\alpha\beta}_{\hat{M}\hat{N}}\right)^2
-\frac{m^2}{4}\Big((\Pi^{-1})_{i}^{\,\,\,\alpha}S_{\hat{M}\hat{N}\hat{O}\,\alpha}\Big)^2
\right]\right.+\\
&&\left.\frac{}{}+\frac{m}{96} \epsilon^{\hat{M}\hat{N}\hat{O}\hat{P}\hat{Q}\hat{R}\hat{S}}S_{\hat{M}\hat{N}\hat{O}\,\alpha}
\delta^{\alpha\beta}F^{(S)}_{\hat{P}\hat{Q}\hat{R}\hat{S}\,\beta}\right.+ \nonumber\\
&&\left.\frac{}{}
-\frac{i}{32\sqrt{3}} \epsilon^{\hat{M}\hat{N}\hat{O}\hat{P}\hat{Q}\hat{R}\hat{S}}
\epsilon_{\a\b\gamma\d\eta}
\delta^{\a\a^{\prime}}S_{\hat{M}\hat{N}\hat{O}\,\a^{\prime}}
\delta^{\b\b^{\prime}}{\cal F}_{\hat{P}\hat{Q}\,\b^{\prime}}^{\,\,\,\,\,\,\,\,\,\,\,\,\,\,\gamma}
\delta^{\d\d^{\prime}}{\cal F}_{\hat{R}\hat{S}\,\d^{\prime}}^{\,\,\,\,\,\,\,\,\,\,\,\,\,\,\eta}
\right\}+C.S.
\nonumber
\,,
\eeqs
where the last term refers to the two Chern-Simons topological interactions,
explicitly written in Ref.~\cite{Pernici:1984xx}.
The objects appearing in the action are defined as follows:\footnote{
Complete anti-symmetrisation is normalised so that $[n_1n_2\cdots n_p]\equiv
 \frac{1}{p!}(n_1n_2\cdots n_p-n_2n_1\cdots n_p+\cdots)$.}
\beqs
P_{\hat{M}\,i}^{\,\,\,\,\,\,\,\,\,\,j}&\equiv&\frac{1}{2}\left((\Pi^{-1})_i^{\,\,\,\a}{\cal D}_{\hat{M}\,\a}^{\,\,\,\,\,\,\,\,\,\,\b}\Pi_{\b}^{\,\,\,\,j}
+(i\leftrightarrow j)\right)\,,\\
{\cal D}_{\hat{M}\,\a}^{\,\,\,\,\,\,\,\,\,\,\b}&\equiv& \delta_{\a}^{\,\,\,\b}\partial_{\hat{M}}
+i g {\cal A}_{\hat{M}\,\a}^{\,\,\,\,\,\,\,\,\,\,\,\b}\,,\\
T_{ij}&\equiv& (\Pi^{-1})_{i}^{\,\,\,\a} (\Pi^{-1})_{j}^{\,\,\,\b}\delta_{\a\b}\,,\\
{\cal F}_{\hat{M}\hat{N}\,\a}^{\,\,\,\,\,\,\,\,\,\,\,\,\,\,\,\,\b}&\equiv&
2\left(\partial_{[\hat{M}}{\cal A}_{\hat{N}]\,\a}^{\,\,\,\,\,\,\,\,\,\,\,\,\,\b}\,+\,i g {\cal A}_{[\hat{M}\,\a}^{\,\,\,\,\,\,\,\,\,\,\,\,\gamma}
{\cal A}_{\hat{N}]\gamma}^{\,\,\,\,\,\,\,\,\,\,\b}\right)\,,\\
F^{(S)}_{\hat{M}\hat{N}\hat{P}\hat{Q}\,\a}&\equiv&4\left(\partial_{[\hat{M}}S_{\hat{N}\hat{P}\hat{Q}]\a}\,+\,i 
g {\cal A}_{[\hat{M}\,\a}^{\,\,\,\,\,\,\,\,\,\,\,\,\,\b}S_{\hat{N}\hat{P}\hat{Q}]\b}\right)\,.
\eeqs
The mass scale $m$ is related to the gauge coupling $g$ by $g=2m$~\cite{Pernici:1984xx}.
In the applications, we set $m=1$.  We will return at the appropriate moment to the fact that 
the gauge symmetry acting on  the $3$-forms is not manifest in
Eq.~(\ref{Eq:Pernici}).

\subsection{$SO(5)$ to $SO(4)$ breaking}
\label{Sec:breaking}

We parameterise the scalar manifold by making
explicit use of the envisaged  breaking $SO(5)\rightarrow SO(4)$,
and  decompose  the $\Pi_{\alpha}^{\,\,\,i}$ scalar  fields 
into irreducible representations of $SO(4)$, as $14=1\oplus 9\oplus 4$. We denote the resulting multiplets
with matrix-valued fields $\lambda_{i}^{\,\,\,j}\sim 1$, $\Delta_{i}^{\,\,\,j}\sim 9$, 
and $O_{\alpha}^{\,\,\,i}\sim 4$.
The matrix $\Pi_{\alpha}^{\,\,\,i}$ is given by
\beqs
\Pi&\equiv&O\,
\Delta\,\lambda\,\,\,\,(=\,O\,
\lambda\,\Delta)\,.
\label{Eq:decompose}
\eeqs
The diagonal matrix $\lambda$ has the following form
\beqs
\lambda&=&{\rm diag}\left(e^{-\frac{\phi}{2\sqrt{5}}}\,,
e^{-\frac{\phi}{2\sqrt{5}}}\,,
e^{-\frac{\phi}{2\sqrt{5}}}\,,
e^{-\frac{\phi}{2\sqrt{5}}}\,,
e^{2\frac{\phi}{\sqrt{5}}}\right)\,,
\eeqs
where $\phi$ is the field responsible for the 
breaking $SO(5)\rightarrow SO(4)$---if  $\langle \phi \rangle \neq 0$ in the vacuum.
The symmetric matrix $\Delta$ commutes with $\lambda$. It can be written as
\beqs
\Delta&\equiv&\left(\begin{array}{c|c}
{\Delta}_{(4)} & 0 \cr
\hline
0 & 1
\end{array} \right)\,
\eeqs
with the $4 \times 4$ matrix-valued field satisfying  ${\Delta}_{(4)}= {\Delta}_{(4)}^{T}$ and ${\rm det}(\Delta_{(4)})=1$,
hence making it explicit that  ${\Delta}_{(4)}$ transforms as the $9$ of $SO(4)$.
Finally,
 the  matrix $O$ is written as
\beqs
O&\equiv&\exp\left[i \sum_{\hat{A}}\pi^{\hat{A}}t^{\hat{A}}\right]\,\mathbb{1}_5\,,
\eeqs
where  we factorise $O$ in the exponential, living  in  $SO(5)$, and 
$\mathbb{1}_5$, with elements $\delta_{\alpha}^{\,\,\,\,i}$  carrying  indexes both in $SO(5)$ and $SO(5)_c$.
We restrict the  matrices $t^{\hat{A}}$, with $\hat{A}=1\,,\cdots\,, 4$, to be the hermitian and traceless (broken) generators 
of $SO(5)$ such that $t^{\hat{A}}\,\lambda+\lambda\, t^{\hat{A}\,T}\neq 0$. The $\pi^{\hat{A}}$ fields 
describing the  $SO(5)/SO(4)$ coset
are associated with  the pNGBs.

The unbroken generators, denoted as $t^{\bar{A}}$, with $\bar{A}=5\,,\cdots\,,10$,
obey the relations $t^{\bar{A}}\,\lambda+\lambda\, t^{\bar{A}\,T} = 0$.
We adopt the
 normalisations $\Tr (t^{\bar{A}} t^{\hat{B}}) \,=\,0$, $\Tr (t^{\bar{A}} t^{\bar{B}}) \,=\,\frac{1}{2}\delta^{\bar{A}\bar{B}}$
and 
$\Tr (t^{\hat{A}} t^{\hat{B}}) \,=\,\frac{1}{2}\delta^{\hat{A}\hat{B}}$.
The $1$-forms decompose in terms of vector $A_{\hat{M}}^{\,\,\,\,\bar{A}}$ and axial-vector  $A_{\hat{M}}^{\,\,\,\,\hat{A}}$ fields:
\beqs
{\cal A}_{\hat{M}\,\alpha}^{\,\,\,\,\,\,\,\,\,\,\,\,\beta}&\equiv & A_{\hat{M}}^{\,\,\,\,\bar{A}} \left( t^{\bar{A}}\right)_{\alpha}^{\,\,\,\,\beta}+
A_{\hat{M}}^{\,\,\,\,\hat{A}}\left(t^{\hat{A}}\right)_{\alpha}^{\,\,\,\,\beta}\,.
\eeqs
For the $3$-forms, the $SO(5)/SO(4)$ decomposition is simpler, as $5=4\oplus 1$,
and we denote the $4$ as $S_{\hat{M}\hat{N}\hat{O}\,\bar{\alpha}}$,
with  $\bar{\alpha}=1\,,\cdots 4$, while $S_{\hat{M}\hat{N}\hat{O}\,5}$ is the $SO(4)$ singlet.

A minimal amount of algebra leads to the simplifying relation
\beqs
\Tr\Big[\left( \lambda^{-1}\partial_{\hat{M}}\lambda\right) \left(\lambda^{-1}\partial_{\hat{N}}\lambda\right)\Big]
&=&\partial_{\hat{M}}\phi\,\partial_{\hat{N}}\phi\,.
\eeqs
Making use of Eq.~(\ref{Eq:decompose}), further algebraic manipulations allow to rewrite the action as follows:
\beqs
{\cal S}_7&=&\int \di^7 x \left\{\sqrt{-\hat{g}_7}\left[\frac{{\cal R}_7}{4}
-\frac{1}{4}g^{\hat{M}\hat{N}} \partial_{\hat{M}}\phi\partial_{\hat{N}}\phi+
\right.\right.
\\
&&\left.\left.\frac{}{}\nonumber
- \frac{1}{8}g^{\hat{M}\hat{N}} \Tr\Big( \Delta^{-1} \partial_{\hat{M}}\Delta \left\{ \Delta^{-1}, \partial_{\hat{N}} \Delta \right\} +
\right.\right.
\\
&&\left.\left.\frac{}{}\nonumber
+ O^{-1}{\cal D}_{\hat{M}} O
\left[O^{-1}{\cal D}_{\hat{N}} O\,,\,\Delta^{2}\lambda^{2}\right]\Delta^{-2}\lambda^{-2}
+\right.\right.\\
&&
\left.\left.\frac{}{}\nonumber
 + 4\Delta^{-1} \lambda^{-1} \partial_{\hat M} \Delta \partial_{\hat N} \lambda + O^{-1} \mathcal D_{\hat M} O \left[ \partial_{\hat N} \Delta^2, \Delta^{-2} \right] \Big) +
 \right.\right.\\
&&
\left.\left.\frac{}{}\nonumber
-\frac{m^2}{8}\left(2\,\Tr\left(\Delta^{-4}\lambda^{-4}\right)
-\left(\Tr\left(\Delta^{-2} \lambda^{-2}\right)\right)^2\right)+
\right.\right.
\\
&&\left.\left.\frac{}{}\nonumber
-\frac{1}{8}g^{\hat{M}\hat{R}}g^{\hat{N}\hat{S}}\,\Tr\Big(\Delta^2\lambda^2O^{-1}F_{\hat{M}\hat{N}}O\Delta^2\lambda^2 O^{-1}F_{\hat{R}\hat{S}}O\Big)
+\right.\right.\\
&&
\left.\left.\frac{}{}\nonumber
-\frac{m^2}{4}g^{\hat{M}\hat{P}}g^{\hat{N}\hat{Q}}g^{\hat{O}\hat{R}}\,
\left(S_{\hat{M}\hat{N}\hat{O}}^{\,\,\,\,\,\,\,\,\,\,\,\,\,\,\,T}O
\Delta^{-2}\lambda^{-2}O^{-1}S_{\hat{P}\hat{Q}\hat{R}}\right)
\right]\right.+\\
&&\left.\frac{}{}+\frac{m}{96} \epsilon^{\hat{M}\hat{N}\hat{O}\hat{P}\hat{Q}\hat{R}\hat{S}}S_{\hat{M}\hat{N}\hat{O}\,\alpha}
\delta^{\alpha\beta}F^{(S)}_{\hat{P}\hat{Q}\hat{R}\hat{S}\,\beta}
\right.+\nonumber
\\
&&\left.\frac{}{}\nonumber
-\frac{i}{32\sqrt{3}} \epsilon^{\hat{M}\hat{N}\hat{O}\hat{P}\hat{Q}\hat{R}\hat{S}}
\epsilon_{\a\b\gamma\d\eta}
\delta^{\a\a^{\prime}}S_{\hat{M}\hat{N}\hat{O}\,\a^{\prime}}
\delta^{\b\b^{\prime}}{\cal F}_{\hat{P}\hat{Q}\,\b^{\prime}}^{\,\,\,\,\,\,\,\,\,\,\,\,\,\,\gamma}
\delta^{\d\d^{\prime}}{\cal F}_{\hat{R}\hat{S}\,\d^{\prime}}^{\,\,\,\,\,\,\,\,\,\,\,\,\,\,\eta}
\right\}+C.S.
\nonumber
\,.
\eeqs

To make contact with earlier studies (e.g., Refs.~\cite{Campos:2000yu,Elander:2013jqa,Elander:2020csd,Elander:2020fmv}),
we  observe that it would be consistent to truncate the theory by setting $\Delta=\mathbb{1}_5=O$,
and ${\cal A}_{\hat{M}\,\alpha}^{\,\,\,\,\,\,\,\,\,\,\,\beta}=0=S_{\hat{M}\hat{N}\hat{O}\,\alpha}$ identically. The action of the reduced theory becomes
\beqs
{\cal S}_7^{(0)}&=&\int \di^7 x \sqrt{-\hat{g}_7}\left(\frac{{\cal R}_7}{4}-\frac{1}{4} \hat{g}^{\hat M \hat N}
\partial_{\hat M} \phi \partial_{ \hat N} \phi -{\cal V}_7(\phi)\right)\,,
\eeqs
where the potential is given by
\beqs
{\cal V}_7(\phi)&=&\frac{m^2}{8}\Big(2\,\Tr \lambda^{-4}-(\Tr \lambda^{-2})^2\Big)\,=\,
\frac{m^2}{8}\left(e^{-\frac{8}{\sqrt{5}}\phi}-8e^{-\frac{3}{\sqrt{5}}\phi}-8e^{\frac{2}{\sqrt{5}}\phi}\right)\,,
\eeqs
in agreement with Eqs.~(1--2) of Ref.~\cite{Elander:2020fmv}.

\subsection{Truncating the action to quadratic order}
\label{Sec:2}

We adopt the milder assumptions that in the vacuum $\langle \Delta \rangle =\mathbb{1}_5=\langle O \rangle$, and
$\langle {\cal A}_{\hat{M}}\rangle=0=\langle S_{\hat{M}\hat{N}\hat{O}}\rangle$. We will ultimately be interested in computing spectra, which only requires considering the action to second order in fluctuations around a given background. We hence retain 
the full dependence of the action on $\phi$ and the metric,
but expand at the quadratic order in the fields 
that are trivial in the vacuum (and their derivatives).  
We start by rewriting the matrix-valued $\Delta_{(4)}$ as follows:
\beqs
\Delta_{(4)}&\equiv&\mathbb{1}_4 +s^{\tilde{A}}{\cal T}^{\tilde{A}}
+\frac{1}{2}s^{\tilde{A}}s^{\tilde{B}}{\cal T}^{\tilde{A}}{\cal T}^{\tilde{B}}+\cdots\,,
\eeqs
where we omit  higher powers of the fields $s^{\tilde{A}}$. In this expression, which is just the quadratic-order
approximation of the exponential, the nine traceless and symmetric matrices 
${\cal T}^{\bar{A}}$ (with $\tilde{A}=1\,,\cdots\,,9$)
obey the relation $\Tr ({\cal T}^{\tilde{A}}{\cal T}^{\tilde{B}}) =\frac{1}{2}\delta^{\tilde{A}\tilde{B}}$.

We hence arrive to the action that we use in the rest of the paper,
in which we retain the full dependence on the metric,
and on the scalar field $\phi$, but expand and truncate 
to quadratic order in all other fields:
\beqs
{\cal S}_7^{(2)}&=&\int \di^7 x \left\{\sqrt{-\hat{g}_7}\left[\frac{{\cal R}_7}{4}
-\frac{1}{4}\hat{g}^{\hat{M}\hat{N}}\left(\partial_{\hat{M}}\phi\partial_{\hat{N}}\phi
+\frac{1}{2}\sum_{\tilde{A}=1}^{9}
\partial_{\hat{M}}s^{\tilde{A}}\partial_{\hat{N}}s^{\tilde{A}}\right)
\label{Eq:quad}
\right.\right.+
\\
&&\left.\left.\frac{}{}\nonumber
-\frac{1}{8}\sinh^2\left(\frac{\sqrt{5}}{2}\phi\right) \sum_{\hat{A}=1}^{4}
\left(\partial_{\hat{M}}\pi^{\hat{A}}+g A_{\hat{M}}^{\,\,\,\,\hat{A}}\right)\hat{g}^{\hat{M}\hat{N}}
\left(\partial_{\hat{N}}\pi^{\hat{A}}+g A_{\hat{N}}^{\,\,\,\,\hat{A}}\right)
\nonumber
+\right.\right.\\
&&\nonumber
\left.\left.\frac{}{}
-\frac{m^2}{8}
\left(e^{-\frac{8}{\sqrt{5}}\phi}-8e^{-\frac{3}{\sqrt{5}}\phi}-8e^{\frac{2}{\sqrt{5}}\phi}\right)+
\frac{m^2}{4}e^{-\frac{3}{\sqrt{5}}\phi} \sum_{\tilde{A}=1}^{9}s^{\tilde{A}}s^{\tilde{A}} +
\right.\right.
\\
&&\left.\left.\frac{}{}\nonumber
-\frac{1}{16}\hat{g}^{\hat{M}\hat{R}}\hat{g}^{\hat{N}\hat{S}} \, \left( e^{\frac{3}{\sqrt{5}}\phi} \frac{}{}\sum_{\hat{A}=1}^4
F_{\hat{M}\hat{N}}^{\,\,\,\,\,\,\,\,\,\,\hat{A}} F_{\hat{R}\hat{S}}^{\,\,\,\,\,\,\,\,\,\,\hat{A}}
+ e^{-\frac{2}{\sqrt{5}}\phi} \frac{}{}\sum_{\bar{A}=5}^{10}
F_{\hat{M}\hat{N}}^{\,\,\,\,\,\,\,\,\,\,\bar{A}} F_{\hat{R}\hat{S}}^{\,\,\,\,\,\,\,\,\,\,\bar{A}} \right)
+\right.\right.\\
&&
\left.\left.\nonumber\frac{}{}
-\frac{m^2}{4}\hat{g}^{\hat{M}\hat{P}}\hat{g}^{\hat{N}\hat{Q}}\hat{g}^{\hat{O}\hat{R}}\,
\left(\sum_{\bar{\alpha}=1}^4\frac{}{}e^{\frac{1}{\sqrt{5}}\phi}S_{\hat{M}\hat{N}\hat{O}\,\bar{\alpha}}
S_{\hat{P}\hat{Q}\hat{R}\,\bar{\alpha}}+e^{-\frac{4}{\sqrt{5}}\phi}
S_{\hat{M}\hat{N}\hat{O}\,5}
S_{\hat{P}\hat{Q}\hat{R}\,5}
\right)
\rule{-0.05cm}{0.75cm}
\right]\right.+\\
&&\left.\frac{}{}+\frac{m}{24} \epsilon^{\hat{M}\hat{N}\hat{O}\hat{P}\hat{Q}\hat{R}\hat{S}}\left(
\sum_{\bar{\alpha}=1}^4
S_{\hat{M}\hat{N}\hat{O}\,\bar{\alpha}}
\partial_{\hat{P}}S_{\hat{Q}\hat{R}\hat{S}\,\bar{\alpha}}
+
S_{\hat{M}\hat{N}\hat{O}\,5}
\partial_{\hat{P}}S_{\hat{Q}\hat{R}\hat{S}\,5}
\right)\nonumber
\rule{-0.05cm}{0.75cm}
\right\}\,.
\eeqs
The topological terms have been omitted, as they appear only at higher orders in the field expansion.
The self-duality condition is evident from the last two terms:
along the equations of motion,
 the differential of the 3-form must be proportional to the 3-form itself.

\section{Toroidal reduction on $S^1\times S^1$}
\label{Sec:torus}

We recover the  (low energy) five-dimensional duals of four-dimensional field theories by
considering solutions that satisfy the  following ansatz:
\beqs
\di s^2_7 &=& e^{-2\chi}\di s_5^2+e^{3\chi-2\omega}\left(\frac{}{}\di \eta +\chi_{M}\di x^{M}\right)^2
+e^{3\chi+2\omega}\left(\frac{}{}\di \zeta +\omega_{M}\di x^{M}+\omega_6 \di \eta\right)^2\,,
\label{Eq:7}
\eeqs
where $\chi_M$ and $\omega_M$ are naturally defined as covariant fields, having lower (curved) space-time 
indexes 
$M=0, 1, 2, 3, 5$ in five dimensions.
The compact coordinates $0\leq \eta, \zeta < 2\pi$ describe  the torus.
For all fields, we restrict attention to the case in which derivatives with respect to 
$\eta$ and $\zeta$ vanish identically, dimensionally reducing the model.
The anti-periodic boundary conditions for all the fermions
along the $\eta$ direction, combined with the dimensional reduction ansatz,
 sets all the fermionic fields to vanish identically.

\begin{table}
\caption{The field content of the model, organised in terms of the irreducible representations of the symmetries in 
$D=7$ dimensions ($SO(5)$ multiplets), as well as $D=5$ and $D=4$ dimensions
($SO(4)$ multiplets). In the case of the language in  $D=4$ dimensions, we indicate the field content in terms of 
the massive representations of the Poincar\'e group, and keep into account the
gauge-invariant combinations only.
The irreducible representations for which we indicate $N_{\rm dof}= -$  refer to cases where
either the degrees of freedom combine into gauge invariant combinations with other fields, or
have already been included due to the self-duality of the $3$-forms.
}
\label{Fig:Fields}
\tiny
\begin{center}
\begin{tabular}{|c|c|c||c|c|c||c|c|c|}
\hline\hline
\multicolumn{3}{|c|}{{
$D=7$, $SO(5)$,
}} &
\multicolumn{3}{|c|}{{
$D=5$, $SO(4)$,
}} &
\multicolumn{3}{|c|}{{
$D=4$, $SO(4)$,
}}
\cr
\multicolumn{3}{|c|}{{
{\rm massless irreps.}
}} &
\multicolumn{3}{|c|}{{
{\rm massless irreps.}
}} &
\multicolumn{3}{|c|}{{
{\rm massive irreps.}
}}
\cr
\hline\hline
{\rm Field} & $SO(5)$ & $N_{\rm dof}$  
&{\rm Field} & $SO(4)$ & $N_{\rm dof}$  
&{\rm Field} & $SO(4)$ & $N_{\rm dof}$  \cr
\hline\hline
$\hat{g}_{\hat{M}\hat{N}}$ & $1$ & $14$ &
$g_{MN}$ & $1$ & $5$ &
$g_{\mu\nu}$ & $1$ & $5$ \cr
 & &  &
 & &  &
$g_{\mu5}$ & $1$ & $-$ \cr
 & &  &
 & &  &
$g_{55}$ & $1$ & $-$ \cr
 & &  &
$\chi_M$ & $1$ & $3$ &
$\chi_{\mu}$ & $1$ & $3$ \cr
 & &  &
 & &  &
$\chi_{5}$ & $1$ & $-$ \cr
 & & &
$\omega_{M}$ & $1$ & $3$ &
$\omega_{\mu}$ & $1$ & $3$ \cr
 & &  &
 & &  &
$\omega_{5}$ & $1$ & $-$ \cr
 & & &
$\omega_6$ & $1$ & $1$ &
$\omega_6$ & $1$ & $1$ \cr
 & & &
$\chi$ & $1$ & $1$ &
$\chi$ & $1$ & $1$ \cr
& & &
$\omega$ & $1$ & $1$ &
$\omega$ & $1$ & $1$ \cr
\hline
$\Pi_{\alpha}^{\,\,\,i}$ & $14$ & $14$ &
$\phi$ & $1$ & $1$ &
$\phi$ & $1$ & $1$ \cr
&&&
$\pi^{\hat{A}}$ & $4$ & $4$ &
$\pi^{\hat{A}}$ & $4$ & $4$ \cr
&&&
$s^{\tilde{A}}$ & $9$ & $9$ &
$s^{\tilde{A}}$ & $9$ & $9$ \cr
\hline
${\cal A}_{\hat{M}}^{\,\,\,A}$ & $10$ & $50$ &
$A_{M}^{\,\,\,\hat{A}}$  & $4$ & $12$ &
$A_{\mu}^{\,\,\,\hat{A}}$  & $4$ & $12$ \cr
&&&
&&&
$A_{5}^{\,\,\,\hat{A}}$  & $4$ & $-$ \cr
&&&
$A_{6}^{\,\,\,\hat{A}}$  & $4$ & $4$ &
$A_{6}^{\,\,\,\hat{A}}$  & $4$ & $4$ \cr
&&&
$A_{7}^{\,\,\,\hat{A}}$  & $4$ & $4$ &
$A_{7}^{\,\,\,\hat{A}}$  & $4$ & $4$ \cr
&&&
$A_{M}^{\,\,\,\bar{A}}$  & $6$ & $18$ &
$A_{\mu}^{\,\,\,\bar{A}}$  & $6$ & $18$ \cr
&&&
&&&
$A_{5}^{\,\,\,\bar{A}}$  & $6$ & $-$ \cr
&&&
$A_{6}^{\,\,\,\bar{A}}$  & $6$ & $6$ &
$A_{6}^{\,\,\,\bar{A}}$  & $6$ & $6$ \cr
&&&
$A_{7}^{\,\,\,\bar{A}}$  & $6$ & $6$ &
$A_{7}^{\,\,\,\bar{A}}$  & $6$ & $6$ \cr
\hline
$S_{\hat{M}\hat{N}\hat{O}\,\alpha}$ & $5$ & $50$ &
$S_{6MN\,\bar{\beta}}\sim B_{MN\,\bar{\beta}}$ & $4$ & $12$ &
$S_{6\mu\nu\,\bar{\beta}}\sim B_{\mu\nu\,\bar{\beta}}$ & $4$ & $12$ \cr
&&&
&&&
$S_{6\mu 5\,\bar{\beta}}\sim X_{\mu\,\bar{\beta}}$ & $4$ & $12$ \cr
&&&
$S_{6MN\,5}\sim B_{MN\,5}$ & $1$ & $3$ &
$S_{6\mu\nu\,5}\sim B_{\mu\nu\,5}$ & $1$ &$3$ \cr
&&&
&&&
$S_{6\mu 5\,5}\sim X_{\mu\,5}$ & $1$ & $3$ \cr
&&&
$S_{7MN\,\bar{\beta}}\sim B^{\prime}_{MN\,\bar{\beta}}$ & $4$ & $12$ &
$S_{7\mu\nu\,\bar{\beta}}\sim B^{\prime}_{\mu\nu\,\bar{\beta}}$ & $4$ & $-$ \cr
&&&
&&&
$S_{7\mu 5\,\bar{\beta}}\sim X^{\prime}_{\mu\,\bar{\beta}}$ & $4$ & $-$ \cr
&&&
$S_{7MN\,5}\sim B^{\prime}_{MN\,5}$ & $1$ & $3$ &
$S_{7\mu\nu\,5}\sim B^{\prime}_{\mu\nu\,5}$ & $1$ &$-$ \cr
&&&
&&&
$S_{7\mu 5\,5}\sim X^{\prime}_{\mu\,5}$ & $1$ & $-$ \cr
&&&
$S_{67M\,\bar{\beta}}$ & $4$ & $12$ &
$S_{67\mu\,\bar{\beta}}$ & $4$ & $12$ \cr
&&&
&&&
$S_{675\,\bar{\beta}} (+\varphi_{\bar{\beta}})$ & $4$ & $4$ \cr
&&&
$S_{67M\,5}$ & $1$ & $3$ &
$S_{67\mu\,5}$ & $1$ & $3$ \cr
&&&
&&&
$S_{675\,5} (+\varphi_5)$ & $1$ & $1$ \cr
&&&
$S_{MNO\,\bar{\beta}}$ & $4$ & $4$ &
$S_{\mu\nu\sigma\,\bar{\beta}}$ & $1$ & $-$ \cr
&&&
&&&
$S_{\mu\nu5\,\bar{\beta}}$ & $1$ & $-$ \cr
&&&
$S_{MNO\,5}$ & $1$ & $1$ &
$S_{\mu\nu\sigma\,5}$ & $1$ & $-$ \cr
&&&
&&&
$S_{\mu\nu5\,5}$ & $1$ & $-$ \cr
\hline\hline
\end{tabular}
\end{center}
\end{table}

The field-strength tensors for $\chi_M$ and $\omega_M$, obtained by expanding the 
Einstein-Hilbert action with the ansatz in Eq.~(\ref{Eq:7}), are given  by
\beqs
F^{\,\,\,\,(\chi)}_{MN}&\equiv&\partial_M\chi_N-\partial_N\chi_M\,,\\
F^{\,\,\,\,(\omega)}_{MN}&\equiv&\partial_M\omega_N-\partial_N\omega_M
+\left(\chi_M\partial_N\omega_6-\chi_N\partial_M\omega_6\right)\,.
\label{Eq:omegaF}
\eeqs
In the following, consistently with our treatment of the other fields,
 we ignore the last two terms in Eq.~(\ref{Eq:omegaF}), because we assume that in the vacuum
$\langle \chi_M\rangle = \langle \omega_M\rangle=\langle \omega_6 \rangle=0$, and we retain 
in the action only terms up to quadratic in the
fields that have vanishing VEV.

While we allow $\langle \chi \rangle \neq 0 \neq \langle \omega \rangle$,
 we consider  background solutions of the form:
\beqs
\di s^2_5 &=&e^{2A(r)}\di x_{1,3}^2 + \di r^2\,.
\eeqs
We further  restrict the background solutions to obey the constraint $A=\frac{5}{2}\chi +\omega$,
so that the factors in front of the $\di x_{1,3}^2$ and $\di \zeta^2$ terms in the metric
in Eq.~(\ref{Eq:7}) are identical.
By defining ${\cal A}\equiv A-\chi=\frac{3}{2}\chi+\omega$ 
and $\di \r \equiv e^{-\chi} \di r$, the background
metric  reads
\beqs
\di s^2_7 &=& \di \r^2 +e^{2{\cal A}}\left(\di x_{1,3}^2+e^{-4\omega} \di \eta^2
+\di \zeta^2\right)\,,
\eeqs
making it visible that domain-wall (DW) solutions preserving Poincar\'e invariance in six dimensions
have $\langle \omega \rangle=0$ and AdS$_7$ solutions have constant $\partial_{\rho} {\cal A}$.

We rewrite the $128$ bosonic degrees of freedom 
and their action
in terms of fields in five dimensions.
We start this exercise with some counting.
The whole field content is enumerated in Table~\ref{Fig:Fields}.
The $14$ scalars remain unchanged, but for the fact that they depend only on the $x^M$ coordinates.
Yet, for reasons that we discussed when writing the quadratic action in Eq.~(\ref{Eq:quad}),
we find it convenient to treat separately the field $\phi$,
the $4$ pNGBs $\pi^{\hat{A}}$, and the $9$ scalars $s^{\tilde{A}}$. 
The $14$ degrees of freedom of the graviton decompose as $14=5\oplus 3\oplus 3\oplus 1\oplus 1 \oplus 1$,
in terms of massless representations of the Poincar\'e group in five dimensions.
These include  the graviton in five dimensions, beside the aforementioned two vectors 
$\chi_M$ and $\omega_M$, and three real scalars $\chi$, $\omega$, and $\omega_6$---as defined  in Eq.~(\ref{Eq:7}).
The additional $U(1)^2$ gauge  symmetry---beside $SO(5)$---realizes  the isometries of the torus.

The decomposition of the fifty degrees of freedom carried by the $1$-forms
 is straightforward, as $10 \otimes 5=(4\oplus 6)\otimes (3\oplus 1\oplus 1)$, where the first factor refers to
 the decomposition of $SO(5)$ representations  onto  $SO(4)$ ones, while the second to the (massless) Poincar\'e  representations 
 decomposed from seven to
  five dimensions.
The $10$ vector fields ${A}_{M}^{\,\,\,A}$ are supplemented by
$10+10$ additional real scalars ${A}_6^{\,\,\,A}$ and ${ A}_7^{\,\,\,A}$, 
 split into $10=4 \oplus 6$ of $SO(4)$.

Some  details about the decomposition of the $3$-forms can be found in Appendix~\ref{Sec:3}.
It  is useful to start with massive representations in the counting exercise.
The $SO(5)$ indexes follow the decomposition 
$5=1\oplus 4$ of $SO(4)$. A
 massive $3$-form in seven dimensions would decompose as $20=4\oplus 6 \oplus 6 \oplus 4$,
where the two $4$s denote massive vectors, while the two $6$s refer to massive $2$-forms.\footnote{
While massless $1$-forms are equivalent to massless $2$-forms in $D=5$ dimensions, 
 this is not so for massive $1$-forms and $2$-forms, that are distinct representations of the Poincar\'e group in five dimensions.
  The massive vector has the same number of propagating degrees of freedom as a massless vector together with a scalar, 
 while the massive $2$-form is obtained by soldering two massless vectors (see for instance Refs.~\cite{Noronha:2003vp,Samtleben:2008pe}).}
 Imposing the self-duality condition  identifies the pairs of $4$s and $6$s, and 
 yields, for each $3$-form,
 one massive $1$-form and one massive $2$-form ($10=4\oplus 6$).

We can as well perform the counting exercise in terms of (on-shell) massless fields.
We have one graviton ($5$ degrees of freedom), 
 twenty-seven massless vectors ($3$ degrees of freedom each),  and 
forty-two scalars, for a grand total of $128$
propagating bosonic degrees of freedom, matching the field content of
 maximal supergravity in $D=5$ dimensions.

The action of Eq.~\eqref{Eq:Pernici}---and hence its truncated version in Eq.~(\ref{Eq:quad})---is written in a particular gauge. We now write an action in $D=5$ dimensions, which captures all the degrees of freedom, and restores the gauge invariance (for details, see Appendix~\ref{Sec:3}).
For convenience, we isolate the three active scalars $\Phi^a=\{\phi,\omega,\chi\}$
that have non-trivial background profiles, from the other scalar fields $\Phi^{(0) a}$, 
to write the action:
\beqs
\label{Eq:Action5}
{\cal S}_5&=&\int \di^5 x \sqrt{-g_5}
\left\{ \frac{{ R}}{4} -\frac{1}{2}g^{MN}G_{ab}\partial_M\Phi^a\partial_N\Phi^b
 - {\cal V}_5(\Phi^a)
 +\right.\\
 &&\nonumber
 \left.
 -\frac{1}{2}g^{MN}G^{(0)}_{ab}\partial_M\Phi^{(0)a}\partial_N\Phi^{(0)b}
 -\frac{1}{2}m^{(0)2}_{ab}\Phi^{(0)a}\Phi^{(0)b}
  +\right.\\
 &&\nonumber
 \left.
 -\frac{1}{2} g^{MN} G^{(1)}_{AB}
 {\cal H}_{\ \ \ \ M}^{(1)A} {\cal H}^{(1)B}_{\ \ \ \ N}
  -\frac{1}{4} g^{MO} g^{NP} H^{(1)}_{AB}F_{\,\,\,\,MN}^{A}F^{B}_{\,\,\,\,OP}
  +\right.\\
 &&\nonumber
 \left.
 -\frac{1}{4} g^{MO} g^{NP} H^{(2)}_{\Gamma\Delta} {\cal H}^{(2)\Gamma}_{\,\,\,\,\,\,\,\,\,\,MN}{\cal H}^{(2)\Delta}_{\,\,\,\,\,\,\,\,\,\,\,OP}\,
-\frac{1}{12} g^{MP} g^{NQ} g^{OR} K^{(2)}_{\Gamma\Delta}\, {\cal H}^{(3)\Gamma}_{\,\,\,\,\,\,\,\,\,\,MNO}{\cal H}^{(3)\Delta}_{\,\,\,\,\,\,\,\,\,\,\,PQR}
 \right\}\,.
\eeqs
We now describe in detail each of the terms  in Eq.~(\ref{Eq:Action5}), and provide
explicit forms for all the entries. $R$ is the Ricci scalar in five dimensions, defined with the conventions 
described in Appendix~\ref{Sec:Formalism}.
The sigma-model metric for the three active scalars $\Phi^a=\{\phi,\omega,\chi\}$ is
 \beqs
 G&=&\left(\begin{array}{ccc}
 \frac{1}{2}&&\cr
 &1&\cr
 &&\frac{15}{4}\cr
 \end{array}\right)\,,
 \eeqs
 where (here and in the following) we conventionally leave blank the vanishing entries
 in the matrices, in order to lighten the notation.
 The potential  is ${\cal V}_5(\Phi^a)=e^{-2\chi}{\cal V}_7(\phi)$, and depends only 
 on $\phi$ and $\chi$, while $\omega$ has no potential. 
 
 The thirty ordinary scalars that have trivial background values are denoted by
 $\Phi^{(0)a}=\left\{s^{\tilde{A}}, \omega_6, A_6^{\,\,\,\,\hat{A}}, A_7^{\,\,\,\,\hat{A}}, 
 A_6^{\,\,\,\,\bar{A}}, A_7^{\,\,\,\,\bar{A}}\right\}$, and they have sigma-model metric
 that we find convenient to write in block-diagonal form as follows:
\beqs
G^{(0)}=\frac{1}{4}\left(\begin{array}{c|c|c|c|c|c}
&&&&&\cr
\mathbb{1}_{9}&&&&&\cr
&&&&&\cr
\hline
&&&&&\cr
&
e^{4\omega}&&&&\cr
&&&&&\cr
\hline
&&&&&\cr
&&
e^{\frac{3\phi}{\sqrt{5}}-3\chi+2\omega}\,\mathbb{1}_{4}&&&\cr
&&&&&\cr
\hline
&&&&&\cr
&&&
e^{\frac{3\phi}{\sqrt{5}}-3\chi-2\omega}\,\mathbb{1}_{4}&&\cr
&&&&&\cr
\hline
&&&&&\cr
&&&&
e^{-\frac{2\phi}{\sqrt{5}}-3\chi+2\omega}\,\mathbb{1}_{6}&\cr
&&&&&\cr
\hline
&&&&&\cr
&&&&&
e^{-\frac{2\phi}{\sqrt{5}}-3\chi-2\omega}\,\mathbb{1}_{6}\cr
&&&&&\cr
\end{array}\right),
\label{Eq:probesG}
\eeqs
where $\mathbb{1}_9$, $\mathbb{1}_4$, and $\mathbb{1}_6$ are $9\times 9$, $4\times 4$, and $6\times 6$ 
identity matrices, respectively.
The matrix of the squares of the masses is diagonal as well, and is written as follows:
\beqs
\label{Eq:probesm}
\frac{m^{(0)\,2}}{m^2}=\left(\begin{array}{c|c|c|c|c|c}
&&&&&\cr
-\frac{e^{-2\chi-\frac{3\phi}{\sqrt{5}}}}{2}\,\mathbb{1}_{9}&&&&&\cr
&&&&&\cr
\hline
&0&&&&\cr
\hline
&&&&&\cr
&&
{e^{-5\chi+2\omega}}{}\sinh^2\left(\frac{\sqrt{5}}{2}\phi\right)\,\mathbb{1}_{4}
&&&\cr
&&&&&\cr
\hline
&&&&&\cr
&&&{e^{-5\chi-2\omega}}\sinh^2\left(\frac{\sqrt{5}}{2}\phi\right)\,\mathbb{1}_{4}&&\cr
&&&&&\cr
\hline
&&&&&\cr
&&&&\mathbb{0}_{6}&\cr
&&&&&\cr
\hline
&&&&&\cr
&&&&&\mathbb{0}_{6}\cr
&&&&&\cr
\end{array}\right),
\eeqs
where $m^2=1$ is the mass appearing in Eq.~(\ref{Eq:Pernici}), and $\mathbb{0}_6$ 
is a vanishing $6\times 6$ matrix.

For the seventeen 1-form fields, we adopt  the convenient choice of 
 basis given by $V^A_{\,\,\,M}=\{\chi_M, \omega_M, 
A^{\bar{A}}_{\,\,\,M}, A^{\hat{A}}_{\,\,\,M}, S_{67M\,\bar{\beta}}, S_{67M\,5}\}$.
Disregarding the self-interactions,
the field strengths are defined by $F^A_{\,\,\,MN}\equiv 2\partial_{[M}V^{A}_{\,\,\,N]}$, and
the kinetic terms are controlled by the following:
\beqs
H^{(1)}=\frac{1}{4}\left(\begin{array}{c|c|c|c|c|c}
&&&&&\cr
e^{5\chi-2\omega}&&&&&\cr
&&&&&\cr
\hline
&&&&&\cr
&e^{5\chi+2\omega}&&&&\cr
&&&&&\cr
\hline
&&&&&\cr
&&
e^{-\frac{2\phi}{\sqrt{5}}+2\chi}\,\mathbb{1}_{6}
&&&\cr
&&&&&\cr
\hline
&&&&&\cr
&&&
e^{\frac{3\phi}{\sqrt{5}}+2\chi}\,\mathbb{1}_{4}&&\cr
&&&&&\cr
\hline
&&&&&\cr
&&&&4e^{-4\chi-\frac{\phi}{\sqrt{5}}}\mathbb{1}_{4}&\cr
&&&&&\cr
\hline
&&&&&\cr
&&&&&4e^{-4\chi+\frac{4\phi}{\sqrt{5}}}\cr
&&&&&\cr
\end{array}\right).
\label{Eq:H(1)}
\eeqs

Via the Higgs mechanism, nine among the $1$-forms acquire a mass (we will refer to these as
axial-vector fields),
by eating up as many pseudoscalar fields.
We hence define  gauge-invariant combinations of 1-forms and 
derivatives of the pseudoscalars:
\beqs
{\cal H}^{(1)\,\,A}_{\,\,\,\,\,\,\,M}
&\equiv&
\left\{
0,
0,
0,
\frac{\partial_M\pi^{\hat{A}}}{2}+m A^{\hat{A}}_{\,\,\,\,M},
\partial_M\varphi_{\bar{\beta}}+m S_{67M\,\bar{\beta}},
\partial_M\varphi_{5}+m S_{67M\,5}
\right\}\,.
\label{Eq:Hdef}
\eeqs
The four fields $\pi^{\hat{A}}$  have been introduced to parameterise the matrices $U$.
We made use of the relation $g=2m$,
in writing the first non-trivial entry in Eq.~(\ref{Eq:Hdef}). 
We introduce here five additional scalars $\varphi_{\alpha}$,
that had been gauge-fixed away in Eq.~(\ref{Eq:Pernici})---we remind the Reader 
the action has been lifted from
Ref.~\cite{Pernici:1984xx}. By doing so, we reinstate manifest gauge invariance.
 It follows that the mass matrix for the 1-forms  is governed by the following:
\beqs
\label{Eq:G(1)}
G^{(1)}=\left(\begin{array}{c|c|c|c|c|c}
0&&&&&\cr
\hline
&0&&&&\cr
\hline
&&&&&\cr
&&
\mathbb{0}_{6}
&&&\cr
&&&&&\cr
\hline
&&&&&\cr
&&&\sinh^2\left(\frac{\sqrt{5}}{2}\phi\right)\,\mathbb{1}_{4}
&&\cr
&&&&&\cr
\hline
&&&&&\cr
&&&&e^{-6\chi+\frac{\phi}{\sqrt{5}}}\mathbb{1}_{4}&\cr
&&&&&\cr
\hline
&&&&&\cr
&&&&&e^{-6\chi-\frac{4\phi}{\sqrt{5}}}\cr
&&&&&\cr
\end{array}\right).
\eeqs

The treatment of the massive $2$-form fields is provided by defining the following objects:
\beqs
{\cal H}^{(2)\Gamma}_{\,\,\,\,\,\,\,\,MN}&=&{\cal F}^{\Gamma}_{\,\,\,\,MN} + m B^{\Gamma}_{\,\,\,\,MN}\,,\\
{\cal F}^{\Gamma}_{\,\,\,\,MN}&=&2 \partial_{[M}{\cal A}^{\prime\,\Gamma}_{\,\,\,\,N]}\,,\\
{\cal H}^{(3)\Gamma}_{\,\,\,\,MNO}&=&3\partial_{[M}B^{\Gamma}_{\,\,\,\,NO]}\,,
\eeqs
where $B^{\Gamma}_{\,\,\,\,MN}\equiv \{S_{6MN\,\bar{\beta}}, S_{6MN\,5}\}$.
In analogy with $\varphi_{\bar{\beta}}$ and $\varphi_5$,  the vectors
${\cal A}^{\prime}_{M\,\bar{\beta}}$ and ${\cal A}^{\prime}_{M\,5}$
had been set to zero in Eq.~(\ref{Eq:Pernici}), as in
Ref.~\cite{Pernici:1984xx}, and their introduction 
 reinstates manifest gauge invariance.
The action of the massive $2$-forms is then determined by the two matrices
\beqs
\label{Eq:H(2)}
H^{(2)}&=&\left(\begin{array}{c|c}
&\cr
e^{-\chi+2\omega+\frac{\phi}{\sqrt{5}}}\,\mathbb{1}_{4}&\cr
&\cr
\hline
&\cr
&e^{-\chi+2\omega-\frac{4\phi}{\sqrt{5}}}\cr
&\cr
\end{array}\right),\\
\label{Eq:K(2)}
K^{(2)}&=&\left(\begin{array}{c|c}
&\cr
e^{\chi+2\omega-\frac{\phi}{\sqrt{5}}}\,\mathbb{1}_{4}&\cr
&\cr
\hline
&\cr
&e^{\chi+2\omega+\frac{4\phi}{\sqrt{5}}}\cr
&\cr
\end{array}\right).
\eeqs

\subsection{Background solutions}
\label{Sec:solutions}

In this paper we generalise the study of fluctuations
to encompass the whole spectrum of bosonic fluctuations
around the backgrounds presented in Ref.~\cite{Elander:2020fmv}. 
We hence report in this short section 
only the essential elements that  identify the backgrounds, while 
 more extended discussions can be found in
 Ref.~\cite{Elander:2020fmv} (and references therein).

In seven dimensions, the scalar potential  admits two critical  points. 
One with $\langle \phi \rangle =\phi_{UV}\equiv 0$ plays a central role in this paper,
and corresponds to the strongly-coupled fixed point that defines the dual field theory in six dimensions. 
At this critical point, with the current normalisations,  
one finds that ${\cal V}_7(\phi_{UV})=-\frac{15}{8}$.
The other  critical point   has
$\langle \phi\rangle=\phi_{IR}\equiv -\frac{1}{\sqrt{5}}\log (2)$, 
and   ${\cal V}_7(\phi_{IR})=-\frac{5}{2^{7/5}}$,
but is known to be perturbatively unstable~\cite{Pernici:1984zw}.

Having implemented the toroidal compactification described 
in the previous section, we consider background solutions for which 
 $\lim_{\r\rightarrow +\infty}\phi=\phi_{UV}=0$, obtained by solving the equations of motion~\eqref{eq:backgroundEOM1} and~\eqref{eq:backgroundEOM2}. In the asymptotic regime of the geometry,
one recovers (locally)
Poincar\'e invariance
 in six dimensions.
These solutions are written as the following  power
series expansion in  the small coordinate $z\equiv e^{-\r/2}$:
\beqs
\phi(z)&=& \label{Eq:UV1}
\phi_2 z^2+ \left(\phi_4-\frac{18 \phi_2^2 \log (z)}{\sqrt{5}}\right)z^4
+   \\
   && \nonumber
+
   \left(\frac{162}{5} \phi_2^3 \log (z)-\frac{637 \phi_2^3}{30}-\frac{9 \phi_2
   \phi_4}{\sqrt{5}}\right)z^6+   \\
   && \nonumber
     +\frac{1}{600} z^8 \left(11921 \sqrt{5} \phi_2^4+2480
   \phi_2^2 \phi_4-180 \sqrt{5} \phi_4^2\right)+   \\
   && \nonumber
   +\frac{6}{25} z^8 \phi_2^2 \log (z) \left(45 \phi_4-62 \sqrt{5} \phi_2^2\right)
    -\frac{486 z^8 \phi_2^4 \log ^2(z)}{5 \sqrt{5}}
   +{\cal O}(z^{10})\,,
\\
   \omega(z)&= &\label{Eq:UV2}
\omega_{U}+\omega_{6}z^6+{\cal O}(z^{10})
\,,\\
\chi(z)&=&   \label{Eq:UV3}
 \chi_{U}-\frac{2}{3}\log (z)-\frac{\phi_2^2 z^4}{30}\,
 +   \\
   && \nonumber
+\frac{2  }{675}
  \left(\frac{675}{2}\chi_6-150\,\omega_{6}+72 \sqrt{5} \phi_2^3 \log (z)-6 \sqrt{5} \phi_2^3-20
   \phi_2 \phi_4\right)z^6
 +   \\
   && \nonumber
     +\frac{z^8}{1200} \left(1355
   \phi_2^4+108 \sqrt{5} \phi_2^2 \phi_4-40 \phi_4^2\right) +   \\
   && \nonumber
   +\frac{3}{50} z^8 \phi_2^2 \log  (z) \left(4 \sqrt{5} \phi_4-27 \phi_2^2\right)
      -\frac{54}{25} z^8 \phi_2^4 \log ^2(z)
   +{\cal O}(z^{10})
\,,\\
A(z)&=&   \label{Eq:UV4}
A_{U}-\frac{5}{3}\log (z)-\frac{\phi_2^2
   z^4}{12}
+   \\
   && \nonumber
   +\frac{ 1}{270} \left(\frac{135}{2}\chi_6-30\,\omega_{6}+144 \sqrt{5} \phi_2^3 \log (z)-12
   \sqrt{5} \phi_2^3-40 \phi_2 \phi_4\right)z^6+   \\
   && \nonumber
   +z^8 \left(\frac{271 \phi_2^4}{96}+\frac{9 \phi_2^2 \phi_4}{8 \sqrt{5}}
   -\frac{\phi_4^2}{12}\right)+   \\
   && \nonumber
   +z^8 \log (z) \left(\frac{3 \phi_2^2 \phi_4}{\sqrt{5}}-\frac{81 \phi_2^4}{20}\right)
      -\frac{27}{5} z^8 \phi_2^4 \log ^2(z)
   +{\cal O}(z^{10})
\,.
\eeqs
This expansion is characterised by seven integration constants: 
$\phi_2$, $\phi_4$, $\omega_U$, $\omega_6$, $\chi_U$,
$\chi_6$, and $A_U$. 

The  DW solutions form a  subclass with $\omega_U=\omega_6=\chi_6=0$
 and $\chi_U=\frac{2}{5}A_U$, leaving $A_U$, $\phi_2$ and $\phi_4$ 
 as independent non-trivial free parameters.
As anticipated, we impose the milder constraint $A=\frac{5}{2}\chi+\omega$:
what we call confining solutions have $\chi_6=0$, and $A_U=\frac{5}{2}\chi_U+\omega_U$.
We may further require that $A_U=0=\chi_U$, via a rescaling of the coordinates $x^{\mu}$
and a shift of $\r$, in such a way as to identify
 physically equivalent solutions. This would leave $\phi_2$, $\phi_4$ 
and $\omega_6$ as free parameters.

The confining solutions form a 1-parameter family:
 two of  the three parameters are fixed (non-trivially)
  by requiring that the solutions be regular and smooth at the end of space,
  as
the geometry  closes off at some finite value $\r_o$.
While the value of $\phi(\r)$ evolves, either towards either positive or negative values,
the circle parametrised by $\eta$ shrinks to zero size when $\r\rightarrow \r_o$, so
that $\phi(\r)$ is finite for any $\r_o\leq \r < \infty$.

The IR expansion of 
the confining  solutions depends on two harmless constants $\chi_I$ and $\omega_I$ (besides
 $\r_o$) and on the physically meaningful  parameter $\phi_I$. It reads as
 follows:
\beqs
\phi(\r)&=&
   \label{Eq:IR1}
\phi_I
   -\frac{1}{2 \sqrt{5}}(\r-\r_o)^2
   e^{-\frac{8 \phi_I}{\sqrt{5}}} \left(-3 e^{\sqrt{5} \phi_I}+2 e^{2
   \sqrt{5} \phi_I}+1\right)+\cdots
\,,\\
\omega(\r)&=&
   \label{Eq:IR2}
\omega_I-\frac{\log
   (\r-\r_o)}{2}
   +\frac{1}{40} (\r-\r_o)^2 e^{-\frac{8 \phi_I}{\sqrt{5}}} \left(8
   e^{\sqrt{5} \phi_I}+8 e^{2 \sqrt{5} \phi_I}-1\right) +\cdots
   \,,\\
\chi(\r)&=&
   \label{Eq:IR3}
\chi_I   +\frac{\log (\r-\r_o)}{3}
+   \\
   && \nonumber
  +\frac{1}{6000}(\r-\r_o)^4 e^{-\frac{16 \phi_I}{\sqrt{5}}} 
 \left(32 e^{\sqrt{5} \phi_I}-56 e^{2 \sqrt{5} \phi_I}+224 e^{3 \sqrt{5} \phi_I}+32 e^{4 \sqrt{5} \phi_I}-7\right)+\cdots
\,,
\eeqs
while $A(\r)=\frac{5}{2}\chi(\r)+\omega(\r)$.

As we restrict attention to solutions flowing from the UV critical point, we must require
$\phi_I>\phi_{IR}$, but without upper bounds on  $\phi_I$.
As explained in Ref.~\cite{Elander:2020fmv}, the invariants ${\cal R}_7$,
  ${\cal R}_{7\,\hat{M}\hat{N}}{\cal R}_{7}^{\,\,\,\hat{M}\hat{N}}$, and 
${\cal R}^{\,\,\,\,\,\hat{P}}_{7\,\,\,\,\,\hat{M}\hat{N}\hat{Q}}{\cal R}_{7\,\hat{P}}^{\,\,\,\,\,\,\,\,\,\hat{M}\hat{N}\hat{Q}}$ are 
 finite.
The constraint $\omega_I=\frac{3}{2}\chi_I$ removes a possible conical singularity, and the internal angles 
have periodicity $2\pi$.

\begin{figure}[t]
\begin{center}
\includegraphics[width=14.2cm]{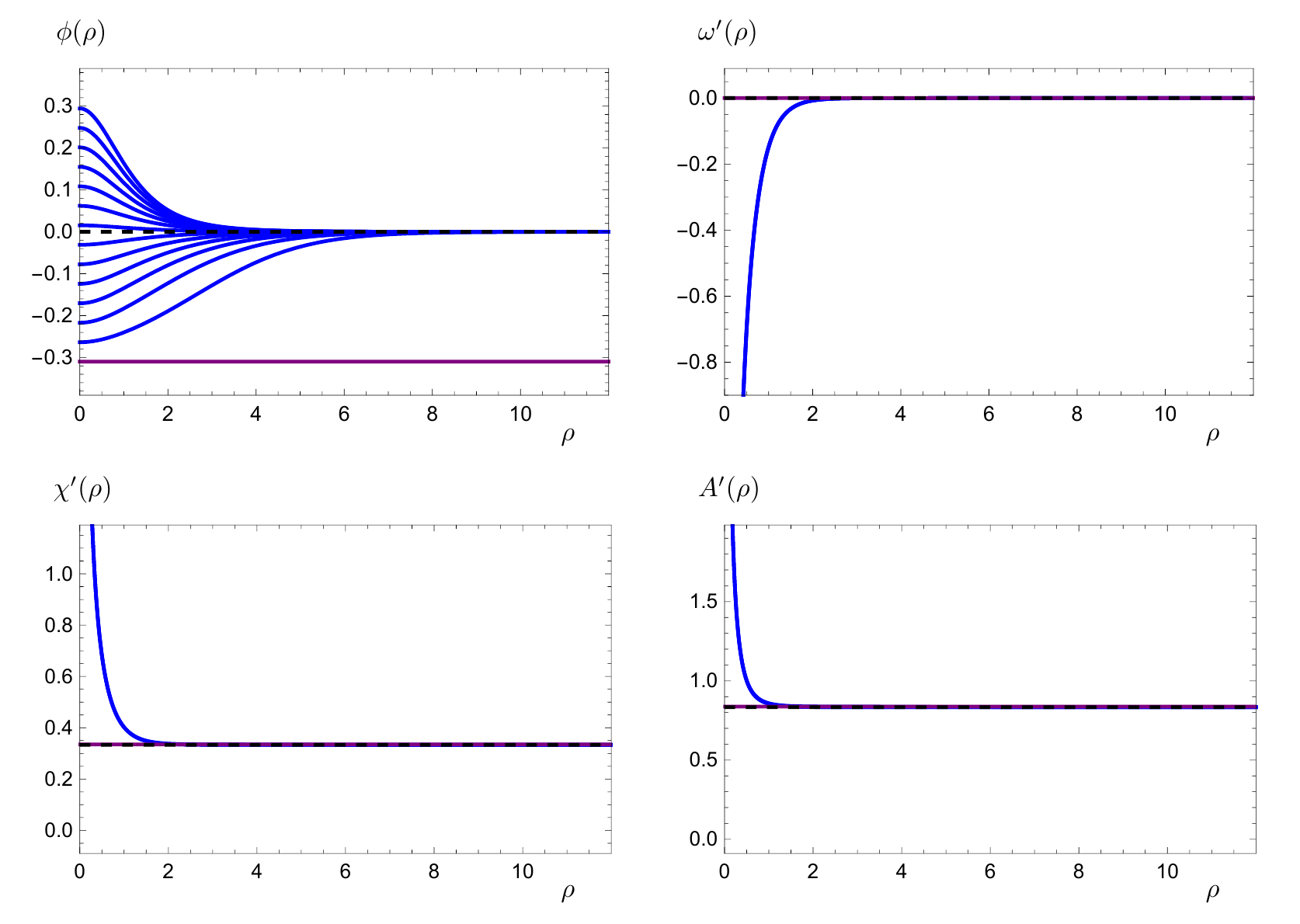}
\caption{Background fields in illustrative examples of confining solutions with $\r_o=0$.  
Left to right and top to bottom:
 the functions $\phi(\r)$,
$\omega^{\prime}(\r)$, $\chi^{\prime}(\rho)$, and $A^{\prime}(\r)$.
For comparison, the horizontal lines represent the DW solutions at the $\phi_{UV}$ (black, dashing) and 
$\phi_{IR}$ (purple, continuous).
In confining solutions, $\omega$, $\chi$, and $A$ diverge at the end of space, but $\phi$ is finite.
 }
\label{Fig:Bgs}
\end{center}
\end{figure}

These solutions are called confining, because the shrinking of the circle parametrised 
by $\eta$ is completely smooth, and, 
upon lifting the theory to type IIA, it is possible to apply the standard prescription that 
allows to compute the expectation value of a rectangular Wilson loop in the boundary theory, 
from which one can obtain a static potential between probe 
quarks that reproduces the expected asymptotic linear dependence on the quark separation---see Appendix~\ref{Sec:Lift} and references therein.
The $\phi=0$ solution  is the background discussed by Witten in Ref.~\cite{Witten:1998zw},
and the generalisations to the full class discussed here have been presented in Refs.~\cite{Elander:2013jqa} 
and~\cite{Elander:2020fmv}.

We depict examples of both confining and DW solutions in Fig.~\ref{Fig:Bgs}.
It is worth noticing that, while in the plot showing $\phi(\r)$ is it clearly possible 
to distinguish each non-trivial solution, including the DW cases for the two
critical choices $\phi(\r)=\phi_{UV}$ and $\phi(\r)=\phi_{IR}$,
by contrast, on the scale of the plots for the functions $\omega^{\prime}(\r)$, $\chi^{\prime}(\r)$, and $A^{\prime}(\r)$,
the difference between the various confining solutions is too small to be resolved, 
and so is the difference in curvature between the two DW solutions.

\section{Linearised fluctuations and mass spectra}
\label{Sec:spectrum}

We devote this section to the calculation of the mass spectra of bound states 
of the confining field theory, which the dictionary of gauge-gravity dualities identifies
with the spectrum of fluctuations around the gravity background.
We treat the fluctuations of metric and active scalars using the gauge-invariant formalism developed in 
Refs.~\cite{Bianchi:2003ug,Berg:2005pd,Berg:2006xy,Elander:2009bm,Elander:2010wd} 
(and~\cite{Elander:2018aub,Elander:2020csd}).
For the  $p$-forms, 
with $p=0, 1, 2$, we implement the 
$R_{\xi}$ gauge and focus only on physical combinations of the fluctuated fields that do not depend on the 
gauge-fixing ($\xi$) parameters, following the general principles 
enunciated in Ref.~\cite{Elander:2018aub}.
Extended details about the treatment of the $3$-form, in particular in reference to
the self-duality conditions, are  shown in Appendix~\ref{Sec:3}.
The technical details on how we perform the calculations 
are relegated to Appendix~\ref{Sec:Formalism}, as well as details about the notation, 
where appropriate, while in this section we discuss only the physical results.

In order to perform the numerical calculations yielding the spectrum, we introduce 
a regulating procedure, and the extrapolation to the physical results is obtained by
a process that resembles what in the lattice literature is referred to as improvement.
We define two boundaries $\r_1$ and $\r_2$ to the holographic coordinate 
so that $\r_1\leq \r \leq \r_2$, and add boundary-localised terms in the action, which determine the boundary conditions
obeyed by the fluctuations. We then compute the spectrum of  small fluctuations that 
obey such boundary conditions, identifying the discrete values of $M^2=-q^2$ (where $q^\mu$ is the four-momentum of the fluctuations) for which the system admits solutions.
The physical spectrum is recovered in the
 limit $\r_1\rightarrow \r_o$ and $\r_2\rightarrow +\infty$.
We could perform the calculations explicitly for finite $\r_1$ and $\r_2$, and then repeat the calculations
and extrapolate towards the physical limits, as was done for example in Ref.~\cite{Elander:2020fmv}.
Instead of doing so, we apply the boundary conditions to the asymptotic expansions 
of the fluctuations---see Appendix~\ref{Sec:ExpansionsUV} and~\ref{Sec:ExpansionsIR}---and then 
use what results in order to set up the boundary conditions
in the numerical study.
By doing so, the convergence of the spectra computed at finite cutoffs is much faster, and as
we shall see we obtained improved results in respect to the literature.
We notice in passing that for generic (non fine-tuned) choices of boundary terms this process 
selects the subleading behaviors in the solutions of the linearised equations,
in agreement with standard procedures of the gauge-gravity dictionary.
We will return to the case of special, fine-tuned choices, and their consequences,
in Section~\ref{Sec:bc}.

When we look at the spectrum of states of the boundary theory, by studying the fluctuations around the background solutions, it is best to count states in terms of massive representations of the Poincar\'e group.
We expect the following towers of four-dimensional 
composite states to emerge---see also Table~\ref{Fig:Fields}.
\begin{itemize}
\item A tower of spin-2 massive tensors, from the graviton (5 dofs each).
\item Three towers of scalars, related to gauge invariant combinations of the active scalars and the trace of the metric.
\item Thirty inactive scalars giving rise to as many towers, with the 
degeneracies of the $SO(4)$ multiplets they belong to ($9\oplus 1\oplus 4\oplus 4\oplus 6\oplus 6$).
\item Seventeen massive $1$-forms corresponding to towers of  spin-1 states, with degeneracies 
dictated by the representations of $SO(4)$ ($1\oplus 1 \oplus 4 \oplus 6 \oplus 4 \oplus 1$). Eight of these towers are referred to as `vectors' in the following, the other nine being dubbed `axial-vectors.'
\item Nine towers of pseudoscalars, transforming as 
 $4\oplus 4 \oplus 1$ of $SO(4)$---hence reaching the total of $42$ scalars---closely 
 associated with the nine aforementioned axial-vectors by the Higgs mechanism. 
 \item  Thirty dofs represented by the  $2$-forms, yielding 
ten more towers of massive vectors
 ($3$ dofs each), transforming as $4\oplus 1 \oplus 4\oplus 1$ of $SO(4)$.
\end{itemize}
The degeneracies due to $SO(4)$ representations 
imply that beside the tower of spin-2 states ($1$ sequence of mass eigenstates),
there are  $12$ sequences of masses corresponding to the spin-0 towers, 
and $10$ different sequences of masses for the spin-1 particles.
In summary, the $128$ bosonic degrees of freedom organise themselves
 in representations of $SO(4)$ and of the Poincar\'e
group to yield $23$ distinct sequences of mass eigenstates.

\subsection{Tensor and active scalars}
\label{Sec:spectrum1}

In this subsection, we consider the tower of states associated with
 the spin-2 massive graviton, and the three towers of scalars
obtained by fluctuating the three active scalars $(\phi, \chi, \omega)$.
The main results have already been presented elsewhere (see Ref.~\cite{Elander:2020fmv} and references therein),
and this short section serves mostly to make the presentation 
self-contained, as well as to cross-check that the results
be consistent with the literature.

\begin{figure}[t]
\begin{center}
\includegraphics[width=15.2cm]{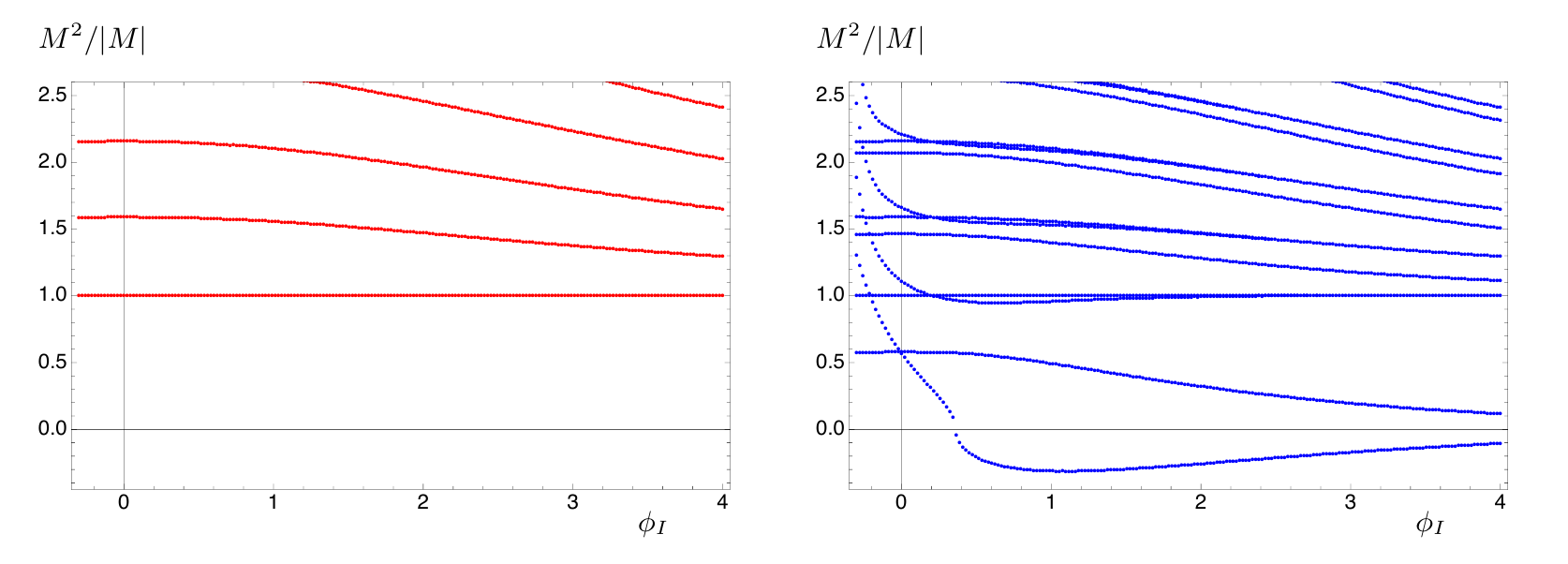}
\caption{Mass spectrum $M^2=-q^2$ of tensor modes (left panel) and active scalars (right panel)
as a function of the parameter $\phi_I$, normalised
 in units of the lightest tensor mode. 
Numerical calculations use $\r_1=10^{-10}$ and $\r_2=10$. 
 }
\label{Fig:SpectrumTensors}
\end{center}
\end{figure}

In Fig.~\ref{Fig:SpectrumTensors} we display the mass spectrum of fluctuations of the metric 
and of the three active scalars. We summarise in the caption 
of Fig.~\ref{Fig:SpectrumTensors} and in Appendix~\ref{Sec:Formalism},
respectively,
the details of the numerical and formal manipulations we implement to compute the mass spectrum.
We notice that while the equations and boundary conditions 
for this sector of the spectrum
are those in  Ref.~\cite{Elander:2020fmv}, the process
by means of which we implemented the boundary conditions improves
the convergence in respect to Ref.~\cite{Elander:2020fmv}, better removing spurious cutoff artefacts,
and hence the results on display in this paper are a numerical improvement 
upon the existing literature.

We normalise the spectra to the mass of the lightest tensor bound state, 
 to remove spurious dependences on  arbitrary additive integration constants in the
background values of $A$, $\chi$ and $\omega$.
The results are in agreement with the literature. 
In particular, we notice the emergence
 of a tachyonic state (with negative $M^2<0$), for backgrounds generated with 
large and positive values of $\phi_I$.
We notice however that while the tachyon appears first at
 $\phi_I\lsim 0.447$ in Ref.~\cite{Elander:2020fmv}, with the improvement 
 we implement this is now happening at $\phi_I\sim 0.36$.
In the region in which this state is light, it is also an approximate dilaton, as in Ref.~\cite{Elander:2020fmv}
it is shown that its composition consists predominantly of the fluctuations of the trace of the metric (which
 holographically corresponds to  the dilatation operator).

\subsection{All other scalars}
\label{Sec:spectrum2}

To the best of our knowledge, the rest of the spectrum has not been computed before for general $\phi_I$.
In particular, we report here the first calculation of the spectrum of all the spin-0 states
that descend from maximal supergravity.

\begin{figure}[t]
\begin{center}
\includegraphics[width=15.2cm]{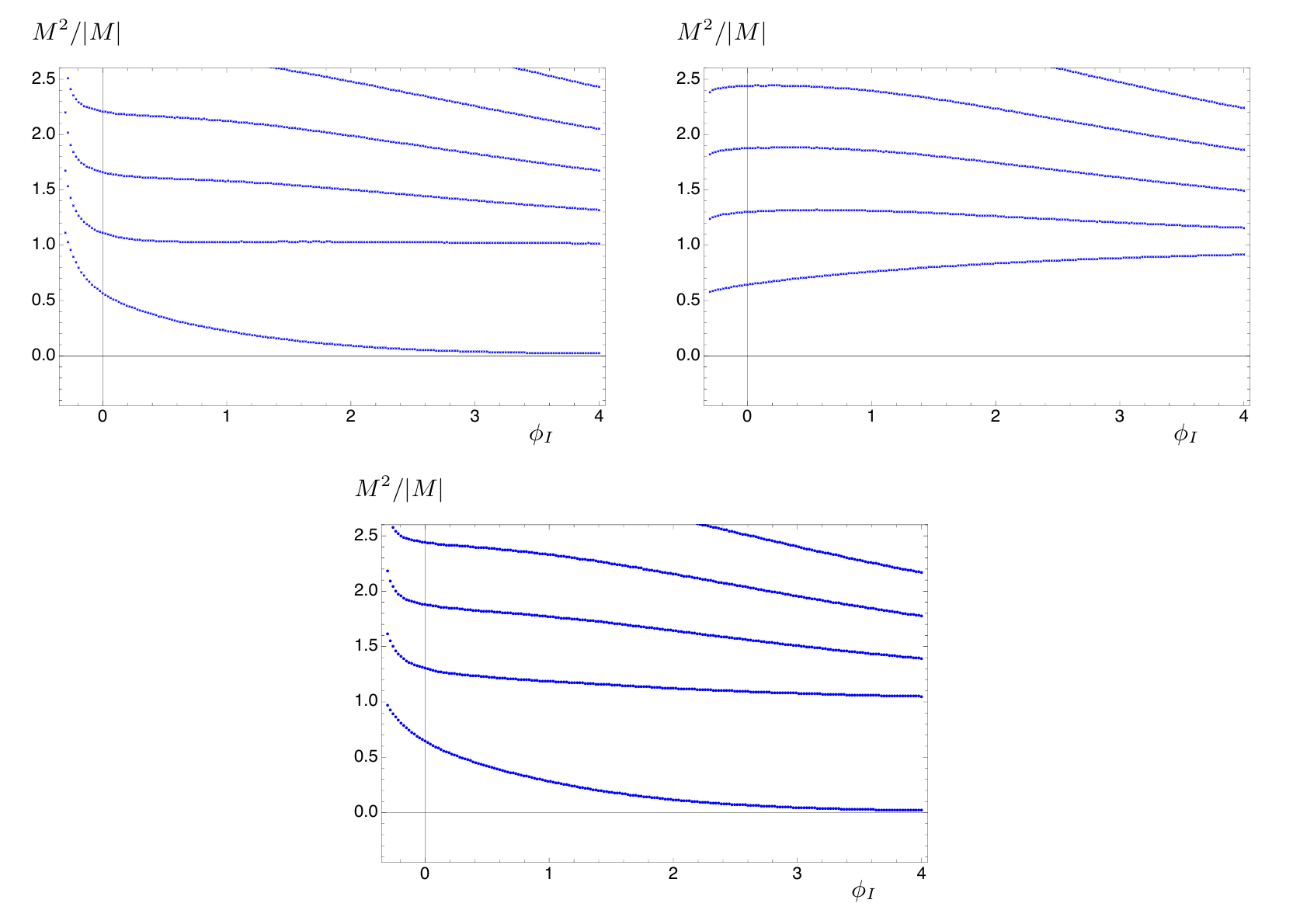}
\caption{
Mass spectrum, with $M^2=-q^2$, of the three groups of pseudoscalar modes
and their excited states as a function of the parameter $\phi_I$, normalised
 in units of the lightest tensor mode. The markers are chosen to match the representations under $SO(4)$.
Left to right and top to bottom: the fourth, fifth and sixth element of ${\cal H}^{(1)\,A}_{\,\,\,M}$ as in Eq.~(\ref{Eq:Hdef}).
 Numerical calculations use $\r_1=10^{-10}$ and $\r_2=10$. 
\label{Fig:Goldstones}}
\end{center}
\end{figure}

We start with the three towers of Goldstone bosons, and we report their mass spectra in 
Fig.~\ref{Fig:Goldstones}. 
As anticipated, they transform as $4 \oplus 4 \oplus 1$ of  $SO(4)$, respectively. 
The first of the towers
describes the $SO(5)/SO(4)$ coset, and the lightest states of this 
sequence correspond to the pNGBs of the dual theory.
They acquire a mass in the presence of explicit symmetry
 breaking in the dual theory. This is signalled by the presence of 
 $\phi_2\neq 0$, in the UV expansion of the background solutions in
 Eq.~(\ref{Eq:UV1}).
 Interestingly, we notice that when $\phi_I$ is positive and large, this 
 group of degenerate bound states becomes parametrically light. We know from Refs.~\cite{Elander:2020fmv,Roughley:2021suu} that this is the limit in which $\phi_4$ is enhanced with respect to $\phi_2$. This is the limit in which one intuitively expects to see Goldstone bosons, and the continuous connection between the region of parameter space with large and small $\phi_I$ is the central element suggesting to interpret the lightest excitations in these towers of states as pNGBs---in spite of their large mass when $\phi=0$.

The other two towers ($5$ dofs) correspond to the breaking of the $SO(6)$ to $SO(5)$. This is somewhat counterintuitive and requires further explanation. We remind the Reader that $SO(6)$ is the global symmetry of the five-sphere $S^5$, and that maximal supergravity in five dimensions indeed has such gauged symmetry. We also remind the Reader that the field content of the ungauged theory is the same as that of the gauged theory. Because there is no $S^5$ in the geometry, the additional $5=4 \oplus 1$ Goldstone bosons have no apparent reason to be light. Yet, interestingly, when $\phi_I$ is large and positive, the $SO(4)$ singlet becomes parametrically light. As a tangential remark, it may be worth reminding the Reader that $SO(6)/SO(5)\sim SU(4)/Sp(4)$ is another coset which has attracted some attention in the literature (see, e.g., Refs.~\cite{Barnard:2013zea,Arthur:2016ozw,Pica:2016zst,Drach:2017btk,Drach:2020wux,Bennett:2017kga,Bennett:2019jzz,Bennett:2019cxd}). While it might be interesting to study models based on this coset, this is clearly beyond the purposes of this paper, and requires a generalisation of the model proposed here.

\begin{figure}[t]
\begin{center}
\includegraphics[width=15.2cm]{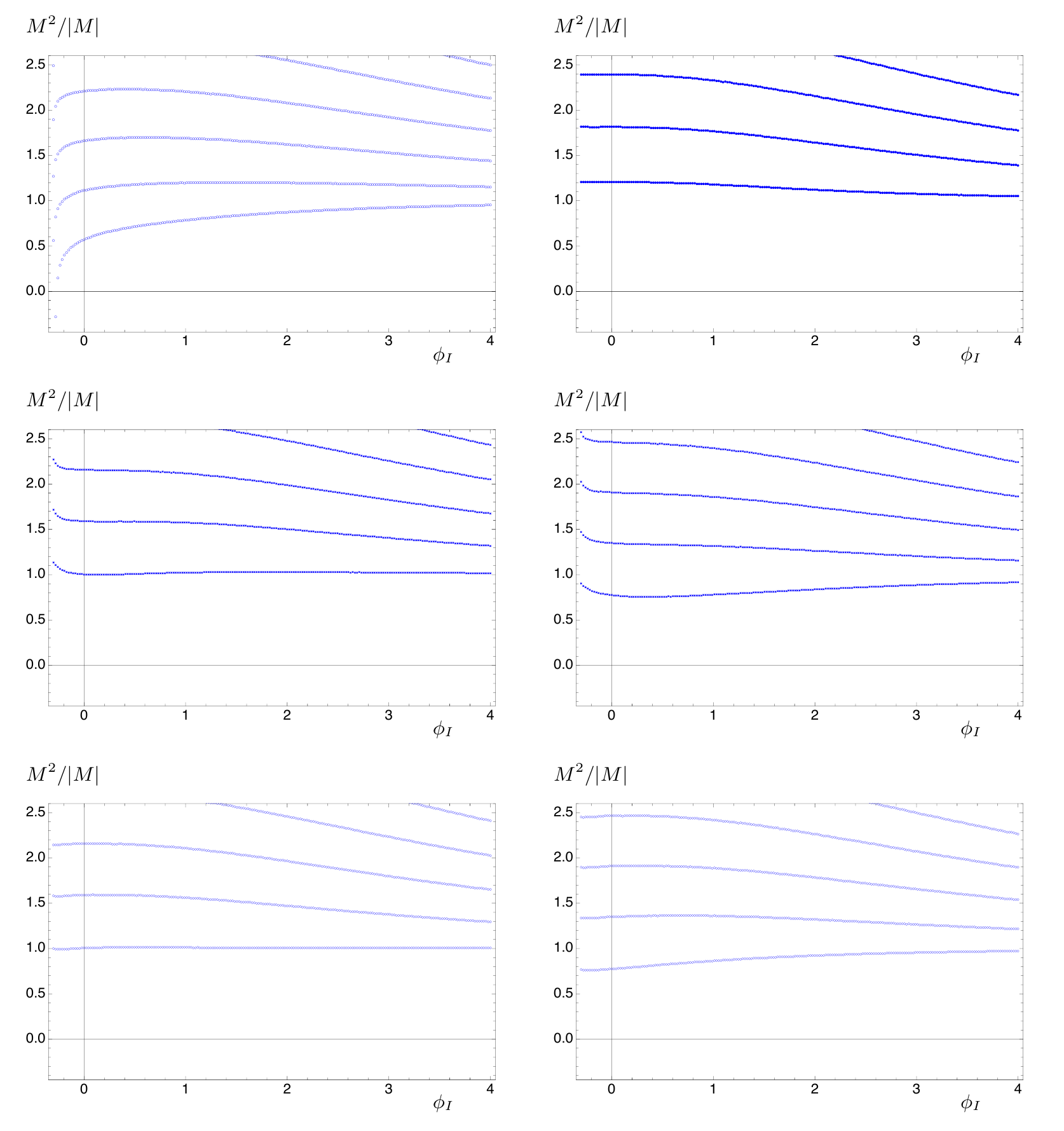}
\caption{
Mass spectrum, with $M^2=-q^2$, of fluctuations of the scalar fields  that vanish on the backgrounds,
as a function of the parameter $\phi_I$. The masses are expressed
 in units of the lightest tensor mode in the spectrum. The markers are chosen to match the representations under $SO(4)$.
Left to right and top to bottom, the nine degenerate scalar modes denoted in the main body 
of the paper as $s^{\tilde{A}}$,
the singlet $\omega_6$, the four degenerate $A_6^{\hat{A}}$, the 
four degenerate $A_7^{\hat{A}}$, the six degenerate  $A_6^{\bar{A}}$
and the six degenerate  $A_7^{\bar{A}}$.
Numerical calculations use $\r_1=10^{-10}$ and $\r_2=10$.
 }
\label{Fig:Scalars6}
\end{center}
\end{figure}

In Fig.~\ref{Fig:Scalars6} we display the masses of the other scalar fluctuations. 
As has been known for quite a long time~\cite{Pernici:1984zw}, the non-supersymmetric critical point
of the seven-dimensional potential is perturbatively unstable. The nine 
scalars $s^{\tilde{A}}$ have mass (in seven dimensions)
 below the unitarity bound, when evaluated as a fluctautation around
 the AdS$_7$ solution with $\langle \phi \rangle =\phi_{IR}$.
In the case of this paper, some of the flows we are studying 
 see the scalar $\phi$ approach this non-supersymmetric critical point,
and hence there is a legitimate concern that  tachyons might appear,
in spite of the fact that the theory is dimensionally reduced, and
 confining solutions do not exhibit local five-dimensional Poincar\'e invariance.
 Direct calculation shows that there are indeed nine degenerate tachyons in the spectrum
of the $s^{\tilde{A}}$ states, which appear for values of the parameter $\phi_I \lesssim -0.26$ close to $\phi_{IR}$.
As in the case of the other dynamically-generated tachyon we discussed in Sec.~\ref{Sec:spectrum1}
(among the fluctuations of the active scalars),
we expect a phase transition to be present, to separate the tachyonic phase from the physical one.
Contrary to the case of the active scalar for large and positive $\phi_I$, though, we know that this
scalar is not a dilaton, as it is associated with fluctuations of a  field that vanishes in the background,
and hence these fluctuations cannot mix with the trace of the metric, which is the bulk field associated to the
boundary dilatation operator---an extensive discussion of this  general 
argument can be found in Ref.~\cite{Elander:2020csd}.

The other five groups of degenerate scalar fluctuations do not display any qualitative nor quantitative features
that deserve further attention. 
The towers of states associated with 
the singlet $\omega_6$, the four degenerate $A_6^{\hat{A}}$, the 
four degenerate $A_7^{\hat{A}}$, the six degenerate  $A_6^{\bar{A}}$
and the six degenerate  $A_7^{\bar{A}}$, all have only mild dependence on $\phi_I$, 
and have  masses that start around the same value of the spin-2 states,
with the $6$ particles associated with $A_6^{\bar{A}}$ showing almost exact degeneracy with the tensors.
We will not discuss these five sequences of masses any further in the following.

\subsection{$1$-forms in five dimensions} 
\label{Sec:spectrum-1}

\begin{figure}[t]
\begin{center}
\includegraphics[width=15.2cm]{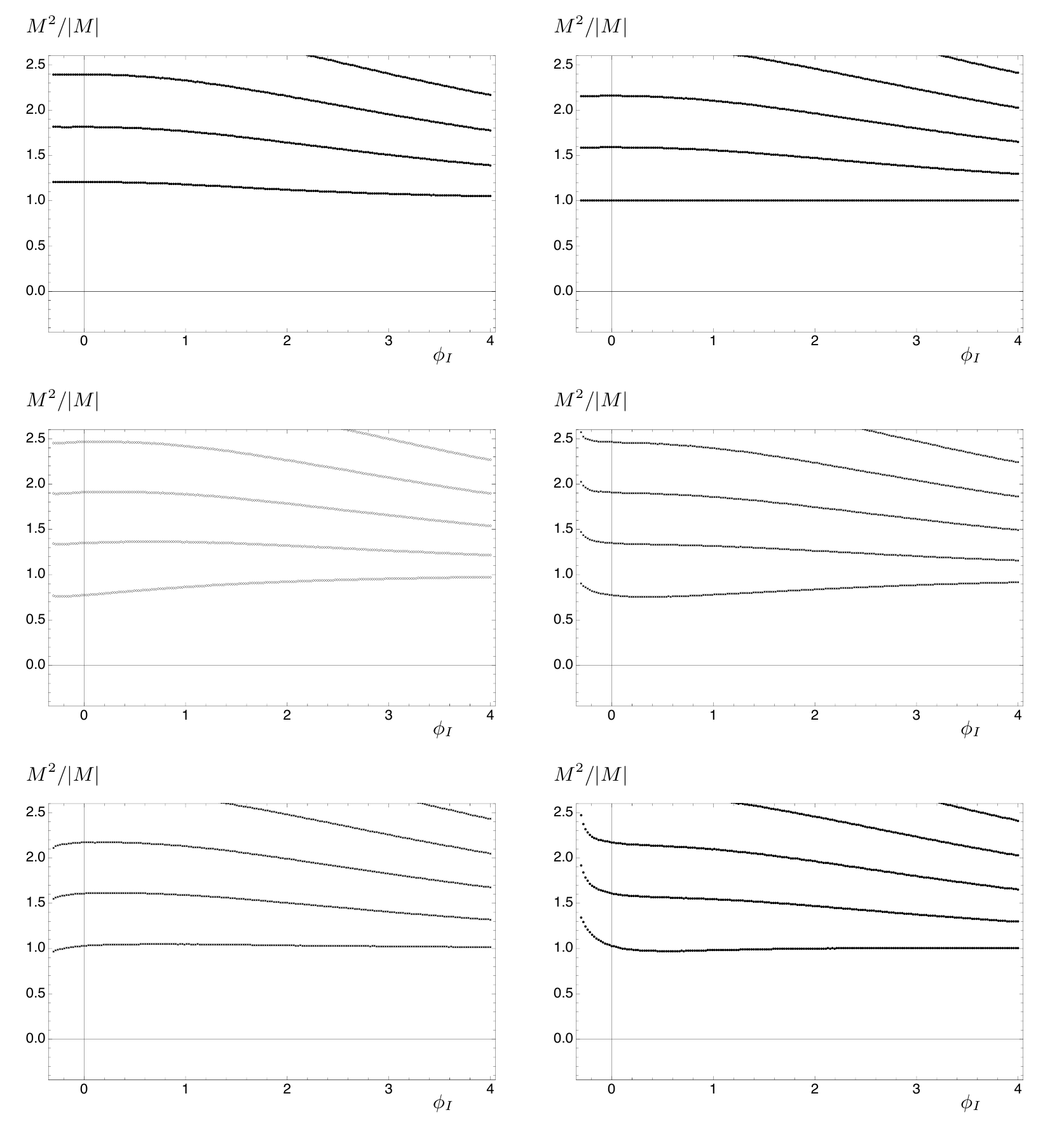}
\caption{
Mass spectrum, with $M^2=-q^2$, of the (gauge-invariant)
transverse fluctuations of the $1$-forms,
as a function of the parameter $\phi_I$. The masses are normalised
 to the lightest tensor mode. The markers are chosen to match the representations under $SO(4)$.
Left to right and top to bottom: $\chi_{\mu}$, $\omega_{\mu}$, 
the six $A^{\bar{A}}_{\mu}$ (all these states are also called vectors elsewhere),
the four $A^{\hat{A}}_{\mu}$,
 the four $S_{67\mu\,\bar{\beta}}$, and $S_{67\mu\,5}$ (the axial-vectors).
 Numerical calculations use $\r_1=10^{-10}$ and $\r_2=10$.
 }
\label{Fig:Vectors6}
\end{center}
\end{figure}

Our numerical results for the six towers of excitations of the $1$-forms (vectors) are depicted in Fig.~\ref{Fig:Vectors6}. Two things are worth noticing. First, the unexpected fact that the spin-1 states corresponding to the axial-vector fields $S_{67\mu\,\bar{\beta}}$ and $S_{67\mu\,5}$, which together transform as a $5$ of $SO(5)$, and span the $SO(6)/SO(5)$ coset, are not significantly heavier than the spin-1 states related to the gauged $SO(5)$. We would have expected the latter, comprising the vectors along the $SO(4)$, and the (axial-)vectors along the $SO(5)/SO(4)$ directions to be the lightest spin-1 states, somewhat separated from the others, as in generic QCD-like theories, in which, aside from the pNGBs, the lightest among the other bound states are the generalisations of the $\r$ and $a_1$ mesons. Such separation is not there, and we will also find additional spin-1 states with comparable masses in the next subsection.

Second, for non-vanishing  $\langle \phi \rangle$, the background breaks $SO(5)$ to $SO(4)$,
and  produces a splitting between the masses of the corresponding  two groups of towers,
as expected.
But we find that the sign of the mass splitting depends on the sign of $\phi_I$.
The six spin-1 states $A_{\mu}^{\bar{A}}$, broadly speaking corresponding to the $\rho$ 
mesons, are lighter than the four $A_{\mu}^{\hat{A}}$ for negative values of $\phi_I<0$.
But when $\phi_I>0$, this ordering is inverted.
This effect, if appearing in the absence of explicit breaking of the global symmetry,
would be a possible signature of
violations of unitarity, as it is not what expected from the analysis of dispersion relations in field theory.
But the holographic interpretation of the backgrounds, in field-theoretical terms, 
indicates that 
  we are in the presence of explicit symmetry breaking, and hence the sign of the splitting is a free parameter.

The other $1$-forms do not display features of particular physical interest.
The fluctuations of $\chi_{\mu}$, $\omega_{\mu}$---or, better,
their gauge-invariant, transverse components---all yield results in which even the lightest
mass eigenvalue is no lighter than the strong coupling scale, which in this
paper we conventionally
associate with the mass of the lightest spin-2 tensor mode. 
The $S_{67O\,{\alpha}}$ fields are related by the self-duality conditions
to the  $S_{MNO\,\alpha}$ fields, and hence we do not need to study the 
equations of motion and boundary conditions obeyed by the latter, as they cannot 
yield additional information (for details, see Appendix~\ref{Sec:3}).

\subsection{Massive $2$-forms in five dimensions}
\label{Sec:spectrum-2}

The treatment of the massive  $2$-forms (in five dimensions) starts from the derivation of the equations of motion 
for $S_{6NO}$ and $S_{7NO}$, which is
summarised in Appendix~\ref{Sec:3}, and yields four additional towers 
of spin-1 particles in the dual four-dimensional theory.
Along the lines of thought we followed
 for the $1$-forms, we  rewrite the five-dimensional action 
in a manifestly gauge-invariant form first, and then implement the
$2$-form generalisation of the  $R_{\xi}$ gauge, by following the formal treatment 
in Ref.~\cite{Elander:2018aub}.
 By doing so, we isolate 
the gauge-independent, transverse component---which we generically denote by $B_{\mu\nu}$ ($B^{\prime}_{\mu\nu}$) and $X_{\mu}$ ($X^{\prime}_{\mu}$)---and hence identify the physical spectrum
 of states without ambiguities. We 
 summarise only the most important final results in Appendix~\ref{Sec:Formalism},
 in particular the implications of the fact that the
 $3$-forms  obey a self-duality condition, which  is necessary for consistency.

\begin{figure}[t]
\begin{center}
\includegraphics[width=15.2cm]{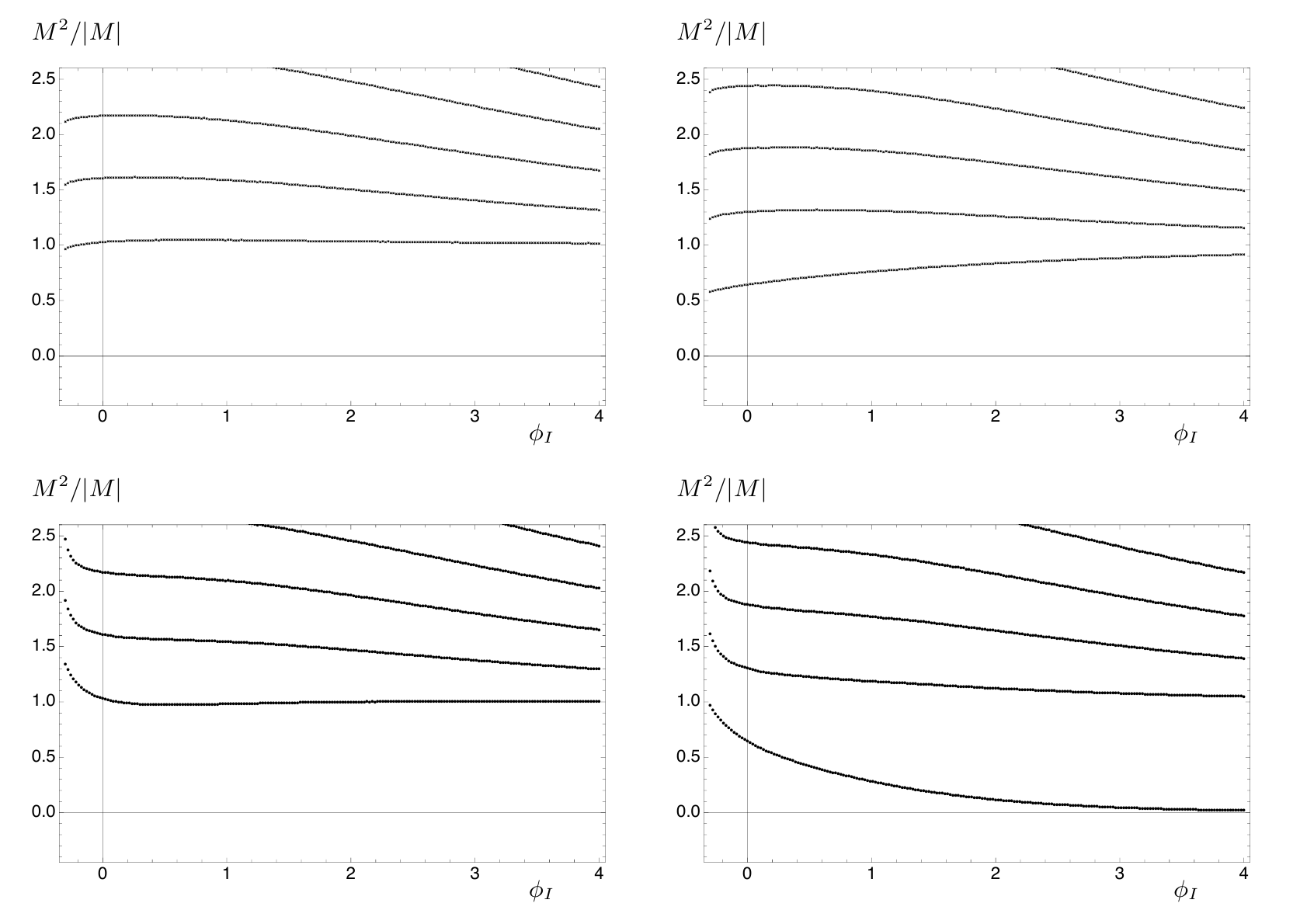}
\caption{Mass spectrum, with $M^2=-q^2$, of the spin-1 particles coming from the $2$-forms in five dimensions,
as a function of the parameter $\phi_I$, and normalised
to the lightest tensor mode. The markers are chosen to match the representations under $SO(4)$.
Left to right and top to bottom: the four $B_{\mu\nu\,\bar{\beta}}$,
the four $X_{\mu\,\bar{\beta}}$,  $B_{\mu\nu\,5}$,
and $X_{\mu\,5}$.
Numerical calculations use $\r_1=10^{-10}$ and $\r_2=10$. }
\label{Fig:Tensors4-Mixed}
\end{center}
\end{figure}

As the resulting equations and boundary conditions are peculiarly 
affected by the self-duality conditions, we exhibit them explicitly in the body of the paper,
while more detail is in Appendix~\ref{Sec:Formalism}.
They are the following:
\beqs
\label{Eq:2f1}
0&=&\left[\frac{}{}
q^2e^{-2A}-\partial_r^2+\frac{\partial_r H^{(2)\prime}_{A}}{H^{(2)\prime}_{A}}\partial_r
+m^2H^{(2)}_{A}H^{(2)\prime}_{A}\right]P^{\mu\rho}P^{\nu\sigma} B^{A}_{\,\,\,\rho\sigma}(q,r)\,,\\
\label{Eq:2f2}
0&=&\left.\left[\frac{}{}\partial_r
+\sqrt{H^{(2)}_{A}H^{(2)\prime}_{A}m^2}\right]P^{\mu\rho}P^{\nu\sigma} B^{A}_{\,\,\,\rho\sigma}(q,r)\right|_{r=r_i}\,,
\\
\label{Eq:2f3}
0&=&\left[\frac{}{}\partial_r^2-\frac{\partial_r H^{(2)}_{A}}{H^{(2)}_{A}}\partial_r -e^{-2A}q^2
-m^2{H^{(2)}_{A}}H^{(2)\prime}_{A}\right] X^{A}_{\,\,\,\mu}(q,r)\,,\\
\label{Eq:2f4}
0&=&\left.\frac{}{}\left[\frac{}{}\partial_r+\sqrt{H^{(2)}_{A}H^{(2)\prime}_{A}m^2}
\right] X^{A}_{\,\,\,\mu}(q,r)\right|_{r=r_i}\,,
\eeqs  
where $A=1\,,\cdots\,,5$, and these equations describe both the degrees of freedom
contained in the $2$-forms denoted by $S_{6NO}$ and $S_{7NO}$. The functions $H^{(2)}_{A}$ and $H^{(2)\prime}_{A} = 1/K^{(2)}_{A}$ are given in Eqs.~(\ref{Eq:H(2)}--\ref{Eq:K(2)}). We use the identification of $B_{\mu\nu}$ (and $X_\mu$) with the gauge-invariant field containing $S_{6NO}$, purely as a conventional choice. As shown in Appendix~\ref{Sec:probes}, the corresponding equations for $B^{\prime}_{\mu\nu}$ and $X^{\prime}_\mu$, associated with $S_{7NO}$, can be brought into the same form after a judicious choice of boundary conditions and the identification of $B^{\prime}_{\mu\nu}$ with $X_\mu$  and vice versa. This is made possible due to the self-dual nature of $S_{MNO}$, and hence the entirety of the associated part of the spectrum can be extracted by considering only the equations for $B_{\mu\nu}$ and $X_\mu$ given in Eqs.~(\ref{Eq:2f1}--\ref{Eq:2f4}).

Once more, we impose these boundary conditions on the 
asymptotic expansions of the fluctuations, and then set up the numerical solver to match the 
resulting asymptotics, which effectively retains only the subleading terms in the expansions themselves.
We display the results in Fig.~\ref{Fig:Tensors4-Mixed}.
As one can see, all four sets of towers of states
show the
 generic expectation that all these
fields, which correspond to spin-1 composite states, are heavy, their masses being of the order
of those of the spin-$2$ tensors or higher.
With one valuable exception:  the lightest 
mode corresponding to  component $X_{\mu\,5}$,
along the broken generator of $SO(5)$, becomes 
parametrically light, for asymptotically large values of $\phi_I$.

\section{Towards composite Higgs models}
\label{Sec:towards}

\begin{figure}[th]
\begin{center}
\begin{picture}(450,530)
\put(10,-10){\includegraphics[height=20.1cm]{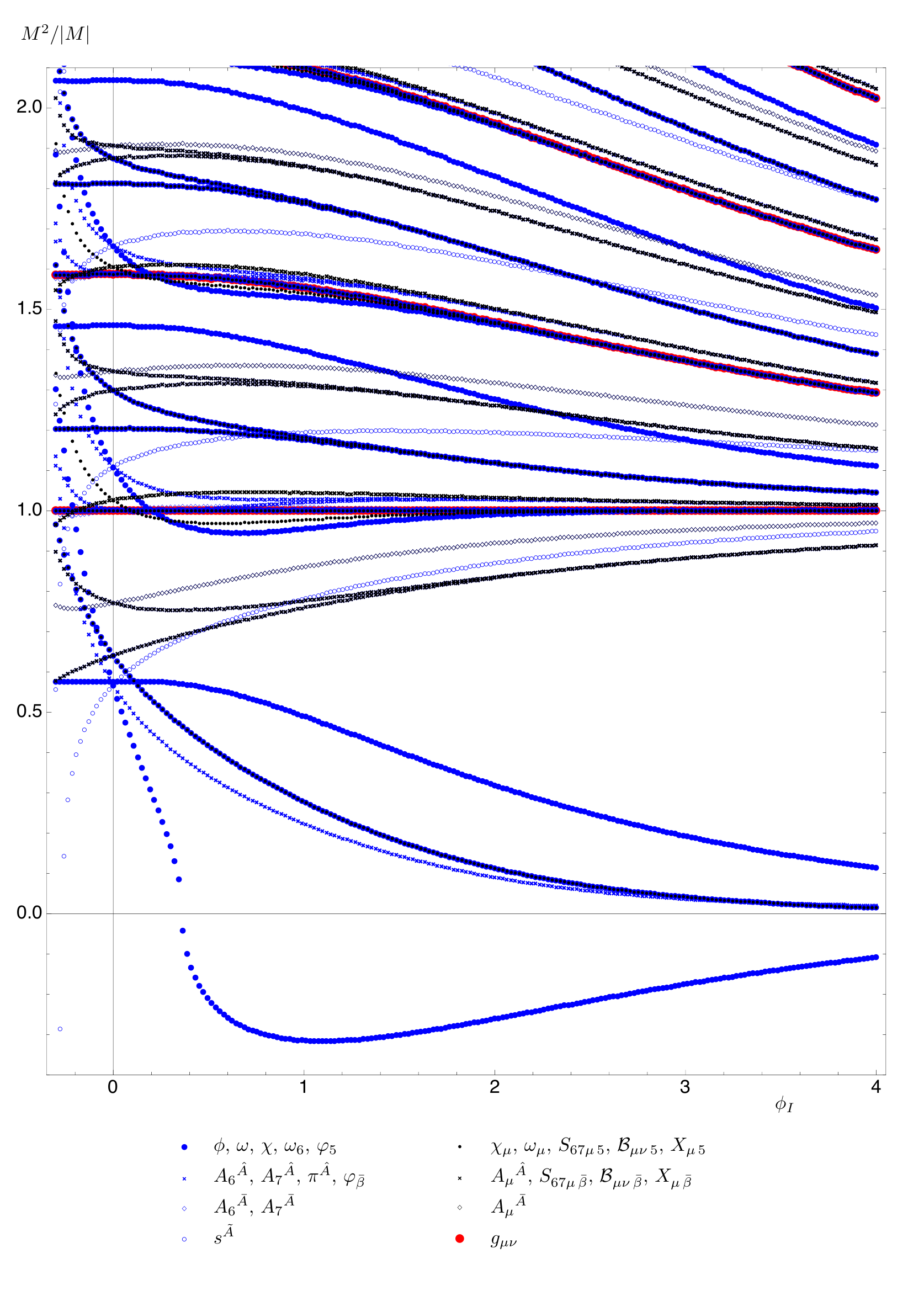}}
\end{picture}
\caption{
Detail of the combined mass spectrum $M^2=-q^2$ of all the bosons,
 as a function of the parameter $\phi_I$, normalised
to the lightest tensor mode, and restricted to the low-mass region. 
In red we depict the spin-2 particles, in black the spin-1, in blue the spin-0.
The markers are chosen to match the representations under $SO(4)$,
while the legend refers back to the notation in Table~\ref{Fig:Fields}.
 Numerical calculations use $\r_1=10^{-10}$ and $\r_2=10$. }
\label{Fig:plotsummaryallofthem}
\end{center}
\end{figure}

So far in this paper, we reported on the results of the calculation of the complete spectrum of 
excitations of the $128$ bosonic degrees of freedom of the theory.
We  start this section by presenting a summary plot of 
the physical results of our extensive study, which are displayed in 
Fig.~\ref{Fig:plotsummaryallofthem}.

The number of bosonic states is so large that the sequence of levels densely fills the
 positive values of the mass above that of the lightest spin-$2$ (tensor) state.
 We remark on the presence of a degeneracy between the spin-2 states with a tower of spin-1 and of spin-0 states, which (for $\phi_I = 0$) had been observed before in the literature~\cite{Brower:2000rp}.
  In general terms, looking at heavy states  with this formalism 
 is not particularly interesting: in spite of retaining 
 a significant number of supergravity fields,
 we are still neglecting the excitations that depend non-trivially on the coordinates on
 the two circles and on the four-sphere in the interior of the geometry,
  and furthermore all the possible other excitations that are not captured by supergravity.
Yet, if we focus our attention on states that are appreciably lighter than the spin-$2$ tensors,
some interesting patterns emerge, which do not depend on the aforementioned 
simplifications and approximations.

We remind the Reader of some of the results already discussed in previous sections. The first observation we make is that two mass eigenstates for spin-0 particles become tachyonic in (separate) parts of 
the parameter space. At negative $\phi_I$,
large enough that the backgrounds approach the non-supersymmetric critical point with
$\langle \phi \rangle = \phi_{IR}$,
these tachyons correspond to the lightest excitations of the nine $s^{\tilde{A}}$, 
and their negative mass is a consequence of 
the instability of such critical point.
Conversely, at large and positive values of $\phi_I$, the state that becomes 
tachyonic is a fluctuation of the background
active scalars, and mixes with the trace of the metric, so that this tachyon is also a dilaton, 
at least approximately~\cite{Elander:2020fmv}.

There is a region of small to moderate values of $\phi_I$ over which such (perturbative) instabilities are absent. 
We do not know what is the current extent of such region: both at positive and negative $\phi_I$, 
a phase transition must be present, separating the stable, physical branch of confining solutions
 from the unstable ones.
This problem was studied in Ref.~\cite{Elander:2020fmv}, which demonstated the existence of a first-order phase transition at a moderate value of $\phi_I>0$. This was supported by establishing the metastable nature of the gravity solutions in the region $0.04\lsim \phi_I \lsim 0.4$.
No analogous study has, to the best of our knowledge, been performed for negative values of $\phi_I$,
and we leave this open question for future investigations.

As expected, the aforementioned two scalar mass eigenstates become exactly degenerate for $\phi_I=0=\phi$, where they are both part of the $14$ of the enhanced $SO(5)$ symmetry. Interestingly, they are also close to degenerate with one of the lightest active states that corresponds to fluctuations of the fields $\{\omega,\chi\}$. It is also worth noticing that a second fluctuation of the active scalars becomes parameterically light when $\phi_I$ is taken to be large, so that in this limit two of the spin-0 $SO(4)$ singlets become massless.

 One of the pseudo-scalar eigenstates, corresponding to the $4$ 
 pNGBs of the $SO(5)\rightarrow SO(4)$  breaking, 
 becomes massless for $\phi_I\rightarrow +\infty$. 
 Interestingly, also one pseudoscalar $SO(4)$ singlet, and the 
 degenerate  spin-1 $SO(4)$ singlet, become massless in this limit.
Asymptotically at large $\phi_I$, we find that the spectrum approaches that of a gapped continuum starting at the threshold set by the mass of the tensors, accompanied by a number of massless states corresponding to the aforementioned two scalars, five pseudo-scalars and one vector.

\subsection{Boundary terms and pNGBs}
\label{Sec:bc}

We complete this section with an exercise that involves the boundary terms in the action.
We focus on the boundary at $\r_2$, and specifically on the boundary term for the four 
pseudo-scalars $\pi^{\hat{A}}$. No explicit mass terms are allowed for these fields,
due to gauge invariance, but the combination $\partial_{\mu}\pi^{\hat{A}}+ 2 m A_{\mu}^{\,\,\,\hat{A}}$
is gauge invariant, and hence a boundary-localised term quadratic in this combination is allowed.
Its coefficient is denoted by $C_2$ in Eqs.~(\ref{Eq:bcUV1}) and~(\ref{Eq:bcUV2}) in the Appendix.
We would like to assess how much the spectrum of the spin-0 and spin-1 states would be affected by 
fine-tuned choices of $C_2\neq 0$.

The reason why this is interesting relates to the connection with
phenomenological CHMs. A non-zero $C_2$ might emerge from the coupling of the pNGBs
to an external sector of the theory, as is envisioned in CHMs, for which this sector is the
 standard model of particle physics---the lightest excitations of the 
 four $\pi^{\hat{A}}$ fields being related to the Higgs fields of the standard model.
 In CHMs, such couplings not only change the dynamics of the pseudoscalars, but must
 induce an instability, that ultimately triggers electroweak symmetry breaking---$SO(4)$
 must break spontaneously to $SO(3)$ in the vacuum.
 Of course, a realistic model would require to also embed the $SU(2)_L\times U(1)_Y$
 gauge group into the $SO(4)$ symmetry, which would require changing also the appropriate $D_2$ terms 
  in Eqs.~(\ref{Eq:bcUV1}) and~(\ref{Eq:bcUV2}). But we leave this model-building task, together with
  other model-building considerations, as well as the whole programme of 
  studying vacuum (mis-)alignment and of assessing the amount of fine-tuning, to future investigations.
  Here we just consider whether we can make the mass of the pNGBs parameterically small, and possibly negative,
  by  dialing $C_2$, without further attention to phenomenological and model-building considerations. 

To be more precise, let us reconsider the numerical procedure we adopted in the treatment of 
this particular multiplet of pseudoscalars.
For $C_2=0$, Eq.~(\ref{Eq:bcUV2}), when taking the limit $\r_2\rightarrow +\infty$,
amounts to selecting the subleading fluctuation in the asymptotic expansion of
$\mathfrak{p}^1$ in Eq.~(\ref{Eq:p1})---effectively restricting the fluctuations to have $\mathfrak{p}^1_{\,\,\,l}=0$.
Explicitly, we see that this is the case 
by making use of the changes of variable $\partial_{r}=e^{-\chi}\partial_{\r}$ and $z=e^{-\r/2}$ to rewrite 
Eq.~(\ref{Eq:bcUV2}) as
\beqs
0&=&\left.\left[\frac{}{}C_2e^{-\chi(\r)}\partial_{\r}+G^{(1)}(\r)\right]X(q,\r)\right|_{\r=\r_2}\nonumber\\
&=&
\left.\left[-\frac{1}{2}C_2e^{-\chi(z)}z\partial_{z}+G^{(1)}(z)\right]X(q,z)\right|_{z=e^{-\r_2/2}}\nonumber\\
&=&
\left.\left[-\frac{1}{2}C_2e^{-\chi(z)}\,\mathfrak{p}^1_{\,\,\,l}
+G^{(1)}(z)\left(\mathfrak{p}^1_{\,\,\,0}+\log(z)\mathfrak{p}^1_{\,\,\,l}\right) + \cdots \right]\right|_{z=e^{-\r_2/2}}\,,
\eeqs
where we made explicit use of the first line of Eq.~(\ref{Eq:p1}),
and where $G^{(1)} = \sinh^2 \left(\sqrt{5}\phi/2\right)$ is the factor appearing in the fourth block of the diagonal
matrix of Eq.~(\ref{Eq:G(1)}).
As the term with the explicit dependence on $\log(z)$ is dominant for $C_2=0$, by taking the limit $\rho_2 \rightarrow +\infty$ we 
are effectively setting $\mathfrak{p}^1_{\,\,\,l}=0$. 

This conclusion holds for any generic choice of $C_2$ such that 
$\lim\limits_{\r_2\rightarrow \infty} \left( \frac{C_2(\r_2) 
e^{-\chi(\r_2)}}{G^{(1)}(\r_2)} + \rho_2 \right) =\pm\infty$.
The other extreme case appears if we allow $C_2$ to be a function of $\r_2$, such that 
when we take $\r_2\rightarrow +\infty$ we find that $\lim\limits_{\r_2\rightarrow \infty} \left( \frac{C_2(\r_2) 
e^{-\chi(\r_2)}}{G^{(1)}(\r_2)} + \rho_2 \right)=0$. In this case, we are fixing $\mathfrak{p}^1_{\,\,\,0}=0$. But we can make a fine-tuned choice, by making $C_2$ a function of $\r_2$,
and requiring that for a given $\r_2$ we dial
\beqs
-\frac{1}{2}C_2(\r_2)e^{-\chi(\r_2)}&\equiv&\frac{1}{2}G^{(1)}(\r_2)\times \left(\frac{}{}\r_2+2c_2\right)\,,
\label{Eq:c2}
\eeqs
with $c_2$ a constant.
By doing so, the boundary condition reduces to
\beqs
0&=&\left.
G^{(1)}(z)\left(\mathfrak{p}^1_{\,\,\,0}+c_2\,\mathfrak{p}^1_{\,\,\,l}\right)\right|_{z=e^{-\r_2/2}}\,.
\eeqs
This process highlights the presence of an additional free parameter, $c_2$,
in the theory, and this free parameter can be dialed to change the spectrum.
Doing so requires fine-tuning, because the scaling of $C_2$ has no a priori relation with
the bulk warp factors, such as $G^{(1)}(\r)$ and $e^{-\chi}$, and hence 
this mechanism amounts to a somewhat contrived exact calcellation, in the $\r_2\rightarrow +\infty$ limit,
between bulk and boundary-localised physical effects.

\begin{figure}[t]
\begin{center}
\includegraphics[width=15.2cm]{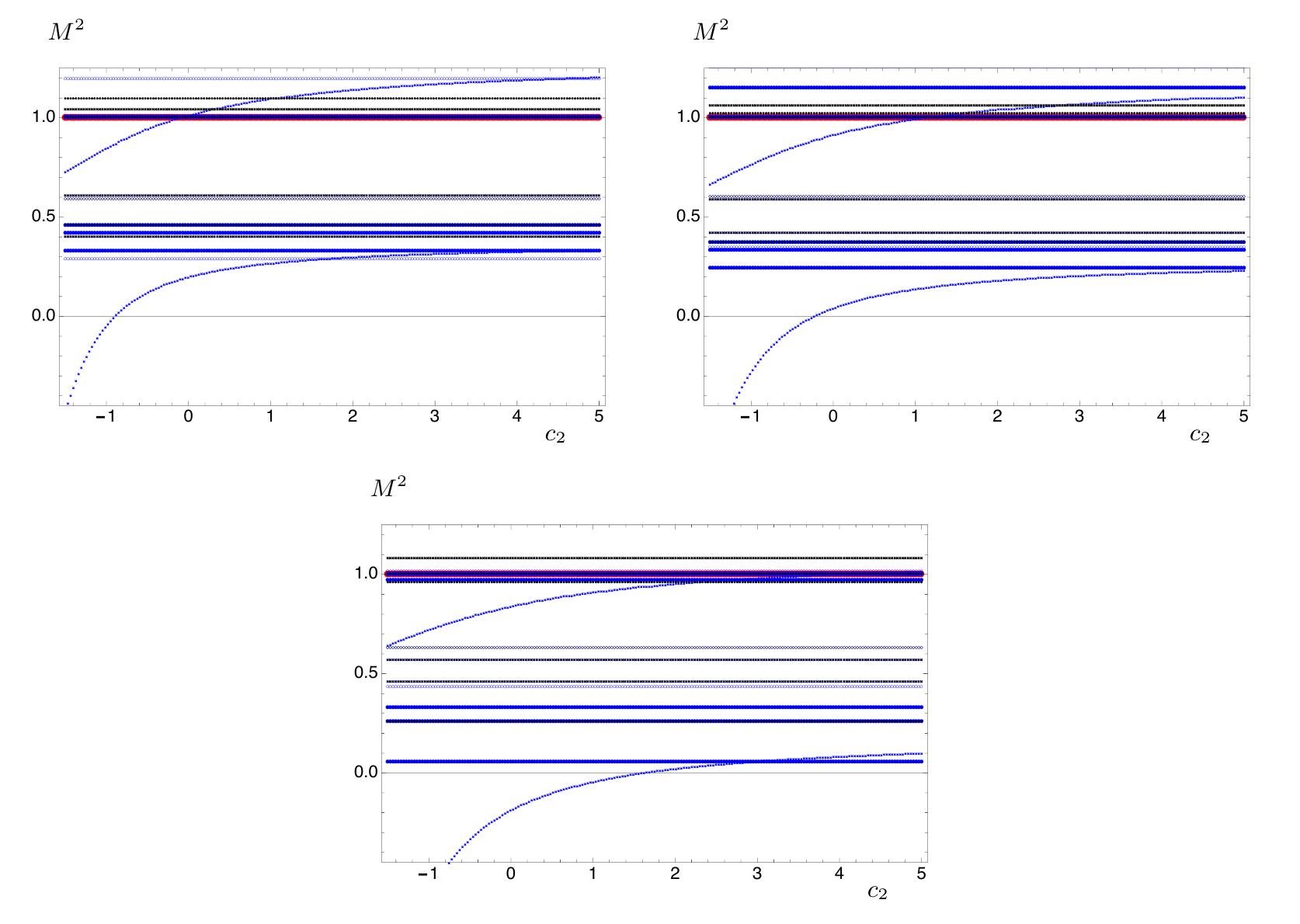}
\caption{
Detail of the combined mass spectrum $M^2=-q^2$ of all the bosons,
 for the choices of $\phi_I=-0.05, 0.05, 0.25$ (top-left, top-right, bottom panels),
 obtained by varying the parameter $c_2$ defined in  Eq.~(\ref{Eq:c2}).
 The colour and symbol coding is the same as in the earlier plots of the mass spectra.
 Numerical calculations use $\r_1=10^{-10}$ and $\r_2=10$. }
\label{Fig:plotc2}
\end{center}
\end{figure}

We show in Fig.~\ref{Fig:plotc2} the dependence of the spectrum, in particular for the four pNGBs,
 on $c_2$, for three representative values of $\phi_I$.
The spectrum in Fig.~\ref{Fig:plotsummaryallofthem} is recovered for $c_2\rightarrow +\infty$.
In all cases, we notice the existence of a narrow range of $c_2$ for which the pseudoscalars
are parametrically light, compared to the rest of the spectrum. This range appears 
 in proximity of a fine-tuned choice of $c_2$ 
below which the pseudoscalar becomes anomalously light and tachyonic.
Depending on the value of $\phi_I$, two different cases can be realised. 
For generic $\phi_I$ (represented here by  $\phi_I=\pm 0.05$,
as in the top panels of Fig.~\ref{Fig:plotc2}), by tuning $c_2$ one can realise a 
hierarchy between the mass of the pseudoscalars
and the rest of the spectrum. 
The little hierarchy of scales that emerges in this way
  is suggestive of the case of interest in composite Higgs models:
in order to trigger electroweak symmetry breaking one needs 
a choice of $c_2$ that makes the pseudoscalar tachyonic, and the instability would indicate 
that the vacuum might further break $SO(4)$ to a $SO(3)$ subgroup.
Interestingly, this scenario is realised both for positive and negative values of $\phi_I$.

For larger and positive values of $\phi_I$, close to the appearance of a tachyon in the scalar spectrum, 
but for which the scalar mass squared is still positive,
one can realise the scenario of composite Higgs models in which the low energy effective theory is the dilaton EFT. In this case one can envisage a low energy spectrum that 
contains also a pseudo-dilaton, together with the pNGBs.
But to do so requires dialing to special values both $\phi_I$ and $c_2$.
We defer the detailed realisation of a phenomenologically viable model of this type to future work.

Let us return to the rest of the bosonic spectrum.
If, as suggested above, we modify $C_2$ for one of the pseudoscalar multiplets, the boundary term affects the 
 associated axial-vector field as well, as the same parameter 
enters Eq.~(\ref{Eq:bcUV1}) as~(\ref{Eq:bcUV2}).
For $D_2=0$, we find that the UV boundary condition is the following:
\beqs
0&=&\left.e^{2A(z)}\left[ -\frac{1}{2} H^{(1)}(z)e^{-\chi(z)} z\partial_z+m^2C_2\right]
P^{\mu\nu}A^{\hat A}_{\,\,\,\mu}(q,z)\right|_{z=e^{-\r_2/2}}\,,
\eeqs
where $H^{(1)}(z) = \frac{1}{4} e^{\frac{3\phi}{\sqrt{5}} + 2\chi}$ is the factor appearing in the fourth block of the diagonal matrix of Eq.~(\ref{Eq:H(1)}).
By replacing the aforementioned, fine-tuned choice of $C_2$, as well as the expansion in Eq.~(\ref{Eq:ax}),
we find
\beqs
	0 &=& \left[ z\partial_z+\frac{2m^2e^{2\chi(z)}G^{(1)}(z)(\r_2+2c_2)}{H^{(1)}(z)} \right] \left[ \frac{}{}\mathfrak{v}^4_{\,\,\,0}\left(1+z^2M^2+\cdots\right) + \mathfrak{v}^4_{\,\,\,4}z^4 \right] \Big|_{z=e^{-\r_2/2}} \nonumber \\
	&=& \left[ z\partial_z + 10\phi_2^2(\r_2+2c_2) z^4 \right] \left[ \frac{}{}\mathfrak{v}^4_{\,\,\,0}\left(1+z^2M^2+\cdots\right) + \mathfrak{v}^4_{\,\,\,4}z^4 \right] \Big|_{z=e^{-\r_2/2}} \,.
\eeqs
Even for the aforementioned, fine-tuned choices of $C_2$, these boundary conditions are equivalent, asymptotically, to setting the coefficient $\mathfrak{v}^4_{\,\,\,0}=0$
for the leading term emerging from the fluctuations, and hence
this choice does not affect the spectrum of the axial vectors.
In particular, all the 1-forms are massive, with $M^2\neq 0$.

\section{Conclusions and Outlook}
\label{Sec:conclusion}

We reconsidered a holographic model of confining dynamics based upon 
a 1-parameter family of  background solutions in maximal supergravity in seven dimensions, compactified on a 2-torus.
The backgrounds are completely regular and smooth. 
The holographic  calculation of the Wilson loop yields the static quark-antiquark 
potential expected in linear confinement.
Furthermore, the 1-parameter family 
describes the breaking $SO(5) \rightarrow SO(4)$ of the global symmetry in the dual field  theory. We presented two main, interrelated, sets of new results.

At first, we computed the spectrum of fluctuations of these models. The complete bosonic mass
spectrum is interpreted in terms of bound states  of the dual confining four-dimensional
 field theory. For the $SO(4)$ singlets, we improved the numerical treatment compared to
  earlier papers, and found agreement with other studies---when the results
  for the same states are available. One original part of this paper is that we extended the literature to include also states that are not singlets of the $SO(4)$ symmetry, and which had previously been ignored.
We found that some of these  multiplets are lighter than the singlets.
We presented all the details of the calculations, from the decomposition of all the dimensionally reduced fields, 
to the treatment of the boundary conditions, taking special care to identify the physical, gauge-invariant degrees of freedom.

The second part of this study concerned the connection of this well established supergravity, and its background solutions, to the, superficially remote, CHM context.
Having observed that the $SO(5)/SO(4)$ coset, characterising the field-theory dual of the 
present theory, also coincides with the one deployed in the construction of 
minimal CHMs, we focused attention on
 the mass spectrum of the four pNGBs associated with $SO(5)\rightarrow SO(4)$ breaking.
In a complete CHM, the four fields providing the low-energy description of 
the pNGBs would become the four components of the Higgs doublet.
As explained elsewhere (see for instance Ref.~\cite{Agashe:2004rs}, or reviews of CHMs such as Ref.~\cite{Panico:2015jxa}), 
the coupling to the standard model fields, within the
 low energy dynamics, can trigger electroweak symmetry breaking, via vacuum misalignment.
We restricted ourselves to consider a more limited question of principle.
We showed how, by dialing specific boundary-localised terms, 
that are allowed by the symmetries and would be naturally generated by coupling
the theory to an external weakly-coupled sector,
it is indeed possible to realise a spectrum that resembles that of minimal CHMs,
and ultimately trigger an instability, which appears at scales lower than that of the strong coupling dynamics.
In passing, we also found that, for non-trivial parameter choices, the low energy spectrum may 
 include also an approximate dilaton,
besides the four pNGBs. 

Within the language of gauge-gravity dualities, it is worth noticing that 
by dialing the two parameters denoted as $\phi_I$ and $c_2$ in the body of the paper,
we gain  the freedom to change the balance between explicit and spontaneous breaking of $SO(5)$.
Let us try to make this statement more precise. Let us start from  the case in which we do not include boundary localised terms,
and in which $\phi_I=0$. The $SO(5)$ symmetry is exact, all of the spectrum is organised in $SO(5)$ multiplets,
there are no pNGBs (see Appendix~\ref{Sec:Brower}).
When we turn on $\phi_I\neq 0$, $SO(5)$ is broken in the background, and  the analysis  of the UV expansions
shows the presence of both explicit and spontaneous breaking, via the coupling and 
vacuum expectation value of the operator dual to the field $\phi$.
The parameter $\phi_I$ itself controls the balance of the two effects.
In general, one can dial $\phi_I$ to large values, and hence recover a set of four
parametrically light pNGBs in this extreme case. 
But this appears possible only at the price of exploring a region of parameter space where a
 tachyonic instability is present.
 In the non-tachyonic region of parameter space, the four pNGBs are not particularly light,
 their masses being just marginally smaller compared to other bosons.
 
 The addition of the boundary-localised term (with finite $c_2$) allows to change the balance 
 between explicit and spontaneous breaking. As a consequence of 
 a cancellation between intrinsic breaking of $SO(5)$ in the strongly-coupled theory, and 
 additional explicit breaking due to the weakly-coupled boundary effects,
 the mass spectrum displays light pNGBs.
 The result is not dissimilar from what emerges in other CHMs currently under investigation 
 on the lattice and with
  EFT techniques---see the ample list of references in the introduction---in which the little hierarchy
 between the strong-coupling scale and the mass of the lightest scalars emerges in similar ways,
from a cancellation requiring some moderate tuning.

Our results show that this very special theory provides a completion, in principle, for minimal CHMs. 
But let us qualify this statement. This theory provides quantitative information about the spectrum of states other than the pNGBs, such as the many
scalars, spin-1 and spin-2 states we studied.
Yet, to build a realistic model there are at least two major additional steps to take, before one can consider 
phenomenological implications and direct testability.
First, one has to embed the $SU(2)\times U(1)$ symmetry of the standard model into $SO(4)$,
and gauge it (weakly) by adding appropriate boundary-localised kinetic terms for the gauge bosons,
in such a way that the dual field theory has a gauged, rather than global, symmetry.
One then must study vacuum (mis-)alignment, and explicitly show that by dialling 
the appropriate boundary-localised term (generalising the aforementioned $c_2$ parameter
to the realistic scenario) one can 
trigger electroweak symmetry breaking. 
After that,  one can proceed to describe the full phenomenology of the resulting model, including the 
process of  mass generation for the SM fermions, the study of the masses and couplings of
the heavier states, and the calculation of precision observables both in the gauge and scalar sectors of the resulting CHM.
 The field content of the supergravity theory, and hence the  bound states of its dual,
  includes towers of particles with non-trivial $SO(4)$ quantum numbers, and a calculable mass spectrum,
  making it potentially quite intriguing and well worth further investigation.
This paper establishes the basic tools needed to start carrying out all these ambitious tasks in the near future.

\begin{acknowledgments}
The work of MP is supported in part by the STFC Consolidated Grants  ST/P00055X/1
and ST/T000813/1. MP has also received funding from the European Research Council (ERC) under the
European Union Horizon 2020 research and innovation programme under grant agreement No 813942. DE was supported in part by the OCEVU Labex (ANR-11-LABX-0060) and the A*MIDEX project (ANR-11-IDEX-0001-02) funded by the ``Investissements d'Aveni'' French government program managed by the ANR.
\end{acknowledgments}

\appendix
\section{Of self-dual massive $3$-forms}
\label{Sec:3}

In this Appendix, we match the action of the self-dual massive $3$-forms in seven dimensions
 to the action of massive $1$-forms and $2$-forms that we study in the body of the paper.
Let us start from restricting our attention to the quadratic part of the action for the $3$-forms,
borrowed from Eq.~(\ref{Eq:quad}):
\beqs
{\cal S}_7^{(2)}&=&\int \di^7 x \left\{\sqrt{-\hat{g}_7}\left[
-\frac{m^2}{4}\hat{g}^{\hat{M}\hat{P}}\hat{g}^{\hat{N}\hat{Q}}\hat{g}^{\hat{O}\hat{R}}\,
\left(\frac{}{}e^{\frac{1}{\sqrt{5}}\phi}S_{\hat{M}\hat{N}\hat{O}\,\bar{\alpha}}\delta^{\bar{\alpha}\bar{\beta}}
S_{\hat{P}\hat{Q}\hat{R}\,\bar{\beta}}
+\right.\right.\right.\\
&&\nonumber\left.\left.\left.
+e^{-\frac{4}{\sqrt{5}}\phi}
S_{\hat{M}\hat{N}\hat{O}\,5}
S_{\hat{P}\hat{Q}\hat{R}\,5}
\right)
\right]
+\frac{m}{24} \epsilon^{\hat{M}\hat{N}\hat{O}\hat{P}\hat{Q}\hat{R}\hat{S}}
S_{\hat{M}\hat{N}\hat{O}\,\alpha}\delta^{\alpha\beta}
\partial_{\hat{P}}S_{\hat{Q}\hat{R}\hat{S}\,\beta}\nonumber
\right\}\,.
\eeqs
Taking the variation of the action with respect to the fields, one obtains the equations of motion:
\beqs
0&=&-\frac{m^2}{2}\sqrt{-\hat{g}_7}\,\hat{g}^{\hat{M}\hat{P}}\hat{g}^{\hat{N}\hat{Q}}\hat{g}^{\hat{O}\hat{R}}\,
e^{\frac{1}{\sqrt{5}}\phi}
S_{\hat{P}\hat{Q}\hat{R}\,\bar{\beta}}
+\frac{m}{12} \epsilon^{\hat{M}\hat{N}\hat{O}\hat{P}\hat{Q}\hat{R}\hat{S}}
\partial_{\hat{P}}S_{\hat{Q}\hat{R}\hat{S}\,\bar{\beta}}\,,\\
0&=&-\frac{m^2}{2}\sqrt{-\hat{g}_7}\,\hat{g}^{\hat{M}\hat{P}}\hat{g}^{\hat{N}\hat{Q}}\hat{g}^{\hat{O}\hat{R}}\,
e^{-\frac{4}{\sqrt{5}}\phi}
S_{\hat{P}\hat{Q}\hat{R}\,5}
+\frac{m}{12} \epsilon^{\hat{M}\hat{N}\hat{O}\hat{P}\hat{Q}\hat{R}\hat{S}}
\partial_{\hat{P}}S_{\hat{Q}\hat{R}\hat{S}\,5}\,,
\eeqs
where we separated the $SO(4)$ multiplet from the singlet, as they have different $\phi$-dependent mass terms.
These equations implement the self-duality conditions that are necessary for massive $3$-forms 
to propagate $10$ degrees of freedom on-shell (and not $20$).
The bulk equations are of first order, and relate the forms to their first derivative.

We reduce the equations to five dimensions with
the ansatz for the metric in Eq.~(\ref{Eq:7}).
The derivatives with respect to $\eta$ and $\zeta$  vanish
for all the fields. We decompose the $3$-forms in their $S_{MNO}$
$S_{67O}$,  $S_{6NO}$, and $S_{7NO}$ components, and make use of the fact that $\sqrt{-\hat{g}_7}=e^{-2\chi}\sqrt{-g_5}$, as well as that the indexes of the antisymmetric Levi-Civita symbols $\epsilon^{\hat M \hat N \hat O \hat P \hat Q \hat R \hat S}$ and $\epsilon^{MNOPQ}$ are lowered using the metrics in $D=7$ and $D=5$ dimensions, respectively. This exercise yields the following system of first-order coupled equations:
\beqs
S_{67M\,\bar{\beta}}
&=&\frac{1}{6m} \frac{e^{6\chi-\frac{1}{\sqrt{5}}\phi}}{\sqrt{-g_5}}
\epsilon_M^{\ \ NOPQ} \partial_N S_{OPQ \,\bar{\beta}}\,,\\
S_{6MN \,\bar{\beta}}
&=& - \frac{1}{2m} \frac{e^{\chi - 2\omega -\frac{1}{\sqrt{5}}\phi}}{\sqrt{-g_5}}
\epsilon_{MN}^{\ \ \ \ \, OPQ} \partial_O S_{7PQ \,\bar{\beta}}\,,\\
S_{7MN\,\bar{\beta}}
&=&\frac{1}{2m} \frac{e^{\chi + 2\omega -\frac{1}{\sqrt{5}}\phi}}{\sqrt{-g_5}}
\epsilon_{MN}^{\ \ \ \ \, OPQ}
\partial_O S_{6PQ \,\bar{\beta}}\,,\\
S_{{MNO}\,\bar{\beta}}
&=&\frac{1}{m} \frac{e^{-4\chi-\frac{1}{\sqrt{5}}\phi}}{\sqrt{-g_5}}
\epsilon_{MNO}^{\ \ \ \ \ \ \, PQ}
\partial_P S_{67Q \,\bar{\beta}}\,,\\
S_{67M\,5}
&=&\frac{1}{6m} \frac{e^{6\chi+\frac{4}{\sqrt{5}}\phi}}{\sqrt{-g_5}}
\epsilon_M^{\ \ NOPQ} \partial_N S_{OPQ \,5}\,,\\
S_{6MN \,5}
&=& - \frac{1}{2m} \frac{e^{\chi - 2\omega +\frac{4}{\sqrt{5}}\phi}}{\sqrt{-g_5}}
\epsilon_{MN}^{\ \ \ \ \, OPQ} \partial_O S_{7PQ \,5}\,,
\label{Eq:6and7}\\
S_{7MN\,5}
&=&\frac{1}{2m} \frac{e^{\chi + 2\omega +\frac{4}{\sqrt{5}}\phi}}{\sqrt{-g_5}}
\epsilon_{MN}^{\ \ \ \ \, OPQ}
\partial_O S_{6PQ \,5}\,,\\
S_{{MNO}\,5}
&=&\frac{1}{m} \frac{e^{-4\chi+\frac{4}{\sqrt{5}}\phi}}{\sqrt{-g_5}}
\epsilon_{MNO}^{\ \ \ \ \ \ \, PQ}
\partial_P S_{67Q \,5}\,,
\eeqs

We can decouple the system into second-order equations, by resolving the mixing of $S_{67M}$ with $S_{MNO}$, and of $S_{6MN}$ with $S_{7MN}$. A substantial amount of algebra relies on the use of the following relations, that hold in five dimensions:
\beqs
\epsilon^{SMNOP}\epsilon_{SM^{\prime}N^{\prime}O^{\prime}P^{\prime}}&=&
(1!4!)g_5\frac{1}{4!}\left(\delta^{M}_{\,\,\,M^{\prime}}\delta^{N}_{\,\,\,N^{\prime}}
\delta^{O}_{\,\,\,O^{\prime}}\delta^{P}_{\,\,\,P^{\prime}}
-\delta^{M}_{\,\,\,M^{\prime}}\delta^{N}_{\,\,\,N^{\prime}}
\delta^{O}_{\,\,\,P^{\prime}}\delta^{P}_{\,\,\,O^{\prime}}+\cdots
\right)
,\\
\epsilon^{RSNOP}\epsilon_{RSN^{\prime}O^{\prime}P^{\prime}}&=&
(2!3!)g_5\frac{1}{3!}\left(\delta^{N}_{\,\,\,N^{\prime}}
\delta^{O}_{\,\,\,O^{\prime}}\delta^{P}_{\,\,\,P^{\prime}}
-\delta^{N}_{\,\,\,N^{\prime}}
\delta^{O}_{\,\,\,P^{\prime}}\delta^{P}_{\,\,\,O^{\prime}}+\cdots
\right)
,\\
\epsilon^{QRSOP}\epsilon_{QRSO^{\prime}P^{\prime}}&=&
(3!2!)g_5\frac{1}{2!}\left(
\delta^{O}_{\,\,\,O^{\prime}}\delta^{P}_{\,\,\,P^{\prime}}
-
\delta^{O}_{\,\,\,P^{\prime}}\delta^{P}_{\,\,\,O^{\prime}}
\right)
.
\eeqs
We also write $[n_1n_2\cdots n_p]\equiv \frac{1}{p!}(n_1n_2\cdots n_p-n_2n_1\cdots n_p+\cdots)$ to denote complete anti-symmetrisation.
The equations of motion are hence written in terms of the tensors $S_{67O}$,
$S_{6NO}$, formulated in the five-dimensional language, and they
read as follows:
\beqs
 \frac{m^2\sqrt{-g_5}}{e^{6\chi-\frac{1}{\sqrt{5}}\phi}}
{S}_{67}{}^{O \,\bar{\beta}}
&=&
2\partial_P \left(\frac{\sqrt{-g_5}}{e^{4\chi+\frac{1}{\sqrt{5}}\phi}}
\partial^{[P}{S}_{67}{}^{O] \,\bar{\beta}}\right)\,,\\
  \frac{m^2\sqrt{-g_5}}{e^{\chi-2\omega-\frac{1}{\sqrt{5}}\phi}}
{S}_{6}{}^{NO \,\bar{\beta}}
&=&
3 \partial_P \left(\frac{\sqrt{-g_5}}{e^{-\chi-2\omega+\frac{1}{\sqrt{5}}\phi}}
{\partial}^{[P}{S}_{6}{}^{NO] \,\bar{\beta}}
\right)\,,\\
 \frac{m^2\sqrt{-g_5}}{e^{6\chi+\frac{4}{\sqrt{5}}\phi}}
{S}_{67}{}^{O \,5}
&=&
2\partial_P \left(\frac{\sqrt{-g_5}}{e^{4\chi-\frac{4}{\sqrt{5}}\phi}}
\partial^{[P}{S}_{67}{}^{O] \,5}\right)\,,\\
  \frac{m^2\sqrt{-g_5}}{e^{\chi-2\omega+\frac{4}{\sqrt{5}}\phi}}
{S}_{6}{}^{NO \,5}
&=&
3 \partial_P \left(\frac{\sqrt{-g_5}}{e^{-\chi-2\omega-\frac{4}{\sqrt{5}}\phi}}
{\partial}^{[P}{S}_{6}{}^{NO] \,5}
\right)\,.
\eeqs
The equations for  $S_{7NO}$ lead to the same states, ultimately
as a consequence of the self-duality conditions, and we could omit them.
They read as follows:
\beqs
  \frac{m^2\sqrt{-g_5}}{e^{\chi+2\omega-\frac{1}{\sqrt{5}}\phi}}
{S}_{7}{}^{NO \,\bar{\beta}}
&=&
3 \partial_P \left(\frac{\sqrt{-g_5}}{e^{-\chi+2\omega+\frac{1}{\sqrt{5}}\phi}}
{\partial}^{[P}{S}_{7}{}^{NO] \,\bar{\beta}}
\right)\,,\\
  \frac{m^2\sqrt{-g_5}}{e^{\chi+2\omega+\frac{4}{\sqrt{5}}\phi}}
{S}_{7}{}^{NO \,5}
&=&
3 \partial_P \left(\frac{\sqrt{-g_5}}{e^{-\chi+2\omega-\frac{4}{\sqrt{5}}\phi}}
{\partial}^{[P}{S}_{7}{}^{NO] \,5}
\right)\,,
\eeqs
where we highlight the  the dependence on $\omega$.
For completeness, we report also the equations for the $S_{MNO\,\alpha}$ fields,
although we do not use them anywhere in the paper:
\beqs
 \frac{{m^2} \sqrt{-g_5}}{e^{-4\chi-\frac{1}{\sqrt{5}}\phi}}
S^{{MNO}\,\bar{\beta}}
&=&4
\partial_P \left(\frac{\sqrt{-g_5}} {e^{-6\chi+\frac{1}{\sqrt{5}}\phi}}
\partial^{[P} S^{MNO] \,\bar{\beta}}
\right)\,,\\
 \frac{{m^2} \sqrt{-g_5}}{e^{-4\chi+\frac{4}{\sqrt{5}}\phi}}
S^{{MNO}\,5}
&=&4 \partial_{{P}} \left(\frac{\sqrt{-g_5}} {e^{-6\chi-\frac{4}{\sqrt{5}}\phi}}
\partial^{[P}S^{MNO] \,5}
\right)
\,.
\eeqs

For the next steps, we borrow the conventions from Eq.~(B.48) of Ref.~\cite{Elander:2018aub}:
\beqs
{\cal S}_5^{(2)}&=&\int\di^5x \sqrt{-g_5}\left\{-\frac{1}{4}H^{(2)} {\cal H}^{(2)}_{MN}{\cal H}^{(2)\,MN}\,
-\frac{1}{12}\,K^{(2)}\, {\cal H}^{(3)}_{MNO}{\cal H}^{(3)\,MNO}\right\}\,,
\label{Eq:Action2}
\eeqs
where the gauge-invariant combinations are defined as follows:
\beqs
{\cal H}^{(2)}_{MN}&=&{\cal F}_{MN} + m B_{MN}\,,\\
{\cal F}_{MN}&=&2 \partial_{[M}{\cal A}_{N]}\,,\\
{\cal H}^{(3)}_{MNO}&=&3\partial_{[M}B_{NO]}\,.
\eeqs

For the $1$-form we follow the notation in Eq.~(B.27) of Ref.~\cite{Elander:2018aub}:
\beqs
\label{Eq:Action1}
{\cal S}_5^{(1)}&=&\int \di^5 x \sqrt{g_5}\left\{-\frac{1}{4}H^{(1)}F_{MN}F^{MN}
+\right.\\
&&\nonumber\left.-\frac{1}{2} G^{(1)}
(\partial_M\pi+m A_M)(\partial^M\pi+m A^M)\right\}\,,
\eeqs
with $F_{MN}=2\partial_{[M}A_{N]}$.
In both cases, the Lagrangian possessed a gauge symmetry, and we can set $ {\cal A}_M=0=\pi$, in what we 
may call the 
unitary gauge.
In this  gauge, the equations of motion for the $2$-forms derived from Eq.~(\ref{Eq:Action2}) take the form:
\beqs
m^2H^{(2)}\sqrt{-g_5}\,B^{NO}
&=&
3 \partial_M \left(K^{(2)}\sqrt{-g_5} \, \partial^{[M}B^{NO]} \right) \,,
\eeqs
and the equations of motion for the $1$-forms, derived from Eq.~(\ref{Eq:Action1}),  take the form:
\beqs
m^2G^{(1)}\sqrt{-g_5}\,A^{N}
&=&
2 \partial_M\left(H^{(1)}\sqrt{-g_5} \, \partial^{[M}A^{N]}\right)\,.
\eeqs

We introduce now the 2-forms $B_{MN}$ and $B_{MN}^{\prime}$
associated with $S_{6MN}$ and $S_{7MN}$, respectively, and the $1$-form $A_{M}$ 
associated with $S_{67M}$.
By direct comparison between the two sets of equations of motion,  
we find that we can cast the action of the massive $2$-forms and $1$-forms
originating in seven dimensions
as in Eqs.~(\ref{Eq:Action2}) and~(\ref{Eq:Action1}), with the identification
with five $2$-forms and five $1$-forms
written as matrices in $SO(5)$ space:
\beqs
B_{MN\,\alpha}&=&\left\{S_{6MN\,\bar{\beta}}\,,S_{6MN\,5}\right\}\,,\\
H^{(2)}&=&{\rm diag}\,\left(e^{-\chi+2\omega+\frac{1}{\sqrt{5}}\phi}\,,e^{-\chi+2\omega-\frac{4}{\sqrt{5}}\phi}
\right)\,,\\
K^{(2)}&=&{\rm diag}\,\left(e^{\chi+2\omega-\frac{1}{\sqrt{5}}\phi}\,,e^{\chi+2\omega+\frac{4}{\sqrt{5}}\phi}
\right)\,,\\
B^{\prime}_{MN\,\alpha}&=&
\left\{S_{7MN\,\bar{\beta}}\,,S_{7MN\,5}\right\}\frac{}{}\,,\\
H^{(2)\prime}&=&{\rm diag}\,\left(e^{-\chi-2\omega+\frac{1}{\sqrt{5}}\phi}\,,e^{-\chi-2\omega-\frac{4}{\sqrt{5}}\phi}
\right)\,=\,\left(K^{(2)}\right)^{-1}\,,\\
K^{(2)\prime}&=&{\rm diag}\,\left(e^{\chi-2\omega-\frac{1}{\sqrt{5}}\phi}\,,e^{\chi-2\omega+\frac{4}{\sqrt{5}}\phi}
\right)\,=\,\left(H^{(2)}\right)^{-1}\,,\\
A_{M\,\alpha}&=&\left\{S_{67M\,\bar{\beta}}\,,S_{67M\,5}\right\}\,,\\
G^{(1)}&=&{\rm diag}\,\left(
e^{-6\chi+\frac{1}{\sqrt{5}}\phi}\,,e^{-6\chi-\frac{4}{\sqrt{5}}\phi}
\right)\,,\\
H^{(1)}&=&{\rm diag}\,\left(e^{-4\chi-\frac{1}{\sqrt{5}}\phi}\,,e^{-4\chi+\frac{4}{\sqrt{5}}\phi}
\right)\,.
\eeqs
These identifications have been used to arrive to the relevant entries of the matrices 
in Eqs.~(\ref{Eq:H(1)}),  (\ref{Eq:G(1)}), (\ref{Eq:H(2)}), and~(\ref{Eq:K(2)}).

\section{Formalism in five dimensions}
\label{Sec:Formalism}

We write here some general intermediate results, of a technical nature, 
that we use in the body of the paper
for the definition of the background equations and the study of the spectrum of fluctuations, starting 
from the action written in the form of Eq.~(\ref{Eq:Action5}).
We follow closely the notation adopted in Ref.~\cite{Elander:2018aub}, 
and indeed some parts of this Appendix are
repetitious in this respect. Nevertheless, we find it useful to add this Appendix in 
order for the paper to be self-contained, and also to clarify possible ambiguities in the notation.

We start by repeating Eq.~(\ref{Eq:Action5}), and then devote the two subsequent subsections
to the analysis of the system of active scalars coupled to gravity, and of the $p$-forms, respectively. 
We include an extensive 
discussion of the boundary conditions for the $2$-forms, as the self-duality condition affects them 
in a way that was not considered in Ref.~\cite{Elander:2018aub}.
This section is concluded by displaying the asymptotic expansions for 
 the fluctuations,  which are used to impose the boundary conditions.

The action in Eq.~(\ref{Eq:Action5}) is the following:
\beqs
{\cal S}_5&=&\int \di^5 x \sqrt{-g_5}
\left\{ \frac{{ R}}{4} -\frac{1}{2}g^{MN}G_{ab}\partial_M\Phi^a\partial_N\Phi^b
 - {\cal V}_5(\Phi^a)
 +\right.\\
 &&\nonumber
 \left.
 -\frac{1}{2}g^{MN}G^{(0)}_{ab}\partial_M\Phi^{(0)a}\partial_N\Phi^{(0)b}
 -\frac{1}{2}m^{(0)2}_{ab}\Phi^{(0)a}\Phi^{(0)b}
  +\right.\\
 &&\nonumber
 \left.
 -\frac{1}{2} g^{MN} G^{(1)}_{AB}
 {\cal H}_{\ \ \ \ M}^{(1)A} {\cal H}^{(1)B}_{\ \ \ \ N}
  -\frac{1}{4} g^{MO} g^{NP} H^{(1)}_{AB}F_{\,\,\,\,MN}^{A}F^{B}_{\,\,\,\,OP}
  +\right.\\
 &&\nonumber
 \left.
 -\frac{1}{4} g^{MO} g^{NP} H^{(2)}_{\Gamma\Delta} {\cal H}^{(2)\Gamma}_{\,\,\,\,\,\,\,\,\,\,MN}{\cal H}^{(2)\Delta}_{\,\,\,\,\,\,\,\,\,\,\,OP}\,
-\frac{1}{12} g^{MP} g^{NQ} g^{OR} K^{(2)}_{\Gamma\Delta}\, {\cal H}^{(3)\Gamma}_{\,\,\,\,\,\,\,\,\,\,MNO}{\cal H}^{(3)\Delta}_{\,\,\,\,\,\,\,\,\,\,\,PQR}
 \right\}\,.
\eeqs
The first line of the action depends on fields that may have non-trivial profiles in the vacuum, 
and their fluctuations are treated with the gauge-invariant sigma-model formalism.
The subsequent three lines in the action contain the kinetic and mass terms for $p$-forms (with $p=0, 1, 2$),
all of which  have vanishing background profile.

\subsection{Scalars coupled to gravity}
\label{Sec:sigmamodel}

It is convenient to make use of the 
gauge-invariant formalism developed in 
Refs.~\cite{Bianchi:2003ug,Berg:2005pd,Berg:2006xy,Elander:2009bm,Elander:2010wd} 
(and~\cite{Elander:2018aub,Elander:2020csd}).
Borrowing from Refs.~\cite{Berg:2005pd,Elander:2010wd}, consider $n$ 
real scalars $ \Phi^a$, with $a=1\,,\,\cdots\,,\,n$ ($n=3$ in this paper).  The action ${\cal S}_D$
of the sigma-model coupled to gravity in $D$ dimensions is written as follows:
\beqs
{\cal S}_D &=&\int\di^Dx\sqrt{-g}\left[\frac{{ R} }{4}
-\frac{1}{2} G_{ab}g^{MN}\partial_M  \Phi^a\partial_N  \Phi^b - {\mathcal V}( \Phi^a)\right]\,.
\label{Eq:ActionD}
\eeqs
The backgrounds of interest are identified by first introducing the following  ansatz for the metric and scalars
\beqs
\di s^2_D&=&\di r^2 +e^{2{ A}(r)}\, \eta_{\mu\nu}\di x^{\mu}\di x^{\nu}\,,\\
 \Phi^a&=& \Phi^a(r)\,,
\eeqs
which assumes that all the background functions depend only on the radial direction $r$ in the geometry.
Greek indexes  $\mu, \nu$ extend over $D-1$ dimensions.
The metric has signature mostly $+$.
The connection symbols, with our  conventions, are
\beqs
\Gamma^P_{\,\,\,\,MN}&\equiv&\frac{1}{2}g^{PQ}\left(\frac{}{}\partial_Mg_{NQ}+\partial_Ng_{QM}-\partial_Qg_{MN}\right)\,,
\eeqs
while the Riemann tensor is
\beqs
R_{MNP}^{\,\,\,\,\,\,\,\,\,\,\,\,\,\,\,\,\,Q}&\equiv&\partial_N\Gamma^Q_{\,\,\,\,MP}
-\partial_M\Gamma^Q_{\,\,\,\,NP}+\Gamma^S_{\,\,\,\,MP}\Gamma^Q_{\,\,\,\,SN}
-\Gamma^S_{\,\,\,\,NP}\Gamma^Q_{\,\,\,\,SM}\,,
\eeqs
the Ricci tensor is
\beqs
R_{MN}&\equiv&R_{MPN}^{\,\,\,\,\,\,\,\,\,\,\,\,\,\,\,\,\,P}\,,
\eeqs
and finally the Ricci scalar is
\beqs
{R}&\equiv&R_{MN}g^{MN}\,.
\eeqs
The (gravity) covariant derivative for a $(1,1)$-tensor takes the form
\beqs
\nabla_M T^{P}_{\,\,\,\,N}&\equiv&\partial_MT^{P}_{\,\,\,\,N}+\Gamma^P_{\,\,\,\,MQ}T^{Q}_{\,\,\,\,N}-\Gamma^Q_{\,\,\,\,MN}T^{P}_{\,\,\,\,Q}\,,
\eeqs
and generalises to any $(m,n)$-tensors.

The radial direction is bounded, with $r_1\leq r \leq r_2$, 
hence we also  add boundary-localised terms to the action, which take the form
\beqs
{\cal S}_1&=&-\int\di^Dx\sqrt{-\tilde{g}}\delta(r-r_1)\left[\frac{1}{2} K +\lambda_{(1)}(\Phi^a)\right]\,,\\
{\cal S}_2&=&\int\di^Dx\sqrt{-\tilde{g}}\delta(r-r_2)\left[\frac{1}{2} K +\lambda_{(2)}(\Phi^a)\right]\,,
\eeqs
where $K=g^{MN}K_{MN}$ is the extrinsic curvature and $\lambda_{(i)}$ are boundary-localised potentials.
The  signs of the two boundary-localised contributions to the action
reflect  the orientation of the ortho-normalised vector $N^M$, 
which is  parallel to the radial direction $r$, 
and satisfies
\beqs
g^{MN}N_NN_M&=&1\,,\\
\tilde{g}_{MN}N^N&=&0\,,
\eeqs
where $\tilde{g}_{MN}\equiv g_{MN}-N_MN_N$ is the induced metric. The 
second fundamental form  
is defined in terms of the covariant derivative $\nabla_M$,  
as $K_{MN}\equiv \nabla_MN_M$.

The sigma-model connection is defined in a similar fashion to gravity.
It descends from the sigma-model metric $ G_{ab}$ and the
 sigma-model derivative  $\partial_b=\frac{\partial}{\partial \Phi^b}$, to read 
\beqs
 {\cal G}^d_{\,\,\,\,ab}&\equiv& \frac{1}{2} G^{dc}\left(\frac{}{}\partial_a G_{cb}
+\partial_b G_{ca}-\partial_c G_{ab}\right)\,.
\eeqs
The sigma-model Riemann tensor  is the following
\beqs
 {\cal R}^a_{\,\,\,\,bcd}
&\equiv& \partial_c {\cal G}^a_{\,\,\,\,bd}-\partial_d {\cal G}^a_{\,\,\,\,bc}
+ {\cal G}^e_{\,\,\,\,bd} {\cal G}^a_{\,\,\,\,ce}- {\cal G}^e_{\,\,\,\,bc} {\cal G}^a_{\,\,\,\,de}\,.
\eeqs
(The indexes in the conventions for the two Riemann tensors follow a reverse ordering.)
Finally,  the sigma-model covariant derivative is
\beqs
D_b X^d_{\,\,\,\,a}&\equiv& \partial_b X^d_{\,\,\,\,a}+{\cal G}^d_{\,\,\,\,cb}X^c_{\,\,\,\,a}
- {\cal G}^c_{\,\,\,\,ab}X^d_{\,\,\,\,c}\,.
\eeqs

The equations of motion satisfied by the background scalars
 are the following:
\beqs
\label{eq:backgroundEOM1}
\partial_r^2\Phi^a\,+\,(D-1)\partial_r {A}\partial_r\Phi^a\,+\, {\cal G}^a_{\,\,\,\,bc}\partial_r\Phi^b\partial_r\Phi^c\,-\,\mathcal V^a
&=&0\,,
\eeqs
where the sigma-model derivatives are given by $\mathcal V^a\equiv  G^{ab}\partial_b \mathcal V$, and  
$\partial_b \mathcal V\equiv \frac{\partial \mathcal V}{\partial \Phi^b}$. The Einstein equations reduce to
\beqs
\label{eq:backgroundEOM2}
(D-1)(\partial_r { A})^2\,+\,\partial_r^2 { A}\,+\,\frac{4}{D-2} \mathcal V &=&
0\,,\\
(D-1)(D-2)(\partial_r { A})^2\,-\,2 G_{ab}\partial_r\Phi^a\partial_r\Phi^b\,+\,4 \mathcal V&=&0\,.
\eeqs

If the potential ${ \mathcal V}$ can be written in terms of a superpotential $\cal W$ satisfying the following relation:
\beqs
{ \mathcal V}&=&\frac{1}{2} G^{ab}\partial_a {\cal W}\partial_b {\cal W}-\frac{D-1}{D-2}{\cal W}^2\,,
\eeqs
then any solution of the first order system defined by
\beqs
\partial_r { A} &=& -\frac{2}{D-2}{\cal W}\,,\\
\partial_r \Phi^a &=& G^{ab}\partial_b {\cal W}\,,
\eeqs
is also a solution of the equations of motion.

\subsubsection{Fluctuations: tensors and active scalars}
\label{Sec:fluctuations}

The fluctuations around the classical background of the active scalars and gravity
are treated with the gauge-invariant formalism 
in Refs.~\cite{Bianchi:2003ug,Berg:2006xy,Berg:2005pd,Elander:2009bm,Elander:2010wd}.
 The scalar fields can be written  as
\beq
	\Phi^a(x^\mu,r) =  \Phi^a(r) + \varphi^a(x^\mu,r) \,,
\eeq
where $\varphi^a(x^\mu,r)$ are small fluctuations around the background 
solutions $\Phi^a(r)$.
By decomposing the metric according to the ADM formalism, one writes
\beqs
	\dd s_D^2 &=& \left( (1 + \nu)^2 + \nu_\sigma \nu^\sigma \right) \dd r^2 + 2 \nu_\mu \dd x^\mu \dd r 
	+ e^{2 {A}(r)} \left( \eta_{\mu\nu} + h_{\mu\nu} \right) \dd x^\mu \dd x^\nu \,,
	\eeqs
	and 
	\beqs
	h^\mu{}_\nu &=& \mathfrak e^\mu{}_\nu + i q^\mu \epsilon_\nu + i q_\nu \epsilon^\mu 
	+ \frac{q^\mu q_\nu}{q^2} H + \frac{1}{D-2} \delta^\mu{}_\nu h,
\eeqs
where $\mathfrak e^\mu{}_\nu$ is transverse and traceless, $\epsilon^\mu$ is transverse, and the Greek 
indices $\mu$, $\nu$ are raised and lowered by the boundary metric 
$\eta_{\mu\nu}$.  $\nu(x^\mu,r)$, $\nu^\mu(x^\mu,r)$,
 $\mathfrak e^\mu{}_\nu(x^\mu,r)$, $\epsilon^\mu(x^\mu,r)$, $H(x^\mu,r)$, and $h(x^\mu,r)$ are small
  fluctuations around the background metric with the warp factor ${ A}(r)$.

After forming the following gauge-invariant (under diffeomorphisms) combinations:
\beqs
	\mathfrak a^a &=& \varphi^a - \frac{\partial_r  \Phi^a}{2(D-2)\partial_r { A}} h \,, \\
	\mathfrak b &=& \nu - \partial_r \left( \frac{h}{2(D-2)\partial_r {A}} \right) \,, \\
	\mathfrak c &=& e^{-2{ A}} \partial_\mu \nu^\mu - \frac{e^{-2{ A}} q^2 h}{2(D-2) \partial_r A} 
	- \frac{1}{2} \partial_r H \,, \\
	\mathfrak d^\mu &=& e^{-2{A}} P^\mu{}_\nu \nu^\nu - \partial_r \epsilon^\mu \,,
\eeqs
the linearized equations of motion decouple--thanks to the algebraic equations
 for $\mathfrak{b}$ and $\mathfrak{c}$.

The tensorial fluctuations $\mathfrak e^\mu{}_\nu$  are gauge-invariant, and obey the equation of motion
\beq
\label{eq:tensoreom}
	\left[ \partial_r^2 + (D-1) \partial_r {A} \partial_r - e^{-2{ A}(r)} q^2 \right] \mathfrak e^\mu_{\,\,\,\nu} = 0 \,,
\eeq
where $M^2\equiv -q^2$ is the mass squared of the states. They obey the boundary conditions
\beq
\label{eq:tensorbc}
\left.\frac{}{}	\partial_r \mathfrak e ^\mu_{\,\,\,\nu} \right|_{r=r_i}= 0 \, .
\eeq
 Eqs.~(\ref{eq:tensoreom})--(\ref{eq:tensorbc}) allow to compute the spectrum of spin-2 states.
 The equation of motion for $\mathfrak d^\mu$ is algebraic; it does not lead to a spectrum of 
composite states. The equations of motion for $\mathfrak b$ and $\mathfrak c$ are also algebraic, 
and solved in terms of $\mathfrak a^a$, which obey the following equations of motion:
\beqs
\label{eq:scalareom}
	0 &=& \Big[ {\cal D}_r^2 + (D-1) \partial_{r}{A} {\cal D}_r - e^{-2{A}} q^2 \Big] \mathfrak{a}^a \,+\,\\ \nonumber
	&& - \Big[  {\mathcal V}^{\,\,a}{}_{\,|c} - \mathcal{R}^a{}_{bcd} \partial_{r}\Phi^b \partial_{r}\Phi^d + 
	\frac{4 (\partial_{r}\Phi^a  {\mathcal V}^{\,b} +  {\mathcal V}^{\,a} 
	\partial_{r}\Phi^b) G_{bc}}{(D-2) \partial_{r} {A}} + 
	\frac{16  {\mathcal V} \partial_{r}\Phi^a \partial_{r}\Phi^b G_{bc}}{(D-2)^2 (\partial_{r}{A})^2} \Big] \mathfrak{a}^c\,,
\eeqs
while the boundary conditions are given by
\beqs
\label{eq:scalarbc}
 \frac{2  e^{2A}\partial_{r} \Phi^a }{(D-2)q^2 \partial_{r}{ A}}
	\left[ \partial_{r} \Phi^b{\cal D}_r -\frac{4  {\cal V} \partial_{r} \Phi^b}{(D-2) 
	\partial_r { A}} - {\cal V}^b \right] \mathfrak a_b - \mathfrak a^a\Big|_{r_i} = 0 \, .
\eeqs
Here, ${\mathcal V}^a{}_{|b} \equiv \frac{\partial {\mathcal V}^a}{\partial \Phi^b} + \mathcal G^a_{\ bc} {\mathcal V}^c$, 
and the background covariant derivative is defined as 
$\mathcal D_r \mathfrak a^a \equiv \partial_r \mathfrak a^a +
 \mathcal G^a_{\ bc} \partial_r  \Phi^b \mathfrak a^c$.

\subsection{$p$-forms and other scalars}
\label{Sec:probes}

The fluctuations of scalars that do not have a vacuum expectation value, for which
$\langle \partial_r \Phi^{(0)a}\rangle =0$, decouple from gravity. They obey linearised equations that can
 be derived with the same formalism as for the active scalars, but the ultimate result is the much simpler expression:
\beqs
\left[ \partial_r^2 + \left( 4\partial_r A + \frac{\partial_r G^{(0)}_a}{G^{(0)}_a} \right) \partial_r 
- \left( \frac{m^{(0)2}_a}{G^{(0)}_a} + q^2 e^{-2A} \right) \right] \Phi^{(0)a}(q,r)&=&0\,.
\eeqs
The boundary conditions simplify to read 
\beqs
\label{eq:inactivebc}
  \Phi^{(0)a}(q,r)\Big|_{r_i} = 0 \, .
\eeqs

For the $1$-forms and $2$-forms, we adopt the convenient choice of the $R_{\chi}$ gauge, appropriately 
generalised to the relevant cases. The advantage of doing so is that by removing from the classical Lagrangian
mixing between fields of different spin, it becomes possible to build manifestly gauge-invariant combinations,
that can be studied without ambiguities. The Reader can find details about the treatment of these
fields in five dimensions in Ref.~\cite{Elander:2018aub}, and we borrow from there
the main equations used in studying the spectrum.

The action of the generic massive $1$-form $V^A_{\,\,\,M}$ is given by the third line of Eq.~(\ref{Eq:Action5}),
and the spectrum of physical states can be obtained by looking at the equations of motion and boundary conditions
for the transverse part of the four-vector $V^A_{\,\,\,\mu}$, as well as the gauge invariant scalar $X^A$
defined in Eqs.~(B.38) and ~(B.39) of Ref.~\cite{Elander:2018aub}:
\beqs
0&=&\left[q^2H^{(1)}_{B}-\partial_r\left(e^{2A}H^{(1)}_{B}\frac{}{}\partial_r\right)
+e^{2A}m^2G^{(1)}_{B}\right]P^{\mu\nu}V^B_{\,\,\,\mu}(q,r)\,,\\
\label{Eq:bcUV1}
0&=&\left.\left[H^{(1)}_{B}e^{2A}\partial_r+q^2D_i+m^2C_ie^{2A}\right]
P^{\mu\nu}V^B_{\,\,\,\mu}(q,r)\right|_{r=r_i}\,,\\
0&=&\left[\partial_r^2+\left(-2\partial_r A
-\frac{\partial_r G^{(1)}_{B}}{G^{(1)}_{B}}\right)\partial_r 
+\left(-e^{-2A}q^2-m^2\frac{G^{(1)}_{B}}{H^{(1)}_{B}}\right)\right]X^B(q,r)\,,\\
\label{Eq:bcUV2}
0&=&\left.\left[\frac{}{}C_i\partial_r + G^{(1)}_{B}\right]X^B(q,r)\right|_{r=r_i}\,.
\eeqs
The last two of these equations are only applicable in the case when pseudo-scalars are present and mix with the corresponding axial-vectors. In Ref.~\cite{Elander:2018aub}   all the boundary-localised terms are set to $C_i=0=D_i$, which reduces the 
boundary conditions for the vectors to Neumann, and for the scalars to Dirichlet.
A more orthodox choice might be to set
$C_1=0=D_1$ in the IR, but $C_2=+\infty=D_2$ in the UV, hence flipping the Dirichlet and Neumann boundary 
conditions---the position of the poles of the relevant correlation functions is unaffected, 
but this flipped choice improves convergence with the UV regulator $r_2$.

The dynamics of the generic $2$-form $B^{\Gamma}_{\,\,\,MN}$ is governed by the
fourth line of the action of Eq.~(\ref{Eq:Action5}). Adapting from Ref.~\cite{Elander:2018aub},
we find for the transverse polarisations of the $2$-form the equations of motion and boundary conditions take the following form:
\beqs
0&=&\left[\frac{}{}
K^{(2)}_{\Gamma}q^2e^{-2A}-\partial_r\left(\frac{}{}K^{(2)}_{\Gamma}\partial_r\right)
+m^2H^{(2)}_{\Gamma}\right]P^{\mu\rho}P^{\nu\sigma} B^{\Gamma}_{\,\,\,\rho\sigma}(q,r)\,,\\
0&=&\left.\left[\frac{}{}K^{(2)}_{\Gamma}E_iq^2e^{-2A}+K^{(2)}_{\Gamma}\partial_r
+D_iH^{(2)}_{\Gamma}m^2\right]P^{\mu\rho}P^{\nu\sigma} B^{\Gamma}_{\,\,\,\rho\sigma}(q,r)\right|_{r=r_i}\,,
\eeqs
where $\Gamma$ is, in the context of this paper, a multi-index that spans the ten different  $2$-forms obtained
from $S_{6NO\,\alpha}$ and $S_{7NO\,\alpha}$.
The other three degrees of freedom propagated by
each massive $2$-form yield an independent tower of massive vectors.
The physical (transverse) massive vectors $X^{\Gamma}_{\,\,\,\mu}$
 defined in Eqs.~(B.59) and~(B.60) of Ref.~\cite{Elander:2018aub}
obey the following equations and boundary conditions:
\beqs
0&=&\left[\frac{}{}\partial_r^2-\frac{\partial_r H^{(2)}_{\Gamma}}{H^{(2)}_{\Gamma}}\partial_r +\left(-e^{-2A}q^2
-m^2\frac{H^{(2)}_{\Gamma}}{K^{(2)}_{\Gamma}}\right)\right]X^{\Gamma}_{\,\,\,\mu}(q,r)\,,\\
0&=&\left.\frac{}{}\left[\frac{}{}\partial_r+\frac{1}{D_i}\right]X^{\Gamma}_{\,\,\,\mu}(q,r)\right|_{r=r_i}\,.
\eeqs
As discussed in Ref.~\cite{Elander:2018aub}, the other components of the $2$-form fields
do not yield additional physical states, and depend on the $\xi$ parameters 
of the $R_{\xi}$-gauge formulation of the theory.

In Ref.~\cite{Elander:2018aub} the choice was made of setting
 $D_i=0=E_i$, in which case we recover  Neumann boundary conditions
 for the $B^{\Gamma}_{\,\,\,\mu\nu}$ fields and Dirichet 
 boundary conditions for the vectors ${X}^{\Gamma}_{\,\,\,\mu}$.
 But doing so in the context of this paper would not take into consideration the implication of
 the self-duality conditions, and hence we pause here to analyse
  in more detail this  choice of boundary conditions.

  The maximal supergravity theory contains a
  set of $3$-forms that obey a self-duality condition, which  is necessary for consistency.
  After introducing appropriate gauge-invariant fields $B_{NO\,\alpha}$ and $B^{\prime}_{NO\,\alpha}$
  (associated with $S_{6NO}$ and $S_{7NO}$, respectively), and the vectors $X_M$ and $X^{\prime}_{M}$,
    the number of 
  degrees of freedom is superficially doubled. In the original seven-dimensional language,
  the $S_{6NO}$ and $S_{7NO}$ fields are related, as for example evident in Eq.~(\ref{Eq:6and7}).
By looking closer to the equations and boundary conditions
  for these fields, 
the functions appearing in all the equations of motion
obey the following non-trivial relations:   
\beq
K^{(2)\prime}=(H^{(2)})^{-1}\,, \hspace{1cm} K^{(2)}=(H^{(2)\prime})^{-1}\,,
\eeq
both of which are
consequences of the aforementioned self-duality conditions.

By replacing $K^{(2)}$ and $K^{(2)\prime}$, 
the equations and boundary conditions can then be written as follows:
\beqs
0&=&\left[\frac{}{}
q^2e^{-2A}-H^{(2)\prime}_{A}\partial_r\left(\frac{1}{H^{(2)\prime}_{A}}\partial_r\right)
+m^2H^{(2)}_{A}H^{(2)\prime}_{A}\right]P^{\mu\rho}P^{\nu\sigma}B^{A}_{\,\,\,\rho\sigma}(q,r)\,,\\
0&=&\left.\left[\frac{}{}E_iq^2e^{-2A}+\partial_r
+D_iH^{(2)}_{A}H^{(2)\prime}_{A}m^2\right]P^{\mu\rho}P^{\nu\sigma}B^{A}_{\,\,\,\rho\sigma}(q,r)\right|_{r=r_i}\,,\\
0&=&\left[\frac{}{}
q^2e^{-2A}-H^{(2)}_{A}\partial_r\left(\frac{1}{H^{(2)}_{A}}\partial_r\right)
+m^2H^{(2)}_{A}H^{(2)\prime}_{A}\right]P^{\mu\rho}P^{\nu\sigma}B^{\prime\,A}_{\,\,\,\rho\sigma}(q,r)\,,\\
0&=&\left.\left[\frac{}{}E^{\prime}_iq^2e^{-2A}+\partial_r
+D^{\prime}_iH^{(2)}_{A}H^{(2)\prime}_{A}m^2\right]P^{\mu\rho}P^{\nu\sigma}
B^{\prime\,A}_{\,\,\,\rho\sigma}(q,r)\right|_{r=r_i}\,,\\
0&=&\left[\frac{}{}{\partial_r}^2-\frac{\partial_r H^{(2)}_{A}}{H^{(2)}_{A}}\partial_r +\left(-e^{-2A}q^2
-m^2{H^{(2)}_{A}}H^{(2)\prime}_{A}\right)\right]X^{A}_{\,\,\,\mu}(q,r)\,,\\
0&=&\left.\frac{}{}\left[\frac{}{}\partial_r+\frac{1}{D_i}\right]X^{A}_{\,\,\,\mu}(q,r)\right|_{r=r_i}\,,\\
0&=&\left[\frac{}{}{\partial_r}^2-\frac{\partial_r H^{(2)\prime}_{A}}{H^{(2)\prime}_{A}}\partial_r +\left(-e^{-2A}q^2
-m^2{H^{(2)}_{A}}H^{(2)\prime}_{A}\right)\right]X^{\prime\,A}_{\,\,\,\mu}(q,r)\,,\\
0&=&\left.\frac{}{}\left[\frac{}{}\partial_r+\frac{1}{D^{\prime}_i}\right]X^{\prime\,A}_{\,\,\,\mu}(q,r)\right|_{r=r_i}\,,
\eeqs  
with $A=1\,,\cdots\,,5$ the $SO(5)$ index replacing the generic $\Gamma$.

One  key observation  is that
 the equations satisfied by the transverse parts of $B_{\mu\nu}$ and $B^{\prime}_{\mu\nu}$
are  the same as those obeyed by $X^{\prime}_{\mu}$ and $X_{\mu}$ (notice the inverted order).
This is a direct consequence of self duality,
and it is most welcome, as there is no a priori reason to favour the formulation of the problem in terms of 
either $B_{\mu\nu}$ or $B^{\prime}_{\mu\nu}$.
Thanks to this observation, we can either chose to study only the pair $B_{\mu\nu}^A$ and $X_{\mu}^A$
(as we do in the body of the paper),
or only the pair $B_{\mu\nu}^{\prime\,A}$ and $X_{\mu}^{\prime\,A}$, with no loss of generality.

For the self-duality to be manifest and exact in the spectrum, the same interchangeable 
roles we see in the equations of motion must appear in the boundary conditions as well. 
In the first instance, 
this requirement forces us to impose that $E_i=0=E^{\prime}_i$, as there is no
equivalent term admissible in the boundary conditions for $X^{\prime}_{\mu}$ and $X_{\mu}$.
We  must furthermore require also that 
\beqs
\left.\frac{}{} H^{(2)} H^{(2)\prime} m^2 \right|_{r_i}&=&\frac{1}{D_iD^{\prime}_i}\,.
\eeqs

Because we want the expressions to be completely symmetrical
under the exchange $S_{6NO}\leftrightarrow S_{7NO}$, we can further 
impose the condition $D_i=D_i^{\prime}$,
which equivalently requires to impose 
\beqs
\left.\frac{}{} m \sqrt{ H^{(2)} H^{(2)\prime}} \right|_{r_i}&=&\frac{1}{D_i}\,,
\eeqs
results in the final form of the equations
reported in the body of the paper as Eqs.~(\ref{Eq:2f1})-(\ref{Eq:2f4}), and ultimately leads to
 the results on display in Fig.~\ref{Fig:Tensors4-Mixed}.

\subsection{Asymptotic expansions for the fluctuations}
\label{Sec:Expansions}

We report here the asymptotic expansions
of the gauge invariant combinations of the fluctuations of the fields around the confining backgrounds,
both in the UV (large $\r$) and in the IR ($\r\rightarrow \r_o$),
which we obtain by assuming that the background functions are written as the 
 power expansions in Eqs.~(\ref{Eq:UV1})--(\ref{Eq:UV4}) and Eqs.~(\ref{Eq:IR1})--(\ref{Eq:IR3}), respectively.
These expansions have been used in order to implement (and
 improve the convergence of) the numerical calculations of the mass spectra,
when taking the limits $\r_1\rightarrow \r_o$ and $\r_2\rightarrow +\infty$, so as to recover physical results
free of spurious cutoff dependences.

For presentation purposes, in this Appendix we simplify the expressions by setting
$\chi_U=A_U=\omega_U=0=\chi_6$ when we write the UV expansions of the fluctuations,
hence retaining only the dependence on $\phi_2$, $\phi_4$, and $\omega_6$.
Conversely, when presenting the IR expansions we set
 $\r_o=0=\chi_I=\omega_I$, and the expansions depend only on $\phi_I$. We note that these two sets of choices cannot be imposed 
simultaneously on a given background. Finally, anticipating that we will use the expansions in order to set up the boundary conditions for the fluctuations in the computation of spectra, we substitute the dependence on $q^2$ in favour of $M^2 = -q^2$.

\subsubsection{UV expansions}
\label{Sec:ExpansionsUV}

In the case of the UV expansions, 
having set $\chi_U=A_U=\omega_U=0=\chi_6$,
we use the convenient coordinate $z=e^{-\frac{1}{2}\r}$,
and write the expansions in powers of the small  coordinate $z$, truncated at finite order.
The confining solutions are characterised by the parameters $\phi_2$,
$\phi_4$, and $\omega_6$.
But these parameters are not independent, as they are all functions of the IR parameters that determine the background solution.
In addition, although  the equations for the fluctuations are of second order,  yet
 the equations are linear, and hence depend on an arbitrary overall normalisation.
 But we retain both free coefficients in the solutions as  free parameters,  for simplicity.

The asymptotic behaviour of the background solutions
is such as to yield an AdS$_7$ geometry, the dual field theory being governed by a strongly-coupled
fixed point in a six-dimensional theory. For scalar fields, this observation implies that we expect 
the two free parameters first appear in front of the terms ${\cal O}(z^{\Delta_1})$ and ${\cal O}(z^{\Delta_2})$,
in the power expansion of the solutions,
with the constraint $\Delta_1+\Delta_2=6$.  But this is not the general behaviour of all the fluctuations,
as we shall see momentarily: some of the fields originate from the compactification on the torus, so that their asymptotic behaviour is determined by the lower-dimensional geometry.

We start from the tensors, the $\mathfrak{e}^{\mu}_{\,\,\,\nu}$ transverse and traceless fluctuations of the metric.
We find the following expansion:
\beqs
\mathfrak{e}^{\mu}_{\,\,\,\nu}(z)&=&
(\mathfrak{e}_0)^{\mu}_{\,\,\,\nu}+
z^2(\mathfrak{e}_0)^{\mu}_{\,\,\,\nu} \frac{M^2   }{2}
+z^4(\mathfrak{e}_0)^{\mu}_{\,\,\,\nu} \frac{(M^2)^2 }{4}
+z^6(\mathfrak{e}_6)^{\mu}_{\,\,\,\nu}  + \\
&&\nonumber
+z^6 \log (z)(\mathfrak{e}_0)^{\mu}_{\,\,\,\nu} \frac{M^2 }{6}
 \left(\frac{4}{5} \phi_2^2- (M^2)^2\right) 
+{\cal O}(z^8)\,,
\eeqs
where $(\mathfrak{e}_0)^{\mu}_{\,\,\,\nu}$ and $(\mathfrak{e}_6)^{\mu}_{\,\,\,\nu}$ are the leading and subleading
 free parameters, respectively.

The expansion for the active scalars is given by the following expressions:
\beqs
\mathfrak{a}^1(z)
&=&
z^2\mathfrak{a}^{1}_{\,\,\,2} 
+z^4\mathfrak{a}^{1}_{\,\,\,4}
-z^4 \log (z) \mathfrak{a}^{1}_{\,\,\,2}  \,\frac{2}{5}\left(5 M^2
+18 \sqrt{5} \phi_2\right)+
\\
&&\nonumber
+
\frac{z^6}{20} \left[\frac{}{}\mathfrak{a}^{1}_{\,\,\,2} \left(-15 (M^2)^2
-108 \sqrt{5} M^2 \phi_2 
-1278 \phi_2^2 -36 \sqrt{5} \phi_4\right)\right.+ \\
&&\left. \nonumber -2 \mathfrak{a}^{1}_{\,\,\,4} \left(5 M^2+18 \sqrt{5} \phi_2\right)\right]
 +\\
&&\nonumber+
\frac{z^6}{20} \log (z)\mathfrak{a}^{1}_{\,\,\,2}\left[\frac{}{} 20 (M^2)^2 
+144 \sqrt{5} M^2 \phi_2 
+1944 \phi_2^2 \right] 
+{\cal O}(z^8)\,,\\
\mathfrak{a}^2(z)
&=&
\mathfrak{a}^{2}_{\,\,\,0}
+z^2  \mathfrak{a}^{2}_{\,\,\,0}  \frac{M^2}{2} 
+z^4\mathfrak{a}^{2}_{\,\,\,0}   \frac{(M^2)^2}{4}  +\\
&&\nonumber
+z^6 \mathfrak{a}^{2}_{\,\,\,6}
+z^6 \log  (z)   \mathfrak{a}^{2}_{\,\,\,0} \frac{M^2}{30}  \left(\frac{}{}4 \phi_2^2-5 (M^2)^2\frac{}{}\right) 
+{\cal O}(z^8)\,,\\
\mathfrak{a}^3(z)
&=&
\mathfrak{a}^{3}_{\,\,\,0}
+z^2  \mathfrak{a}^{3}_{\,\,\,0}  \frac{M^2}{2} 
+z^4 \mathfrak{a}^{3}_{\,\,\,0} \frac{(M^2)^2}{4}   +\\
&&\nonumber
+z^6 \mathfrak{a}^{3}_{\,\,\,6}
   +z^6  \log (z) \mathfrak{a}^{3}_{\,\,\,0}  \frac{M^2}{30}   \left(\frac{}{}4 \phi_2^2-5 (M^2)^2\frac{}{}\right)+{\cal O}(z^8)\,.
\eeqs
The free parameters are $\mathfrak{a}^{1}_{\,\,\,2}$, $\mathfrak{a}^{1}_{\,\,\,4}$, $\mathfrak{a}^{2}_{\,\,\,0}$, 
$\mathfrak{a}^{2}_{\,\,\,6}$, $\mathfrak{a}^{3}_{\,\,\,0}$, and $\mathfrak{a}^{3}_{\,\,\,6}$. 
The three solutions are decoupled from one another,
at this order. Furthermore, the expansion of the fluctuations 
associated with $\chi$ and $\omega$ 
are identical. Both these observations hold only approximately: terms appearing at higher-orders 
in the expansion in small $z$ introduce mixing.
For example, terms dependent on the combination 
$\omega_6\phi_2(2\mathfrak{a}^2_{\,\,\,0}-3\mathfrak{a}^3_{\,\,\,0})$ appear in the expansion of 
$\mathfrak{a}^1$ at
${\cal O}(z^8)$.

For the three towers of pseudoscalar fields, which
 are the fluctuations of the orthogonal combinations $X^B$ to the
would-be Goldstone bosons higgsed into the massive vectors---defined by 
the non-vanishing entries in Eq.~(\ref{Eq:Hdef})---we find the 
following expansions:
\beqs
\mathfrak{p}^1(z)
\label{Eq:p1}
&=&
\mathfrak{p}^1_{\,\,\,0}+\log (z) \mathfrak{p}^1_{\,\,\,l}
+\\
&&\nonumber
+\frac{z^2}{5 \phi_2} \left[\frac{}{}
\mathfrak{p}^1_{\,\,\,l} \left(5 M^2\phi_2 
+9 \sqrt{5} \phi_2^2 +5 \phi_4\right)
-\mathfrak{p}^1_{\,\,\,0}\left(5\phi_2M^2\right)\right]
+\\
&&\nonumber
+\frac{z^2}{5 \phi_2} \log (z)\left[\frac{}{}
\mathfrak{p}^1_{\,\,\,l} \left(-5 M^2\phi_2 
-18 \sqrt{5} \phi_2^2\right)
\right]
+\\
&&\nonumber
   +\frac{z^4}{80 \phi_2^2}
   \left[\frac{}{}\mathfrak{p}^1_{\,\,\,l} \left(\frac{}{}
-30 (M^2)^2 \phi_2^2
-36 \sqrt{5} \phi_4\phi_2^2
-54 \sqrt{5} M^2\phi_2^3 
 +20 M^2\phi_2 \phi_4 
\right.\right.+\\
&&\left.\left.\nonumber-1057\phi_2^4+20 \phi_4^2\frac{}{}\right)
   +20 \phi_2 \mathfrak{p}^1_{\,\,\,0} \left(
   \frac{}{}(M^2)^2 \phi_2-2 M^2\phi_4+5 \phi_2^3\frac{}{}\right)\right] 
      +\\
      &&\nonumber
   +\frac{z^4}{80 \phi_2^2}\log (z)
   \left[\frac{}{}\mathfrak{p}^1_{\,\,\,l} \left(\frac{}{}
   2 \phi_2^2 \left(\frac{}{}10 (M^2)^2 
-72 \sqrt{5} \phi_4 \right)
-72 \sqrt{5} M^2\phi_2^3 \right.\right.
   +\\
&&\nonumber  \left.\left. 
-40 M^2\phi_2 \phi_4 
+748 \phi_2^4\frac{}{}\right)
   +\mathfrak{p}^1_{\,\,\,0} \left(144
   \sqrt{5} M^2\phi_2^3\right) \right] 
      +\\
      &&\nonumber
   +\frac{z^4}{80 \phi_2^2} \log ^2(z)
   \left[\frac{}{}\mathfrak{p}^1_{\,\,\,l}  \left(\frac{}{}
144 \sqrt{5} M^2\phi_2^3 
+1296\phi_2^4 \frac{}{}\right)
\right] 
      +\\
&&\nonumber+\frac{z^6}{10800\phi_2^2}\left[
 \mathfrak{p}^1_{\,\,\,l} \left( 
 10\phi_2^2 \left(55(M^2)^3-1278 \sqrt{5}(M^2)\phi_4-120
  \omega_6\right)
  +\right.\right.\\
  &&\nonumber\left.\left.
  +10\phi_2^3 \left(99 \sqrt{5}(M^2)^2-10678
  \phi_4\right)-120\phi_2\phi_4 \left(20(M^2)^2+63
   \sqrt{5}\phi_4\right)
  +\right.\right.\\
  &&\nonumber\left.\left.
  -66585(M^2)\phi_2^4-900(M^2)
  \phi_4^2+21066 \sqrt{5}\phi_2^5
 \right)
  +\right.\\
  &&\nonumber\left.
 + \mathfrak{p}^1_{\,\,\,0} \left(
 60\phi_2 \left(-5(M^2)^3\phi_2+30(M^2)^2\phi_4+3
  (M^2) \left(557\phi_2^3
  +60 \sqrt{5}\phi_2 \phi_4\right)
  +\right.\right.\right.\\
  &&\nonumber\left.\left.\left.
  +60 \left(2 \sqrt{5}\phi_2^4+5\phi_2^2 \phi_4\right)\right)
  \right)
\right]+\\
&&\nonumber+\frac{z^6}{900\phi_2}\log (z)\left[
 \mathfrak{p}^1_{\,\,\,l} \left( 
 60\phi_2^2 \left(12 \sqrt{5}(M^2)^2+403\phi_4\right) 
  +150
  (M^2)^2\phi_4
  +\right.\right.\\
  &&\nonumber\left.\left.
    +5   \phi_2 \left(288 \sqrt{5}M^2\phi_4-5(M^2)^3\right)
+27525 M^2 \phi_2^3+32634 \sqrt{5}
  \phi_2^4
 \right)
  +\right.\\
  &&\nonumber\left.
+  \mathfrak{p}^1_{\,\,\,0} \left(
- 540\phi_2^2 \left(\sqrt{5}(M^2)^2+30 M^2\phi_2+10
   \sqrt{5}\phi_2^2\right)
  \right)
\right]+\\
&&\nonumber+{z^6}\log^2 (z)\left[ \mathfrak{p}^1_{\,\,\,l} 
\left(-\frac{3}{25}\phi_2 \left(5 \sqrt{5}(M^2)^2+195 M^2
\phi_2+428 \sqrt{5}\phi_2^2\right)\right)
\right]+{\cal O}(z^8)\,,\\
\mathfrak{p}^2(z)
&=&
\frac{1}{450 z^2}\left[\frac{}{}\mathfrak{p}^2_{\,\,\,-2} 
\left(450+90
   z^2 \left(5 M^2-\sqrt{5} \phi_2\right)
      +\right.\right.\\
   && \left.\left.
   -90 z^4  \log (z)\left(5 (M^2)^2-3 \sqrt{5} M^2 \phi_2-9
   \phi_2^2\right) \nonumber
   +\right.\right.\\
   && \left.\left.\nonumber
   +z^6 \left(50 (M^2)^3 (3 \log (z)-2)+15 \sqrt{5}
   (M^2)^2 \phi_2 (7-12 \log (z))
      +\right.\right.\right.\\
   && \nonumber \left.\left.\left.
   +30 M^2 \left(9 \phi_2^2
   (3-4 \log (z))+2 \sqrt{5} \phi_4\right)-900 \omega_6
      +\right.\right.\right.\\
   && \nonumber \left.\left.\left.
   +6 \sqrt{5}
   \phi_2^3 (143-171 \log (z))+330 \phi_2 \phi_4\right)\right)
         +\right.\\
   && \nonumber \left.
+30 z^4 \mathfrak{p}^2_{\,\,\,+2} 
   \left(15+z^2 \left(3 \sqrt{5} \phi_2-5 M^2\right)\right)\right]
+{\cal O}(z^6)\,,\\
\mathfrak{p}^3(z)
&=&
\frac{1}{450 z^2}\left[\frac{}{}\mathfrak{p}^3_{\,\,\,-2}  
\left( 450 +90 z^2 \left(5 M^2+4 \sqrt{5} \phi_2\right)
   +\right.\right.\\
   && \left.\left.
   -90 z^4 \log (z) \left(5 (M^2)^2+12 \sqrt{5} M^2 \phi_2
+36 \phi_2^2\right)  \nonumber
   +\right.\right.\\
   && \left.\left.\nonumber
   +z^6 \left(50 (M^2)^3 (3 \log (z)-2)+60
   \sqrt{5} (M^2)^2 \phi_2 (12 \log (z)-7)
      +\right.\right.\right.\\
   && \nonumber \left.\left.\left.
   +45 M^2 \phi_2^2
   (216 \log (z)-107)-240 \sqrt{5} M^2 \phi_4-900 \omega_6
         +\right.\right.\right.\\
   && \nonumber \left.\left.\left.
+84
   \sqrt{5} \phi_2^3 (36 \log (z)-23)-120 \phi_2 \phi_4\right)
 \right)
         +\right.\\
   && \nonumber \left.
-30
   z^4 \mathfrak{p}^3_{\,\,\,+2}  \left(-15+z^2 \left(5 M^2+12 \sqrt{5} \phi_2\right)\right)\frac{}{}\right]
+{\cal O}(z^6)\,,
\eeqs
where the free parameters are given by the coefficients $\mathfrak{p}^1_{\,\,\,0}$,
$\mathfrak{p}^1_{\,\,\,l}$, $\mathfrak{p}^2_{\,\,\,+2}$, $\mathfrak{p}^2_{\,\,\,-2}$, $\mathfrak{p}^3_{\,\,\,+2}$,
and $\mathfrak{p}^3_{\,\,\,-2}$.
In the cases of  $\mathfrak{p}^2(z)$ and $\mathfrak{p}^3(z)$, the independent parameters appear 
in the terms of  ${\cal O}(z^{-2})$ and  ${\cal O}(z^2)$,
and we truncate the series expansion at a comparatively lower order.

The  fluctuations of the six types of scalars with trivial background values obey the following UV expansions:
\beqs
\mathfrak{a}^4(z)
&=&
\mathfrak{a}^{4}_{\,\,\,2} z^2+
z^4\left[
\mathfrak{a}^{4}_{\,\,\,4}+
 \log (z)\mathfrak{a}^{4}_{\,\,\,2} \left(\frac{12 \phi_2}{\sqrt{5}}-2 M^2\right)
\right]+
\\
&&\nonumber
+\frac{z^6}{20}\left[\mathfrak{a}^{4}_{\,\,\,2} \left(5 (M^2)^2 (4 \log (z)-3)+12 \sqrt{5} M^2 \phi_2
   (3-4 \log (z))   +
      \right.\right. \\
&&\nonumber \left. \left.
+6 \phi_2^2 (7-12 \log (z))+12 \sqrt{5} \phi_4\right)+
 \right.\\
   &&\nonumber \left.
   +2 \mathfrak{a}^{4}_{\,\,\,4}
   \left(6 \sqrt{5} \phi_2-5 M^2\right)\right]
+{\cal O}(z^8)\,,\\
\mathfrak{a}^5(z)
&=&
\mathfrak{a}^{5}_{\,\,\,0}
+\frac{z^2}{2}  \mathfrak{a}^{5}_{\,\,\,0}M^2
+\frac{z^4}{4}  \mathfrak{a}^{5}_{\,\,\,0}   (M^2)^2+
 \\
&&\nonumber 
+z^6\left[
\mathfrak{a}^{5}_{\,\,\,6}
+ \log (z)\mathfrak{a}^{5}_{\,\,\,0} \frac{M^2}{30}   \left(4 \phi_2^2-5 (M^2)^2\right)
\right]
+{\cal O}(z^8)\,,\\
\mathfrak{a}^6(z)
&=& 
\frac{1}{5} \mathfrak{a}^{6}_{\,\,\,0} \left(5+5 z^2 M^2 +z^4  \log (z) \left(-5 (M^2)^2-3 \sqrt{5} M^2 \phi_2
   +25 \phi_2^2\right)\right)+\nonumber\\
   &&
   +z^4 \mathfrak{a}^{6}_{\,\,\,4}
\frac{}{}+{\cal O}(z^6)\,,\\
\mathfrak{a}^7(z)
&=& 
\frac{1}{5} \mathfrak{a}^{7}_{\,\,\,0} \left(5+5 z^2 M^2 +z^4 \log (z)
\left(-5 (M^2)^2-3 \sqrt{5} M^2 \phi_2   +25 \phi_2^2\right) \right)+\nonumber\\
   &&
+z^4 \mathfrak{a}^{7}_{\,\,\,4}
\frac{}{}+{\cal O}(z^6)\,,\\
\mathfrak{a}^8(z)
&=& 
\mathfrak{a}^{8}_{\,\,\,0} \left(1+z^2 M^2 +  z^4 \log (z)\frac{M^2}{5} \left(2 \sqrt{5} \phi_2-5
   M^2\right) \right)+z^4 \mathfrak{a}^{8}_{\,\,\,4}
\frac{}{}+{\cal O}(z^6)\,,\\
\mathfrak{a}^9(z)
&=& 
\mathfrak{a}^{9}_{\,\,\,0} \left(1+z^2 M^2 +  z^4 \log (z) \frac{M^2}{5} \left(2 \sqrt{5} \phi_2-5
   M^2\right) \right)+z^4 \mathfrak{a}^{9}_{\,\,\,4}
\frac{}{}+{\cal O}(z^6)\,,
\eeqs
where we denoted with $\{\mathfrak{a}^4,\,\cdots,\,\mathfrak{a}^9\}$ the fluctuations associated with the 
fields $\Phi^{(0)a}$, in the same basis as in Eqs.~(\ref{Eq:probesG}) and~(\ref{Eq:probesm}).
The free parameters are $\mathfrak{a}^4_{\,\,\,2}$ and $\mathfrak{a}^4_{\,\,\,4}$,
$\mathfrak{a}^5_{\,\,\,0}$ and $\mathfrak{a}^5_{\,\,\,6}$,
and the four pairs $\mathfrak{a}^i_{\,\,\,0}$ and $\mathfrak{a}^i_{\,\,\,4}$,
with $i=6,\,\cdots,\,9$.

For the linearised fluctuations of the $1$-forms $V^A_{\,\,\,\,M}$, we consider only the transverse polarisations,
and denote them as $\mathfrak{v}^A$, with $A=1,\,\cdots,\,6$, 
ordered in the same basis as in Eq.~(\ref{Eq:H(1)}). We find the following UV expansions.
\beqs
\mathfrak{v}^1(z)
&=&
\frac{1}{60}  \mathfrak{v}^{1}_{\,\,\,0} \left[\frac{}{}60
+15 z^2 M^2  \left(2+z^2 M^2   \right)   +\right.
  \\
&& \nonumber      \left.
   +2 z^6 \log (z) M^2  \left(4 \phi_2^2-5
   (M^2)^2\right) \frac{}{}\right]+z^6  \mathfrak{v}^{1}_{\,\,\,6}
   +{\cal O}(z^8)\,,\\
\mathfrak{v}^2(z)
&=& 
\frac{1}{60} \mathfrak{v}^{2}_{\,\,\,0} \left[\frac{}{}60+15  z^2 M^2\left(2+ 
   z^2 M^2\right)    +\right.
  \\
&& \nonumber      \left.
   +2 z^6 \log (z) M^2  \left(4 \phi_2^2-5
   (M^2)^2\right) \right]+z^6  \mathfrak{v}^{2}_{\,\,\,6}
    +{\cal O}(z^8)\,,\\
\mathfrak{v}^3(z)
&=& 
\mathfrak{v}^{3}_{\,\,\,0} \left(1+ z^2 M^2 + z^4  \log (z) \frac{M^2}{5} \left(2 \sqrt{5} \phi_2-5
   M^2\right)\right)+\nonumber\\
&&
+z^4 \mathfrak{v}^{3}_{\,\,\,4}
      +{\cal O}(z^{6})\,,\\
\mathfrak{v}^4(z)
\label{Eq:ax}
&=& 
\frac{1}{5} \mathfrak{v}^{4}_{\,\,\,0} \left(5+5  z^2M^2+z^4 \log (z)\left(-5 (M^2)^2-3 \sqrt{5} M^2 \phi_2
+25 \phi_2^2\right) \right)+\nonumber\\
&&
+z^4 \mathfrak{v}^{4}_{\,\,\,4}
      +{\cal O}(z^{5})\,,\\
\mathfrak{v}^5(z)
&=& 
\mathfrak{v}^{5}_{\,\,\,-2} \left(\frac{1}{z^2}-\frac{\phi_2}{\sqrt{5}}+M^2+\frac{1}{5} z^2 \log (z)
\left(-5 (M^2)^2+3 \sqrt{5} M^2 \phi_2 +9 \phi_2^2\right) \right)
+\nonumber\\
&&
+z^2 \mathfrak{v}^{5}_{\,\,\,2}
      +{\cal O}(z^{3})\,,\\
\mathfrak{v}^6(z)
&=& 
\mathfrak{v}^{6}_{\,\,\,-2} \left(\frac{1}{z^2}+\frac{4 \phi_2}{\sqrt{5}}+M^2-\frac{1}{5} z^2 \log (z)
\left(5 (M^2)^2+12 \sqrt{5} M^2 \phi_2 +36 \phi_2^2\right) \right)+\nonumber\\
&&
+z^2 \mathfrak{v}^{6}_{\,\,\,2}
      +{\cal O}(z^{3})\,.
\eeqs
In these expressions, the free parameters are the pairs $\mathfrak{v}^{1}_{\,\,\,0}$ and  $\mathfrak{v}^{1}_{\,\,\,6}$,
$\mathfrak{v}^{2}_{\,\,\,0}$,  $\mathfrak{v}^{2}_{\,\,\,6}$, 
 $\mathfrak{v}^{3}_{\,\,\,0}$ and $\mathfrak{v}^{3}_{\,\,\,4}$,
  $\mathfrak{v}^{4}_{\,\,\,0}$ and $\mathfrak{v}^{4}_{\,\,\,4}$,
   $\mathfrak{v}^{5}_{\,\,\,-2}$ and $\mathfrak{v}^{5}_{\,\,\,2}$,
as well as    $\mathfrak{v}^{6}_{\,\,\,-2}$ and $\mathfrak{v}^{6}_{\,\,\,2}$.

 Finally, we list also the UV expansion of the four independent components of the $2$-forms, that are proportional 
 $\mathfrak{t}^i$, with $i=1,\,\cdots, \,4$, and for which we choose the same basis adopted in Eqs.~(\ref{Eq:H(2)}) 
 and~(\ref{Eq:K(2)}). We find the following results:
 \beqs
 \mathfrak{t}^1(z)
&=&
\frac{1}{450 z^2}\left\{\frac{}{}\mathfrak{t}^{1}_{\,\,\,-2} \left[
450+90
   z^2 \left(5 M^2-\sqrt{5} \phi_2\right)     +\right.\right.\\
   &&\nonumber \left.\left.
-90 z^4 \log (z) \left(5 (M^2)^2-3 \sqrt{5} M^2 \phi_2-9
   \phi_2^2\right) 
   +\right.\right.\\
   &&\nonumber \left.\left.
   +z^6 \left(50 (M^2)^3 (3 \log (z)-2)+15 \sqrt{5}
   (M^2)^2 \phi_2 (7-12 \log (z))
     +\right.\right.\right.\\
   &&\nonumber \left.\left.\left.
   +30 M^2 \left(9 \phi_2^2
   (3-4 \log (z))+2 \sqrt{5} \phi_4\right)+900 \omega_6
   +\right.\right.\right.
  \\
&& \nonumber      \left.\left.\left.
   +6 \sqrt{5}
   \phi_2^3 (143-171 \log (z))+330 \phi_2 \phi_4\right)
   \right]     +\right.\\
   &&\nonumber \left.
   +30 z^4 \mathfrak{t}^{1}_{\,\,\,2}
   \left[15+z^2 \left(3 \sqrt{5} \phi_2-5 M^2\right)\right]\right\}
       +{\cal O}(z^{6})\,,\\
        \mathfrak{t}^2(z)
&=&
\frac{1}{450 z^2}\left\{\frac{}{}\mathfrak{t}^{2}_{\,\,\,-2}
 \left[
 450   +90
   z^2 \left(5 M^2-\sqrt{5} \phi_2\right)   +\right.\right.\\
   &&\nonumber \left.\left.
 -90 z^4 \log (z)\left(5 (M^2)^2-3 \sqrt{5} M^2 \phi_2-9
   \phi_2^2\right) 
     +\right.\right.\\
   &&\nonumber \left.\left.
   +z^6 \left(50 (M^2)^3 (3 \log (z)-2)+15 \sqrt{5}
   (M^2)^2 \phi_2 (7-12 \log (z))
     +\right.\right.\right.\\
   &&\nonumber \left.\left.\left.
   +30 M^2 \left(9 \phi_2^2
   (3-4 \log (z))+2 \sqrt{5} \phi_4\right)-900 \omega_6
      +\right.\right.\right.
  \\
&& \nonumber      \left.\left.\left.
+6 \sqrt{5}
   \phi_2^3 (143-171 \log (z))+330 \phi_2 \phi_4\right)\right]
     +\right.\\
   &&\nonumber \left.
   +30 z^4 \mathfrak{t}^{2}_{\,\,\,2}
   \left[15+z^2 \left(3 \sqrt{5} \phi_2-5 M^2\right)\right]\right\}
   +{\cal O}(z^{6})\,,\\
        \mathfrak{t}^3(z)
&=&
\frac{1}{450 z^2}\left\{\frac{}{}\mathfrak{t}^{3}_{\,\,\,-2}
\left[\frac{}{}
450   +90 z^2 \left(5 M^2+4 \sqrt{5} \phi_2\right)
     +\right.\right.\\
   &&\nonumber \left.\left.
   -90 z^4 \log (z)\left(5 (M^2)^2+12 \sqrt{5} M^2 \phi_2+36 \phi_2^2\right) 
  +\right.\right.\\
   &&\nonumber \left.\left.
   +z^6 \left(50 (M^2)^3 (3 \log (z)-2)+60
   \sqrt{5} (M^2)^2 \phi_2 (12 \log (z)-7)
     +\right.\right.\right.\\
   &&\nonumber \left.\left.\left.
   +45 M^2 \phi_2^2
   (216 \log (z)-107)-240 \sqrt{5} M^2 \phi_4
      +\right.\right.\right.
  \\
&& \nonumber      \left.\left.\left.
+900 \omega_6+84
   \sqrt{5} \phi_2^3 (36 \log (z)-23)-120 \phi_2 \phi_4\right)
\right]     +\right.\\
   &&\nonumber \left.
   -30 z^4 \mathfrak{t}^{3}_{\,\,\,2}
   \left[-15+z^2 \left(5 M^2+12 \sqrt{5} \phi_2\right)\right]\right\}
     +{\cal O}(z^{6})\,,\\
        \mathfrak{t}^4(z)
&=&
\frac{1}{450 z^2}\left\{\frac{}{}\mathfrak{t}^{4}_{\,\,\,-2}
\left[\frac{}{}
 450 +90 z^2 \left(5 M^2+4 \sqrt{5} \phi_2\right)
   +\right.\right.\\
   &&\nonumber \left.\left.
-90 z^4  \log (z) \left(5 (M^2)^2+12 \sqrt{5} M^2 \phi_2
+36 \phi_2^2\right)
  +\right.\right.\\
   &&\nonumber \left.\left.
   +z^6 \left(50 (M^2)^3 (3 \log (z)-2)+60
   \sqrt{5} (M^2)^2 \phi_2 (12 \log (z)-7)
     +\right.\right.\right.\\
   &&\nonumber \left.\left.\left.
   +45 M^2 \phi_2^2
   (216 \log (z)-107)-240 \sqrt{5} M^2 \phi_4
      +\right.\right.\right.
  \\
&& \nonumber      \left.\left.\left.
-900 \omega_6+84
   \sqrt{5} \phi_2^3 (36 \log (z)-23)-120 \phi_2 \phi_4\right)
 \right]
     +\right.\\
   &&\nonumber \left.
   -30 z^4 \mathfrak{t}^{4}_{\,\,\,2}
   \left[-15+z^2 \left(5 M^2+12 \sqrt{5} \phi_2\right)\right]\right\}
     +{\cal O}(z^{6})\,,
 \eeqs
where we denote by $\mathfrak{t}^i_{\,\,\,\pm 2}$ the eight free parameters, with $i=1,\,\cdots,\,4$.
Notice that the difference in the sign of the terms depending on $\omega_6$ appears only 
in high-order terms.

\subsubsection{IR expansions}
\label{Sec:ExpansionsIR}

We now consider the background solutions as defined by the small-$\rho$ expansions
displayed in Sec.~\ref{Sec:solutions}. For convenience, in this subsection we set $\rho_o=0$, $\omega_I=0$, and 
$\chi_I=0$, so that the $1$-parameter family of solutions is labelled unambiguously by the choice of 
$\phi_I$, the only remaining integration constant.

We start from the fluctuations of the metric, the traceless and transverse component of which 
is given by
\beqs
(\mathfrak{e}_I)^{\mu}_{\,\,\,\nu}(\rho)&=&
\frac{1}{20} \left[\frac{}{}(\mathfrak{e}_{I,l})^{\mu}_{\,\,\,\nu}\left(\frac{}{}
-5 \log (\rho ) \left(-4+\rho ^2 M^2\right)
+\right.\right.\\
&&\nonumber\left.\left.
+\rho ^2 \left(5 M^2
+e^{-\frac{8 \phi_I}{\sqrt{5}}} \left(1-8 e^{\sqrt{5} \phi_I} 
\left(e^{\sqrt{5} \phi_I}+1\right)\right)\right)\right)
+\right.\\
&&\nonumber\left.
-5
  (\mathfrak{e}_{I,0})^{\mu}_{\,\,\,\nu} \left(-4+\rho ^2 M^2\right)\frac{}{}\right]
+{\cal O}(\rho^4)\,,
\eeqs
where the free parameters are $(\mathfrak{e}_{I,l})^{\mu}_{\,\,\,\nu}$ and $(\mathfrak{e}_{I,0})^{\mu}_{\,\,\,\nu}$.
Here and in the following, we add the subscript $_I$ to avoid potential ambiguities with the 
coefficients of the UV expansions.

For the  three active scalars, the expansion of the relevant gauge-invariant  fluctuations 
is complicated by the mixing terms, and 
reads as follows:
\beqs
\mathfrak{a}^1_{\,\,\,I}(\r)
&=&
\frac{1}{20} e^{-\frac{8 \phi_I}{\sqrt{5}}} \left[\frac{}{}
\mathfrak{a}^{1}_{\,\,\,I,l} \left(\frac{}{}
  -5 e^{\frac{8 \phi_I}{\sqrt{5}}} \left(\frac{}{}\log (\rho ) \left(-4+\rho ^2
   M^2\right)-\rho ^2 M^2\right)\right.
         +\right.\\
   &&\nonumber\left.\left.
   -8 \rho ^2 e^{2 \sqrt{5} \phi_I} \log (\rho )-2
   \rho ^2 e^{\sqrt{5} \phi_I} (9 \log (\rho )-5)+\rho ^2 (16 \log (\rho
   )-15)\frac{}{}\right)
        +\right.\\
   &&\nonumber\left.
 \mathfrak{a}^{1}_{\,\,\,I,0} \left(-18 \rho ^2
   e^{\sqrt{5} \phi_I}-8 \rho ^2 e^{2 \sqrt{5} \phi_I}+16 \rho ^2-5
   e^{\frac{8 \phi_I}{\sqrt{5}}} \left(-4+\rho ^2 M^2\right)\right)
         +\right.\\
   &&\nonumber\left.
+2 \sqrt{5} \rho ^2 \left(-3
   e^{\sqrt{5} \phi_I}+2 e^{2 \sqrt{5} \phi_I}+1\right) \left(\frac{}{}2 (\log
   (\rho )-1) \mathfrak{a}^2_{\,\,\,I,l}
   +\right.\right.\\
   &&\nonumber\left.\left.
   -3 \left((\log (\rho )-1) \mathfrak{a}^3_{\,\,\,I,l}+\mathfrak{a}^{3}_{\,\,\,I,0}\right)+2
  \mathfrak{a}^2_{\,\,\,I,0}\frac{}{}\right)
  \frac{}{}\right]
   +{\cal O}(\rho^4)\,,\\
   \mathfrak{a}^2_{\,\,\,I}(\r)
&=&
\frac{1}{80} e^{-\frac{8 \phi_I}{\sqrt{5}}}  \left[\frac{}{}
 -20 e^{\frac{8 \phi_I}{\sqrt{5}}} \left(\frac{}{}\mathfrak{a}^2_{\,\,\,I,l} \left(\log (\rho ) \left(-4+\rho ^2
   M^2\right)-\rho ^2 M^2\right)
       +\right.\right.\\
   &&\nonumber\left.\left.
   +\mathfrak{a}^2_{\,\,\,I,0} \left(-4+\rho ^2
   M^2\right)\right)
       +\right.\\
   &&\nonumber\left.
   -8 \rho ^2 e^{\sqrt{5}
   \phi_I} \left(3 \sqrt{5} (\log (\rho )-1)\mathfrak{a}^1_{\,\,\,I,l}+2 (5 \log (\rho )-3)
  \mathfrak{a}^2_{\,\,\,I,l}
     +\right.\right.\\
   &&\nonumber\left.\left.
  -15 (\log (\rho )-1)\mathfrak{a}^3_{\,\,\,I,l}+3 \sqrt{5}\mathfrak{a}^1_{\,\,\,I,0}+10\mathfrak{a}^2_{\,\,\,I,0}
  -15\mathfrak{a}^3_{\,\,\,I,0}\right)
     +\right.\\
   &&\nonumber\left.
   +8
   \rho ^2 e^{2 \sqrt{5} \phi_I} \left(2 \sqrt{5} (\log (\rho )-1)
  \mathfrak{a}^1_{\,\,\,I,l}+(6-10 \log (\rho ))\mathfrak{a}^2_{\,\,\,I,l}
     +\right.\right.\\
   &&\nonumber\left.\left.
   +15 \left((\log (\rho )-1)
  \mathfrak{a}^3_{\,\,\,I,l}+a_{3,0}\right)+2 \sqrt{5}\mathfrak{a}^1_{\,\,\,I,0}-10\mathfrak{a}^2_{\,\,\,I,0}\right)
     +\right.\\
   &&\nonumber\left.
   +\rho ^2 \left(8 \sqrt{5}
   (\log (\rho )-1)\mathfrak{a}^1_{\,\,\,I,l}+2 (5 \log (\rho )-3)\mathfrak{a}^2_{\,\,\,I,l}-15 (\log (\rho )-1)\mathfrak{a}^3_{\,\,\,I,l}
      +\right.\right.\\
   &&\nonumber\left.\left.
   +8
   \sqrt{5}\mathfrak{a}^1_{\,\,\,I,0}+10\mathfrak{a}^2_{\,\,\,I,0}-15\mathfrak{a}^3_{\,\,\,I,0}\right)
  \frac{}{}\right]
   +{\cal O}(\rho^4)\,,\\
\mathfrak{a}^3_{\,\,\,I}(\r)
&=&
\frac{1}{200} e^{-\frac{8 \phi_I}{\sqrt{5}}}  \left[\frac{}{}
 -50 e^{\frac{8 \phi_I}{\sqrt{5}}} \left(\frac{}{}\mathfrak{a}^3_{\,\,\,I,l} \left(\log (\rho ) \left(-4+\rho ^2
   M^2\right)-\rho ^2 M^2\right)
      +\right.\right.\\
   &&\nonumber\left.\left.
   +\mathfrak{a}^3_{\,\,\,I,0} \left(-4+\rho ^2
   M^2\right)\frac{}{}\right)
   +\right.\\
   &&\nonumber\left.
   -8 \sqrt{5} \rho ^2 \left(-3
   e^{\sqrt{5} \phi_I}+2 e^{2 \sqrt{5} \phi_I}+1\right)
   \left(\frac{}{} (\log (\rho)-1)\mathfrak{a}^1_{\,\,\,I,l}+\mathfrak{a}^1_{\,\,\,I,0}\right)
      +\right.\\
   &&\nonumber\left.
   +40 \rho ^2 \left(e^{\sqrt{5} \phi_I}+e^{2\sqrt{5} \phi_I}-\frac{1}{8}\right)\times\nonumber\right.\\
   &&\left.\times \nonumber
   \left(\frac{}{}2 (\log (\rho )-1)  \mathfrak{a}^2_{\,\,\,I,l}+(1-3 \log (\rho ))
   \mathfrak{a}^3_{\,\,\,I,l}+2\mathfrak{a}^2_{\,\,\,I,0}-3\mathfrak{a}^3_{\,\,\,I,0}\right)
  \right]
   +{\cal O}(\rho^4)\,,
\eeqs
where the free parameters are $\mathfrak{a}^1_{\,\,\,I,0}$ and $\mathfrak{a}^1_{\,\,\,I,l}$, 
$\mathfrak{a}^2_{\,\,\,I,0}$ and $\mathfrak{a}^2_{\,\,\,I,l}$, and finally
$\mathfrak{a}^3_{\,\,\,I,0}$ and $\mathfrak{a}^3_{\,\,\,I,l}$.

For  the three pseudoscalars, the orthogonal combinations to the 
would-be Goldstone bosons higgsed in the axial-vector fields, we find
the following IR expansions:
\beqs
\mathfrak{p}^1_{\,\,\,I}(\r)
&=&
\frac{1}{2}  \mathfrak{p}^1_{\,\,\,I,0} \left[2-\rho ^2 M^2 \log (\rho )+2 \rho ^2 e^{-\frac{3 \phi_I}{\sqrt{5}}} \log (\rho )
 \left(\cosh \left(\sqrt{5} \phi_I\right)-1\right)\right]+
 \\
 &&\nonumber
 +\rho ^2  \mathfrak{p}^1_{\,\,\,I,2}+{\cal O}(\rho^4)\,,\\
 \mathfrak{p}^2_{\,\,\,I}(\r)
&=&
\frac{1}{40}  \mathfrak{p}^2_{\,\,\,I,l} \left[10 \log (\rho ) \left(\rho ^2 \left(e^{\frac{2 \phi_I}{\sqrt{5}}}
-M^2\right)+4\right)
+\right.\\
&&\left.\nonumber
+\rho ^2 \left(10 M^2+e^{-\frac{8 \phi_I}{\sqrt{5}}} 
\left(14 e^{\sqrt{5} \phi_I}-6 e^{2 \sqrt{5} \phi_I}
-3\right)\right)
\right]
+\\
&&\nonumber
+\frac{1}{4}  \mathfrak{p}^2_{\,\,\,I,0} \left(\rho ^2
   \left(e^{\frac{2 \phi_I}{\sqrt{5}}}-M^2\right)+4\right)
 +{\cal O}(\rho^4)\,,\\
 \mathfrak{p}^3_{\,\,\,I}(\r)
&=&
\frac{1}{40} e^{-\frac{8 \phi_I}{\sqrt{5}}} \left\{\frac{}{} \mathfrak{p}^3_{\,\,\,I,l} \left[\frac{}{}
10 \log (\rho )
   \left(e^{\frac{8 \phi_I}{\sqrt{5}}} \left(4-\rho ^2
   M^2\right)+\rho ^2\right)
   +\right.\right.\\
   &&\nonumber\left.\left.
   +\rho ^2 \left(2
   e^{\sqrt{5} \phi_I} \left(5 M^2 e^{\frac{3 \phi_I}{\sqrt{5}}}
   +12 e^{\sqrt{5} \phi_I}-8\right)-3\right)
   \right]
      +\right.\\
   &&\nonumber\left.
   +10 \ \mathfrak{p}^3_{\,\,\,I,0} \left(e^{\frac{8 \phi_I}{\sqrt{5}}} 
   \left(4-\rho ^2 M^2\right)+\rho ^2\right)\right\}
 +{\cal O}(\rho^4)\,,
\eeqs
where the free parameters are $\mathfrak{p}^1_{\,\,\,I,0} $ and $\mathfrak{p}^1_{\,\,\,I,2}$,
$\mathfrak{p}^2_{\,\,\,I,0} $ and $\mathfrak{p}^2_{\,\,\,I,l}$,
and finally
$\mathfrak{p}^3_{\,\,\,I,0} $ and $\mathfrak{p}^3_{\,\,\,I,l}$.

For the additional scalars, which vanish on the background, the fluctuations obey the following IR expansions:
\beqs
\mathfrak{a}^4_{\,\,\,I}(\r)
&=&
\frac{1}{20} e^{-\frac{8 \phi_I}{\sqrt{5}}} \left[\frac{}{} {-5} e^{\frac{8 \phi_I}
   {\sqrt{5}}} \mathfrak{a}^4_{\,\,\,I,0} \left(\rho ^2
   M^2-4\right)
      +\right.\\
   &&\nonumber \left.
   -5 e^{\frac{8 \phi_I}
   {\sqrt{5}}} \left(\frac{}{}
 \mathfrak{a}^4_{\,\,\,I,l} \left(\log (\rho ) \left(\rho ^2
   M^2-4\right)-\rho ^2 M^2\right)\right)
   +\right.\\
   &&\nonumber \left.
   +\rho ^2 \mathfrak{a}^4_{\,\,\,I,l}
-8 \rho ^2 e^{2 \sqrt{5}
   \phi_I} \mathfrak{a}^4_{\,\,\,I,l}
   -2 \rho ^2 e^{\sqrt{5} \phi_I} \left((5 \log (\rho)-1) \mathfrak{a}^4_{\,\,\,I,l}
   +5 \mathfrak{a}^4_{\,\,\,I,0}\frac{}{}\right)
   \right]
 +{\cal O}(\rho^4)\,,\\
 \mathfrak{a}^5_{\,\,\,I}(\r)
&=&
 \mathfrak{a}^5_{\,\,\,I,0} 
 \left(1-\frac{1}{2} \rho ^2 M^2 \log (\rho )\right)
 + \rho ^2 \mathfrak{a}^5_{\,\,\,I,2}
 +{\cal O}(\rho^4)\,,\\
 \mathfrak{a}^6_{\,\,\,I}(\r)
&=&
\frac{1}{2} 
\mathfrak{a}^6_{\,\,\,I,0} 
\left(\rho ^2 e^{-\frac{8 \phi_I}
{\sqrt{5}}} \log (\rho ) \left(-M^2 e^{\frac{8 \phi_I}
   {\sqrt{5}}}-2 e^{\sqrt{5} \phi_I}+e^{2 \sqrt{5} \phi_I}+1\right)+2\right)
      +\\
   &&\nonumber 
   \rho ^2 \mathfrak{a}^6_{\,\,\,I,2}
 +{\cal O}(\rho^4)\,,\\
 \mathfrak{a}^7_{\,\,\,I}(\r)
&=&
\frac{1}{40} e^{-\frac{8 \phi_I}{\sqrt{5}}} \left[
 -10 e^{\frac{8 \phi_I}{\sqrt{5}}} 
   \left(\frac{}{}\mathfrak{a}^7_{\,\,\,I,0} \left(\rho ^2
   M^2-4\right)
         +\right.\right.\\
   &&\nonumber \left.\left.
   +\mathfrak{a}^7_{\,\,\,I,l}
    \left(\log (\rho ) \left(\rho ^2
   M^2-4\right)-\rho ^2 M^2\right)\frac{}{}\right)
   +
2 \rho ^2 e^{2 \sqrt{5}
   \phi_I} \left((5 \log (\rho )-3) 
   \mathfrak{a}^7_{\,\,\,I,l}
   +5 \mathfrak{a}^7_{\,\,\,I,0}\right)
   +\right.\\
   &&\nonumber \left.
   -2 \rho ^2
   e^{\sqrt{5} \phi_I} \left((10 \log (\rho )+3) \mathfrak{a}^7_{\,\,\,I,l}+10 \mathfrak{a}^7_{\,\,\,I,0}\right)
   +\right.\\
   &&\nonumber \left.
   +\rho^2 \left((10 \log (\rho )-3) \mathfrak{a}^7_{\,\,\,I,l}+10 \mathfrak{a}^7_{\,\,\,I,0}\frac{}{}\right)\frac{}{}
\right]
 +{\cal O}(\rho^4)\,,\\
 \mathfrak{a}^8_{\,\,\,I}(\r)
&=&
\mathfrak{a}^8_{\,\,\,I,0} 
\left(1-\frac{1}{2} \rho ^2 M^2 \log (\rho )\right)
+\rho ^2 \mathfrak{a}^8_{\,\,\,I,2}
 +{\cal O}(\rho^4)\,,\\
 \mathfrak{a}^9_{\,\,\,I}(\r)
&=&
  +
 \mathfrak{a}^9_{\,\,\,I,0}
   \left(1-\frac{\rho ^2 M^2}{4}\right)+
\frac{1}{40} \mathfrak{a}^9_{\,\,\,I,l}
 \left[\frac{}{}-10 \log (\rho ) \left(\rho ^2 M^2-4\right)
    +\right.\\
   &&\nonumber \left.
 \rho ^2 \left(10 M^2+e^{-\frac{8 \phi_I}
{\sqrt{5}}} \left(4 e^{\sqrt{5} \phi_I}-16 e^{2 \sqrt{5} \phi_I}
   -3\right)\right)\right]
 +{\cal O}(\rho^4)\,,
\eeqs
where the free parameters are $\mathfrak{a}^4_{\,\,\,I,0}$ and $\mathfrak{a}^4_{\,\,\,I,l}$,
$\mathfrak{a}^5_{\,\,\,I,0}$ and $\mathfrak{a}^5_{\,\,\,I,2}$,
$\mathfrak{a}^6_{\,\,\,I,0}$ and $\mathfrak{a}^6_{\,\,\,I,2}$,
$\mathfrak{a}^7_{\,\,\,I,0}$ and $\mathfrak{a}^7_{\,\,\,I,l}$,
$\mathfrak{a}^8_{\,\,\,I,0}$ and $\mathfrak{a}^8_{\,\,\,I,2}$, and finally
$\mathfrak{a}^9_{\,\,\,I,0}$ and $\mathfrak{a}^9_{\,\,\,I,l}$.

For the fluctuations of the $1$-form fields, we find the following IR expansions.
\beqs
\mathfrak{v}^1_{\,\,\,I}(\r)
&=&
  -\frac{1}{8} 
   \mathfrak{v}^1_{\,\,\,I,0}
    \left(\rho^2 M^2-8\right)
    + \\
    && \nonumber
    +
\frac{\mathfrak{v}^1_{\,\,\,I,-2} }{8000}
\left[\frac{8000}{\rho ^2}+500 M^2 \log
   (\rho ) \left(\rho ^2 M^2-8\right)
       \right.+ \\
    && \left. \nonumber
    +\rho ^2 \left(-375 (M^2)^2-50 M^2
   e^{-\frac{8 \phi_I}{\sqrt{5}}}+400 M^2 e^{-\frac{3 \phi_I}
   {\sqrt{5}}}
    +400 M^2 e^{\frac{2 \phi_I}{\sqrt{5}}}
       \right.\right.+ \\
    && \left.\left. \nonumber
    -84 e^{-\frac{16
   \phi_I}{\sqrt{5}}}+384 e^{-\frac{11 \phi_I}{\sqrt{5}}}-672
   e^{-\frac{6 \phi_I}{\sqrt{5}}}+2688 e^{-\frac{\phi_I}
   {\sqrt{5}}}+384 e^{\frac{4 \phi_I}{\sqrt{5}}}\right)\right]
 +{\cal O}(\rho^4)\,,\\
 \mathfrak{v}^2_{\,\,\,I}(\r)
&=&
\frac{1}{20} \left[\frac{}{}
-5
   \mathfrak{v}^2_{\,\,\,I,0} 
   \left(\rho ^2 M^2-4\right)+\mathfrak{v}^2_{\,\,\,I,l}
 \left(\frac{}{}-5 \log (\rho ) \left(\rho ^2 M^2-4\right)
        \right.  \right.+ \\
    && \left. \left. \nonumber
+\rho ^2 \left(5 M^2+e^{-\frac{8
 \phi_I}{\sqrt{5}}} \left(1-8 e^{\sqrt{5} \phi_I} \left(e^{\sqrt{5} 
   \phi_I}+1\right)\right)\right)\right)\right]
 +{\cal O}(\rho^4)\,,\\
 \mathfrak{v}^3_{\,\,\,I}(\r)
&=&
   \mathfrak{v}^3_{\,\,\,I,0}
   \left(1-\frac{\rho ^2 M^2}{4}\right)
   +
\frac{1}{40} \mathfrak{v}^3_{\,\,\,I,l}
 \left(\frac{}{}-10 \log (\rho ) \left(\rho ^2 M^2-4\right)
        \right.+ \\
    && \left. \nonumber
 +\rho ^2 \left(10 M^2+e^{-\frac{8 
\phi_I}{\sqrt{5}}} \left(4 e^{\sqrt{5} \phi_I}-16 e^{2 \sqrt{5} 
   \phi_I}-3\right)\right)\right)
 +{\cal O}(\rho^4)\,,\\
 \mathfrak{v}^4_{\,\,\,I}(\r)
&=&
\frac{1}{40} e^{-\frac{8 \phi_I}{\sqrt{5}}} \left[\frac{}{}
-10 e^{\frac{8 
\phi_I}{\sqrt{5}}} \left(\frac{}{}\mathfrak{v}^4_{\,\,\,I,l} 
\left(\log (\rho ) \left(\rho ^2
   M^2-4\right)-\rho ^2 M^2\right)
         \right.  \right.+ \\
    && \left. \left.\nonumber
    +\mathfrak{v}^4_{\,\,\,I,0} \left(\rho ^2
   M^2-4\right)\frac{}{}\right)+2 \rho ^2 e^{2 \sqrt{5} \phi_I}
   \left(\frac{}{}(5 \log
   (\rho )-3) \mathfrak{v}^4_{\,\,\,I,l}+5  \mathfrak{v}^4_{\,\,\,I,0}\right)
              \right.+ \\
    && \left. \nonumber
    -2 \rho ^2 e^{\sqrt{5} \phi_I} 
   \left(\frac{}{}(10
   \log (\rho )+3)  \mathfrak{v}^4_{\,\,\,I,l}+10  \mathfrak{v}^4_{\,\,\,I,0}\right)
                 \right.+ \\
    && \left. \nonumber
    +\rho ^2 \left(\frac{}{}(10 \log (\rho )-3)
 \mathfrak{v}^4_{\,\,\,I,l}+10 
   \mathfrak{v}^4_{\,\,\,I,0}\right)\right)
 +{\cal O}(\rho^4)\,,\\
 \mathfrak{v}^5_{\,\,\,I}(\r)
&=&
\frac{1}{2} \mathfrak{v}^5_{\,\,\,I,0} 
\left(\rho ^2 \log (\rho ) \left(e^{\frac{2 
\phi_I}{\sqrt{5}}}-M^2\right)+2\right)+\rho ^2 
\mathfrak{v}^5_{\,\,\,I,2}
 +{\cal O}(\rho^4)\,,\\
 \mathfrak{v}^6_{\,\,\,I}(\r)
&=&
\frac{1}{2} \mathfrak{v}^6_{\,\,\,I,0} 
\left(\rho ^2 \log (\rho ) \left(e^{-\frac{8 \phi_I}{\sqrt{5}}}
-M^2\right)+2\right)+\rho ^2 
\mathfrak{v}^6_{\,\,\,I,2}
 +{\cal O}(\rho^4)\,,
\eeqs
where the free parameters are
$\mathfrak{v}^1_{\,\,\,I,0}$ and $\mathfrak{v}^1_{\,\,\,I,-2}$,
$\mathfrak{v}^2_{\,\,\,I,0}$ and $\mathfrak{v}^2_{\,\,\,I,l}$,
$\mathfrak{v}^3_{\,\,\,I,0}$ and $\mathfrak{v}^3_{\,\,\,I,l}$,
$\mathfrak{v}^4_{\,\,\,I,0}$ and $\mathfrak{v}^4_{\,\,\,I,l}$,
$\mathfrak{v}^5_{\,\,\,I,0}$ and $\mathfrak{v}^5_{\,\,\,I,2}$, and finally
$\mathfrak{v}^6_{\,\,\,I,0}$ and $\mathfrak{v}^6_{\,\,\,I,2}$.

The physical fluctuations of the  $2$-forms are the following:
\beqs
\mathfrak{t}^1_{\,\,\,I}(\r)
&=&
\frac{1}{2}  \mathfrak{t}^1_{\,\,\,I,0}
 \left(\rho ^2 \log (\rho ) \left(e^{\frac{2 \phi_I}{\sqrt{5}}}-M^2\right)+2\right)+\rho ^2 
 \mathfrak{t}^1_{\,\,\,I,2}
 +{\cal O}(\rho^4)\,,\\
 \mathfrak{t}^2_{\,\,\,I}(\r)
&=&\frac{1}{4} 
   \mathfrak{t}^2_{\,\,\,I,0} 
   \left(\rho ^2
   \left(e^{\frac{2 \phi_I}{\sqrt{5}}}-M^2\right)+4\right)
              + \\
    &&  \nonumber
\frac{1}{40}  \mathfrak{t}^2_{\,\,\,I,l}
 \left[\rho ^2 \left(10 M^2+e^{-\frac{8 \phi_I}{\sqrt{5}}}
 \left(14 e^{\sqrt{5} \phi_I}-6 e^{2 \sqrt{5} 
   \phi_I}-3\right)\right)       
             \right.+ \\
    && \left. \nonumber
    +10 \log (\rho ) \left(\rho ^2 
   \left(e^{\frac{2 \phi_I}{\sqrt{5}}}-M^2\right)+4\right)\right]+{\cal O}(\rho^4)\,,\\
 \mathfrak{t}^3_{\,\,\,I}(\r)
&=&
\frac{1}{2}  \mathfrak{t}^3_{\,\,\,I,0} 
\left(\rho ^2 \log (\rho ) \left(e^{-\frac{8 \phi_I}{\sqrt{5}}}-M^2\right)+2\right)+\rho ^2 
 \mathfrak{t}^3_{\,\,\,I,2}
 +{\cal O}(\rho^4)\,,\\
 \mathfrak{t}^4_{\,\,\,I}(\r)
&=&
\frac{1}{40} e^{-\frac{8 \phi_I}{\sqrt{5}}} \left[10 
   \mathfrak{t}^4_{\,\,\,I,0} 
   \left(\rho ^2+e^{\frac{8 \phi_I}{\sqrt{5}}} \left(4-\rho ^2 M^2\right)\right)
                \right.+ \\
    && \left. \nonumber   
   + \mathfrak{t}^4_{\,\,\,I,l} 
\left(
  10 \log (\rho )
   \left(\rho ^2+e^{\frac{8 \phi_I}{\sqrt{5}}} \left(4-\rho ^2
   M^2\right)\right)    
      \right. \right.   + \\
    && \left.  \left.\nonumber
   +\rho ^2 \left(2
   e^{\sqrt{5} \phi_I} \left(5 M^2 e^{\frac{3 \phi_I}{\sqrt{5}}}+12 e^{\sqrt{5} \phi_I}-8\right)-3\right)\right)  \right]
 +{\cal O}(\rho^4)\,,
\eeqs
with  free parameters 
$\mathfrak{t}^1_{\,\,\,I,0}$ and $\mathfrak{t}^1_{\,\,\,I,2}$,
$\mathfrak{t}^2_{\,\,\,I,0}$ and $\mathfrak{t}^2_{\,\,\,I,l}$,
$\mathfrak{t}^3_{\,\,\,I,0}$ and $\mathfrak{t}^3_{\,\,\,I,2}$, and finally 
$\mathfrak{t}^4_{\,\,\,I,0}$ and $\mathfrak{t}^4_{\,\,\,I,l}$.


\section{Lift to ten and eleven dimensions}
\label{Sec:Lift}

In this Appendix, we report some information about the lift of the background metric and fields
to type-IIA supergravity in ten dimensions, and to maximal gauged supergravity in eleven dimensions.
We also report the result of the holographic calculation of the string tension.
All the material is adapted from the literature, and is of marginal relevance to the main body of the paper,
yet we find it useful to reproduce  it here, not just in order for the paper to be
self contained, but most importantly  to clarify a couple of potential sources of ambiguities in the notation.
While we mostly follow the conventions  in Ref.~\cite{Elander:2013jqa}, there are exceptions, for
 example in the way the constants  $g$ and $m$ appear, which we make  transparent in the following.
 
The metric on the four-dimensional compact internal manifold
is the following:
\beqs
\di \tilde{\Omega}_4^2&\equiv& X^3 \di \xi^2
+\frac{1}{4}X^{-1}\Delta^{-1}\cos^2\xi \left[\frac{}{}\di \theta^2+\sin^2\theta \, \dd \varphi^2+(\di\psi+\cos\theta\di \varphi)^2\right]\,,
\eeqs
where the four angles take values in the intervals
\beqs
0\leq \theta \leq \pi\,,~~~~
0\leq \varphi < 2\pi\,,~~~~
0\leq \psi < 4\pi\,,~~~~
-\frac{\pi}{2}\leq \xi \leq \frac{\pi}{2}\,,
\eeqs
and where the non-trivial functions $X=X(\r,\xi)$ and $\Delta=\Delta(\r,\xi)$ are given in terms of the 
background values of the sigma-model 
scalar field $\phi=\phi(\r)$, via the following expressions:
\beqs
X&=&e^{\phi/\sqrt{5}}\,,\\
\Delta&=&X^4\sin^2\xi+X \cos^2\xi\,.
\eeqs
The expression for $\di \tilde{\Omega}_4^2$
 reduces to the metric on the four-sphere $S^4$ for $\phi=0$ and $X=1=\Delta$, otherwise
the manifold has
 the reduced $SO(4)$ symmetry of $S^3$.

The lift of the metric to eleven dimensions is written as\footnote{The radius of 
$S^4$ is denoted as $g$, rather than $m$, in Refs.~\cite{Cvetic:1999xp,Lu:1999bc,Cvetic:2000ah},
which is not to be confused with $g=2m$. In this paper we adopt the 
conventions of the earlier publication by Pernici et al.~\cite{Pernici:1984xx}.}
\beqs
\di s_{11}^2&=&\Delta^{1/3}\left(\di s_7^2+\frac{1}{m^2}\di\tilde{\Omega}_4^2\right)\,,
\eeqs
where the constant in front of the metric on $S^4$ is the same mass parameter $m=1$ 
that appears in the action of the  gauged supergravity
in seven dimensions.

The alternative, but equivalent, lift to ten dimensions (in type-IIA supergravity)
can be derived from the one in eleven dimensions
 by singling out the circle parametrised by the coordinate $\z$, according to the following definition:
\beqs
\di s^2_{11}&\equiv&e^{-\frac{2}{3}\Phi}\di s_{10,s}^2+e^{\frac{4}{3}\Phi} \di \z^2\,,
\eeqs
where  $\di s_{10,s}^2$ is the metric in string frame, while $\Phi$ is the dilaton
of the ten dimensional theory.
By comparing these two ways of writing  the  metric in eleven dimensions, 
we see that the dilaton $\Phi$ can be identified in the following equality:
\beqs
e^{\frac{4}{3}\Phi}&=&\Delta^{1/3}e^{3\chi+2\omega}\,, 
\eeqs 
where $\chi=\chi(\r)$ and $\omega=\omega(\rho)$ are the background values of the two active scalar
 fields that we introduced when dimensionally reducing the theory from seven to five dimensions.
The dilaton $\Phi$ hence depends on both $\xi$ and $\rho$.

For completeness, we remind the Reader of some general conventions~\cite{Aharony:1999ti}, 
starting from the fact that
 the
metric in string frame is 
\beqs
\di s_{10,s}^2&=&\frac{e^{\Phi/2}}{\sqrt{g_s}}\di s_{10}^2\,.
\eeqs
with  $g_s=e^{\Phi_{\infty}}$  the string tension.
The (Einstein frame) action of type-IIA supergravity  is
\beqs
{\cal S}_{10} &=&\frac{1}{2k^2}\int \di^{10} x \sqrt{-g_{10}}R_{10} +\cdots\,,
\eeqs
where the identities $2k^2 \equiv 2k_0^2g_s^2 \equiv (2\pi)^7\alpha^{\prime\,4}g_s^2 \equiv (2\pi)^7L_P^8 \equiv 16\pi G_{10}$
relate the couplings $k$ and $k_0$ to the string coupling $g_s$, the string tension $\alpha^{\prime}=L_s^2$,
the Planck length $L_P$ and the Newton constant $G_{10}$.
The string-frame action is closely related:
\beqs
{\cal S}_{10,s}&=&\frac{1}{2k_0^2}\int \di^{10} x \sqrt{-g_{10,s}}e^{-2\Phi}\left(R_{10,s}+\frac{}{}\cdots\right)\,.
\eeqs

The action and metric in string frame are used in the calculation of the rectangular Wilson loops,
from 
which one extracts the linear quark-antiquark static potential. We hence write explicitly the relevant parts of the metric:
\beqs
\di s_{10,s}^2&=&
\left(\frac{}{}e^{4\phi/\sqrt{5}}\sin^2\xi+e^{\phi/\sqrt{5}} \cos^2\xi\right)^{1/2}e^{\frac{3}{2}\chi+\omega}
\left[\frac{}{}e^{{3}{}\chi+2\omega}\di x_{1,3}^2 +\di \r^2+\cdots\right]\,.
\label{Eq:stringframe}
\eeqs
We omit the rest of the background fields, expressions for which  can be found in the literature.
Starting from the Nambu-Goto action
\beqs
{\cal S}_{NG}&=&\frac{1}{2\pi\alpha^{\prime}}\int \di \sigma\di \tau \sqrt{-{\rm det}
(g_{s\,MN}\partial_{\alpha}X^M\partial_{\beta}X^N)}\,,
\eeqs
the string tension for the confining dual theories
 can be computed
it terms of the $g_{s\,MN}$ components of the string-frame metric in Eq.~(\ref{Eq:stringframe}),
to find  (see Ref.~\cite{Nunez:2009da} and references therein, for example):
\beqs
\sigma &\propto& \lim_{\r\rightarrow \r_o}g_{xx}
\,=\,
 \lim_{\r\rightarrow \r_o}
e^{3 {\cal A}}\left(\frac{}{}e^{4\phi/\sqrt{5}}\sin^2\xi+e^{\phi/\sqrt{5}} \cos^2\xi\right)^{1/2}\\
&=&e^{\frac{9}{2}\chi_I+3\omega_I}\left(\frac{}{}e^{4\phi_I/\sqrt{5}}\sin^2\xi+e^{\phi_I/\sqrt{5}} \cos^2\xi\right)^{1/2}
\label{Eq:string}
\eeqs
where we used the relation ${\cal A}=\frac{3}{2}\chi+\omega$.
This expression agrees  with Sec.~4.4.1 of Ref.~\cite{Elander:2013jqa},
and explicitly displays the dependence on the angle $\xi$ of solutions with non-vanishing $\phi$.

While a general configuration of the string may have a non-trivial profile in $\xi$,
and is determined by solving a non-trivial coupled system,
we restrict our attention to the two cases with fixed $\xi=0\,,\,\frac{\pi}{2}$.
The two terms inside the bracket in Eq.~(\ref{Eq:string}) are both positive definite, but
there is one important difference between the cases $\phi_I<0$ and $\phi_I>0$.
For positive values of $\phi_I>0$, $\sigma$ is  minimised (and so is the energy of the string
for asymptotically large spatial separation) for $\xi=0$,
which is the equator of the $4$-sphere.
Conversely, for $\phi_I<0$ the energy of a configuration with large spatial configuration
 and string tension are both minimised for $\xi=\pm\frac{\pi}{2}$, 
the poles of the $4$-sphere.
This substantial, important difference suggests the existence of a
phase transition in the system of probe strings, for which the equilibrium configuration 
 as a function of $\phi_I$ changes abruptly at $\phi_I=0$.
 We leave  to future investigations the study of the possible implications of such behaviour.


\section{Comparison with the literature}
\label{Sec:Brower}

\begin{table}
\caption{Numerical results for the spectrum of masses $M_n$
of fluctuations computed in Ref.~\cite{Brower:2000rp} (corresponding to $\phi_I = 0$).
In respect to the original source, besides showing the mass,
rather than the mass square, we normalise the spectrum 
to the lightest spin-$2$ particle---labelled $T_4$.
The labels of the states are explained in Ref.~\cite{Brower:2000rp},
and we put in bracket the spin of the states, highlighting the degeneracies.
All states are $SO(5)$ singlets.
We highlight in red, green and blue, respectively, the $SO(5)$ singlet states that appear also in our own results
(see Table~\ref{Fig:US} for comparison).
The three heavier towers are not part of the spectrum we compute.
}
\label{Fig:Brower}
\tiny
\begin{center}
\begin{tabular}{|c|c|c|c|c|c|}
\hline\hline
$T_4$ & $V_4$ & $S_4$ & $N_4$ & $M_4$ & $L_4$ \cr
$(2/1/0)$ & $(1/0)$ & $(0)$ & $(1/0)$ & $(1/1)$ & (0) \cr
\hline
 {\color{red} 1.00} &  {\color{green} 1.20} & {\color{blue} 0.58} & 1.55 & 1.94 & 2.28 \\
 {\color{red} 1.59} & {\color{green} 1.81} & {\color{blue} 1.46} & 2.23 & 2.55 & 2.93 \\
 {\color{red} 2.15} & {\color{green} 2.39} & {\color{blue} 2.07} & 2.83 & 3.14 & 3.54 \\
 {\color{red} 2.71} & {\color{green} 2.96} & {\color{blue} 2.65} & 3.42 & 3.71 & 4.14 \\
 {\color{red} 3.27} & {\color{green} 3.52} & {\color{blue} 3.22} & 3.99 & 4.28 & 4.72 \\
 {\color{red} 3.83} & {\color{green} 4.08} & {\color{blue} 3.78} & 4.56 & 4.84 & 5.30 \\
 {\color{red} 4.38} & {\color{green} 4.63} & {\color{blue} 4.34} & 5.12 & 5.41 & 5.87 \\
 {\color{red} 4.94} & {\color{green} 5.19} & {\color{blue} 4.90} & 5.68 & 5.97 & 6.43 \\
 {\color{red} 5.49} & {\color{green} 5.74} & {\color{blue} 5.46} & 6.24 & 6.52 & 7.00 \\
 {\color{red} 6.05} & {\color{green} 6.29} & {\color{blue} 6.02} & 6.8 & 7.08 & 7.56 \\
\hline\hline
\end{tabular}
\end{center}
\end{table}

\begin{table}
\caption{Numerical results for the spectrum of masses $M_n$, normalised to the lightest tensor,
of fluctuations computed in this paper, in the case of $SO(5)$ symmetry,
which is 
recovered when $\phi_I = 0$. 
We put in bracket the spin $(J=0,1,2)$ of the states
and we explicitly indicate the $SO(5)$ multiplets $\sim1,5,10,14$.
We highlight in red, green and blue, respectively, the $SO(5)$ singlet states that appear also in our own results
(see Table~\ref{Fig:Brower} for comparison).
Our original contribution is in the non-singlet states.
}
\label{Fig:US}
\tiny
\begin{center}
\begin{tabular}{|c|c|c|c|c|c|c|c|c|c|c|c|c|}
\hline\hline
$g_{\mu\nu}$ & $\chi_{\mu}$ & $\omega_{\mu}$ & $ \omega, \chi$  
& $\omega_6$ &  
$\phi, \pi^{\hat{A}}, s^{\tilde{A}}$ 
& $A_6^{\hat{A}}, A_6^{\bar{A}}$ & $A_7^{\hat{A}}, A_7^{\bar{A}}$ & $\varphi_\alpha$ & $A_{\mu}^{\hat{A}}, A_{\mu}^{\bar{A}}$
& $S_{67\mu\,\alpha}$ 
& $B_{\mu\nu\,\alpha}$ & $X_{\mu\,\alpha}$  \cr
$(2)$ & $(1)$ & $(1)$ & $(0)$ & $(0)$ & $(0)$ & $(0)$ & $(0)$ & $(0)$ & $(1)$ & $(1)$ & $(1)$ & $(1)$ \cr
$\sim 1$ & $\sim 1$ & $\sim 1$ & $\sim 1$ & $\sim 1$ & $\sim 14$ & $\sim 10$ & $\sim 10$ & $\sim 5$ & $\sim 10$ & $\sim 5$ & $\sim 5$ & $\sim 5$  \cr
\hline
& & & & &  & & & & & &  & \cr 
& & & {\color{blue} 0.58} & & 0.57 & & 0.77 & 0.64 & 0.77 & & & 0.64 \cr 
& & & & &  & & & & & & &  \cr 
  {\color{red} 1.00} & {\color{green} 1.20} & {\color{red} 1.00} & {\color{red} 1.00}& {\color{green} 1.20} & 1.11 & 1.00 & 1.35 & 1.30 & 1.35 & 1.03 & 1.03 & 1.30 \cr 
 & & & {\color{blue} 1.46}  & & & & & & & & &  \cr 
 {\color{red} 1.59} & {\color{green} 1.81} & {\color{red} 1.59} & {\color{red} 1.59} & {\color{green} 1.81} & 1.66 & 1.59 & 1.90 & 1.87 & 1.90 & 1.60 & 1.60 & 1.87 \cr 
 & & & {\color{blue} 2.07}& & & & & & & & &  \cr 
   {\color{red} 2.15} & {\color{green} 2.39} & {\color{red} 2.15} & {\color{red} 2.15}& {\color{green} 2.39} & 2.21 & 2.15 & 2.46 & 2.44 & 2.46 & 2.17 & 2.17 &  2.44 \cr 
 & & & {\color{blue} 2.65} & & & & & & & & & \cr 
 {\color{red} 2.71} & {\color{green} 2.96} & {\color{red} 2.71} & {\color{red} 2.71} & {\color{green} 2.96} & 2.75 & 2.71 & 3.01  & 2.99 & 3.01 & 2.72 & 2.72 & 2.99  \cr 
 & & & {\color{blue} 3.22} & & & & & & & & &  \cr 
  {\color{red} 3.27} & {\color{green} 3.52} & {\color{red} 3.27} & {\color{red} 3.27} & {\color{green} 3.52} & 3.30 & 3.27 & 3.57 & 3.55 & 3.57 & 3.28 & 3.28 &  3.55  \cr 
 & & & {\color{blue} 3.78} & & & & & & & & &  \cr 
 {\color{red} 3.83} & {\color{green} 4.08} & {\color{red} 3.83} & {\color{red} 3.83} & {\color{green} 4.08} & 3.85  & 3.83 & 4.12 & 4.10 & 4.12 & 3.83 & 3.83 & 4.10  \cr 
& & & {\color{blue} 4.34} & & & & & & & & &  \cr 
 {\color{red} 4.38} & {\color{green} 4.63} & {\color{red} 4.38} & {\color{red} 4.38} & {\color{green} 4.63} & 4.40 & 4.38 & 4.67 & 4.66 & 4.67 & 4.39 & 4.39 & 4.66  \cr
 & & & {\color{blue} 4.90} & & & & & & & & &  \cr
 {\color{red} 4.93} & {\color{green} 5.18} & {\color{red} 4.93} & {\color{red} 4.93} & {\color{green} 5.19} & 4.95 & 4.93 & 5.22 & 5.21 & 5.22 & 4.94 & 4.94 & 5.21  \cr
 & & & {\color{blue} 5.45} & & & & & & & & &  \cr
 {\color{red} 5.48} & {\color{green} 5.74} & {\color{red} 5.48} & {\color{red} 5.48} & {\color{green} 5.74} & 5.50 & 5.49 & 5.77 & 5.76 & 5.77 & 5.49 & 5.49 & 5.76  \cr
 & & & {\color{blue} 6.01} & & & & & & & & &  \cr
 {\color{red} 6.03} & {\color{green} 6.29} & {\color{red} 6.03} & {\color{red} 6.03} & {\color{green} 6.29} & 6.05 & 6.04 & 6.32 & 6.31 & 6.32 & 6.04 & 6.04 & 6.31  \cr
 & & & & & & & & & & & &  \cr
\hline\hline
\end{tabular}
\end{center}
\end{table}

This Appendix is devoted to comparing our results for the limiting case $\phi_I=0 = \phi$, which represents the $SO(5)$-invariant theory, to the results reported in Ref.~\cite{Brower:2000rp}. In this case, the background solution is known analytically ($A = \frac{5}{2} \chi + \omega$):
\begin{align}
	\omega = - \frac{1}{2} \log \left( \tanh \left( \frac{3\rho}{2}\right) \right) \,, \ \ \
	\chi = \frac{1}{9} \left[ \log \left( \sinh \left( 3\rho \right) \right) + 2\log \left( \tanh \left( \frac{3\rho}{2} \right) \right) \right] \,,
\end{align}
In order to compute the spectrum, we notice that we must use a different parametrisation of the $14$ of $SO(5)$ than that of the scalar manifold described in Sec.~\ref{Sec:breaking}.
We start by reproducing in Table~\ref{Fig:Brower} the numerical results in Table~2 of Ref.~\cite{Brower:2000rp},
while we report our own results in Table~\ref{Fig:US}.
For the three lightest towers of singlet states, we find agreement within 
$1\%$, which is a reasonable estimate of the numerical uncertainty of our own analysis.
The calculation in Ref.~\cite{Brower:2000rp} considers three additional towers, that do not appear in our analysis as they are heavier. For the $SO(5)$-invariant background, the main contribution from this study
 is the spectrum on $SO(5)$ non-singlets, the lightest states of which 
have masses lower than some of the singlets.


\end{document}